\documentclass[a4paper,11pt]{article}
\usepackage{amsmath,amssymb,amsthm}
\pdfoutput=1 

\usepackage{jinstpub} 
\usepackage{colortbl}
\usepackage{textcomp}
\usepackage{mathtools}
\usepackage{eurosym}

\usepackage{hyperref}

\title{Difference of measured proton and He3 EDMs: a reduced systematics test of T-reversal invariance}
\author{Richard M. Talman}
\affiliation{Laboratory for Elementary-Particle Physics, Cornell University, Ithaca, NY, USA}

\emailAdd{richard.talman@cornell.edu}

\abstract{
The upper limit on (time reversal symmetry T-violating) permanent hadron electric dipole moments (EDMs) 
is the PSI neutron EDM value; $d_n = (0.0\pm1.1_{\rm stat}\pm0.2_{\rm sys}\times10^{-26})\,e$\,cm. This paper 
describes an experiment to be performed at a BNL-proposed CLIP project which is to be capable of producing 
intense polarized beams of protons, $p$, helions (He${}^3$ nuclei), h, and other isotopes. 
  The EDM prototype ring PTR (proposed at COSY Lab, Juelich) is expected to measure individual particle 
EDMs (for example ${\rm EDM\_p}$ for the proton) using simultaneous counter-rotating polarized proton beams, 
with statistical error $\pm10^{-30}$e.cm after one year running time, four orders of magnitude less than 
the PSI neutron EDM upper limit, and with comparable systematic error.
  A composite particle, the helion faces T-symmetry constraints more challenging than the proton. Any 
measurably large value of $$\Delta={\rm EDM}_h-{\rm EDM}_p,$$ the difference of helion and proton EDMs, 
would represent BSM physics.
  The plan is to replicate PTR at BNL. The dominant systematic error would be canceled two ways, both 
made possible by phase-locking ``doubly-magic'' 38.6\,MeV proton and 39.2\,MeV helion spin tunes. This 
stabilizes their MDM-induced in-plane precessions, without affecting their EDM-induced out-of-plane 
precessions. The dominant systematic error would therefore cancel in the meaurement of $\Delta$ in a 
fixed field configuration.
  Another systematic error cancellation will come from averaging runs for which both magnetic field and 
beam circulation directions are reversed. Precise magnetic field reversal is made possible by the 
reproducible absolute frequency phase-locking over long runs to eliminate the need for (impractically precise) 
magnetic field measurement. Risk of EDM measurement failure is discussed in a final appendix.
}

\begin{document}
\maketitle
\flushbottom

\section{Introduction}
\emph{This paper is a companion to three other papers, of which only the first is published at present
\ \cite{RT-ICFA}\cite{RT-Positron-q}\cite{RT-LHC2}. Quite long sections appear, modified slightly at most, 
in these papers.}

References to many historically important storage ring EDM papers, including the 2004, Farley et al., original paper\cite{Farley},
are given in reference\ \cite{CYR}. 

The current upper limit on (time reversal symmetry T-violating) permanent nuclear electric dipole moments (EDMs) is 
the PSI neutron EDM value\ \cite{PSI-neutron}; $d_n = (0.0\pm1.1_{\rm stat}\pm0.2_{\rm sys}\times10^{-26})\,e$\,cm

A series of previous papers has shown how a ``prototype ring'' (PTR)\ \cite{PTR-Bad-Honnef}  
can be used to model the performance
of precision experiments of electric (EDM) and magnetic (MDM) dipole moments, with the aim of 
detecting beyond the standard model (BSM) physics.  Until now these proposals have required the use 
of a ``beam accumulator'' ring (BA) to produce the required $10^{10}$ minimum total number of stored 
countercirculating particles to perform the operations needed to measure, for example, ${\rm EDM_p}$ for the
proton or ${\rm EDM_h}$ for the helion individually.  More importantly, it is what is needed to measure 
the ${\rm EDM_p-EDM_h}$ difference, a combination from which the dominant systematic EDM error cancels. 

This paper emphasizes the dramatic improvement in EDM and MDM measurement capability made possible
by the huge increase in isotope source currents the light ion linac (LIL) component of the proposed 
Center for Linac Isotope Project) CLIP project promises.  The paper describes how this increase in source 
current will render bunch accumulation in a BA ring unnecessary, reducing both cost and complexity.

Using (admittedly optimistic) assumptions, the counting statistics EDM error after continuous running for 
one year is expected to be $\pm\,10^{-30}\,e$\,cm. This is four orders of magnitude less than the previous 
upper BSM physics limit derived from measurement of the neutron EDM (as quoted in the abstract)
which includes both systematic and counting statistics errors).  

As originally intended, a major goal for PTR was to measure the proton EDM, in order to guide the estimation 
of and minimization of systematics of a full scale, all-electric EDM ring.  To reduce the PTR cost
(to k\texteuro 16,600 in 2020 euros) many cost saving design measures were taken; led by room temperature 
(non-cryogenic) operation and the use of inexpensive magnetic shielding.  

\begin{figure}[hbt]
\centering
\includegraphics[scale=0.35]{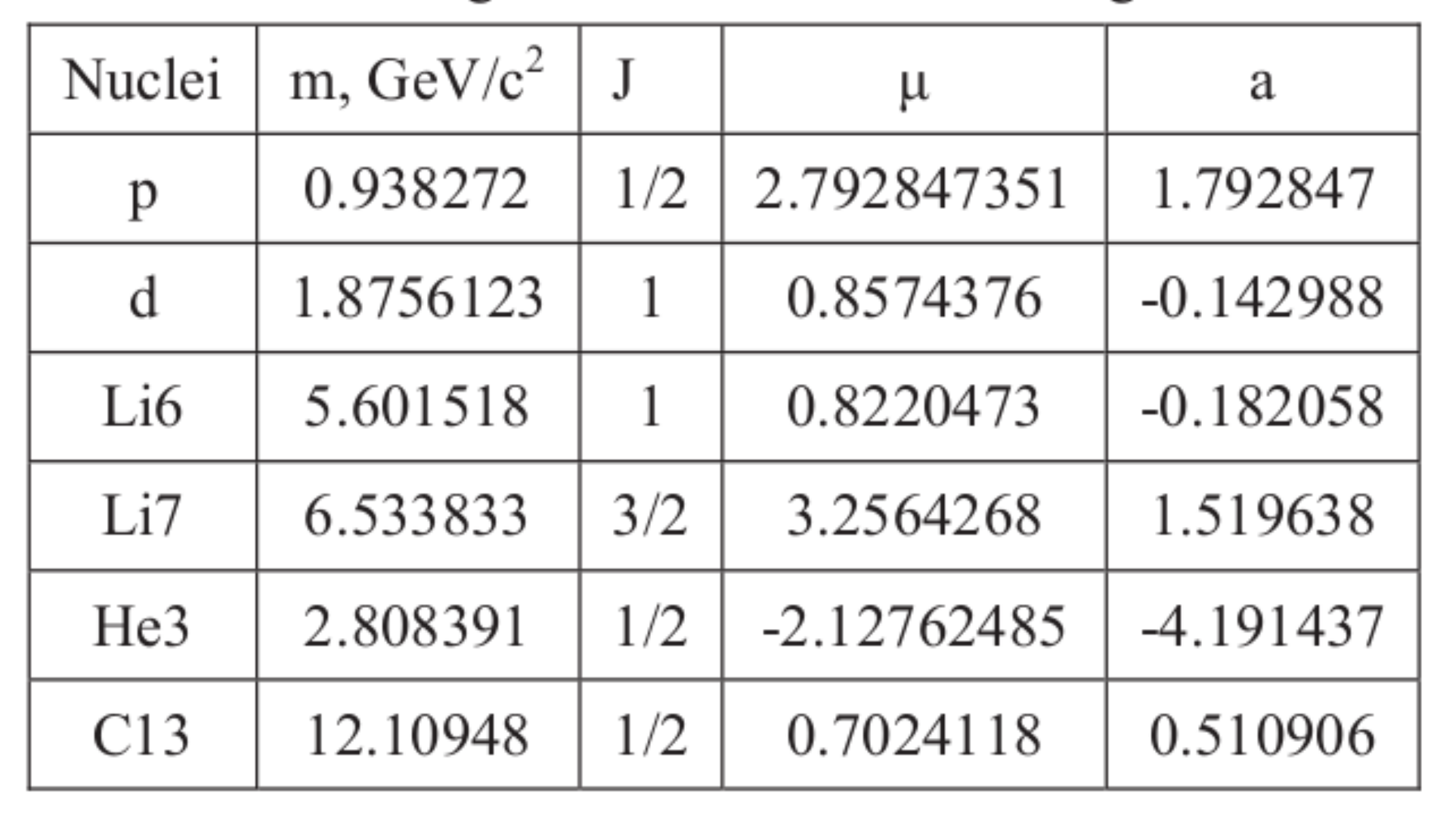}
\caption{\label{fig:KoopParameters}(Exhibited here as the photocopy of the original table to preserve chronology)
this table provides parameters for particle types considered by Koop\ \cite{Koop-different-particles} as possible candidates
for EDM determination. Koop's symbol ``a''\ for anomalous moment is replaced by $G$ in this paper.
A necessary, but not sufficient condition, for a pair of entries to be doubly-magic, is for their 
$G$-values (here $a$-values) to have opposite sign, as is true for $p$ and He3.  Though true also 
for $p$ and $d$ no doubly-magic pairing has, as yet, been found.  The same comment applies to the 
Li6,Li7 and He3,C13 pairings.  Seeming exceptions to this ``opposite sign'' rule are the all-electric
233\,MeV proton and 14.5\,MeV electron EDM rings.  This is discussed in detail in 
Appendix~\ref{sec:FixedPoint}.}
\end{figure}

Using counter-circulating beams of different particle type was first suggested by 
Koop\ \cite{Koop-different-particles}.  Parameters for the baryons Koop discussed are given in 
Figure~\ref{fig:KoopParameters}.
Subsequent studies by the CPEDM group\ \cite{APS-FSU} have shown that many systematic errors can be reduced 
by phase-locked loop, frequency domain exploitation of ``mutual co-magnetometry''. This is epitomized by 
the doubly-magic proton-helion combination emphasized in the present paper.  Because the proposed ring
has electric and magnetic fields superimposed, the term  ``mutual co-magnetometry'' needs to be
interpreted as including the phrase ``in the separate rest frames of the two beams''.  As this usage is
novel, the ``mutual'' modifier is included wherever ``co-magnetometry'' appears in the sequel.

Significant uncertainties remain concerning the systematic errors associated with storage ring operation employing 
superimposed electric and magnetic bending.  Though investigation of these systematic errors will be aided greatly 
by the above mentioned four order of magnitude improvement in counting statistics, the systematic 
${\rm EDM}_p-{\rm EDM}_h$ difference can only be estimated.  Conservatively, the systematic error in individual EDM
measurement will initially be less than $\pm\,10^{-27}\,e$\,cm;\ \cite{TaharCarli} and subject to continuing reduction from 
this value, exploiting the available statistical accuracy. This subject is discussed further in 
Section~\ref{sec:ErrorEstimation}.

Doubly-frozen kinematic conditions can always be found for simultaneous counter-circulation of
arbitrarily different pairs of charged particle types in a storage ring having superimposed electric
and magnetic bending.  It is always possible for at least one or the other of the beams to be said to 
be ``magic'' by virtue of having its spins ``globally frozen''---spins all pointing exactly forward, for example.  
In general, though, spins of the other beam can at most be ``pseudo-frozen''---meaning that, viewed stroboscopically 
at any location the spins ``appear'' to be frozen because their spin tunes are rational fractions, such that the spins
always return to the same orientation at the observation point. For certain spin control procedures (those involving 
spin-phase locking capability) the pseudo-frozen feature is sufficient, but global freezing is necessary for EDM 
precession to increase monotonically for at least several minutes to enable accurate EDM measurement. 

The validity of statements in the previous paragraph (many of which are far from obvious \emph{a priori,})
and similar statements to follow, is based on carefully finding all real roots of the quartic 
Eq.~(\ref{eq:AbbrevFieldStrengths.3-rev}) and quadratic Eqs.~(\ref{eq:Wollnik-TM-params}).  

Because of technical limitations, for example the electric or magnetic fields may be unachievably strong,
or polarimetry may not be efficient enough, or have insufficient analyzing power, conditions that are achievable 
in principle may not be  experimentally practical.

Almost without exception a ring with superimposed electric and magnetic bending is required to achieve the frozen
spins needed for mutual co-magnetometry.  With one of the beam polarizations frozen the magnetic moment of the particles
of the other beam particles can be determined by measuring the spin tune of that beam.

Two leading exceptions that do not require superimposed electric and magnetic fields are 14.5 MeV electrons 
(of either sign) or 233\,MeV protons (of either sign) in purely electric rings\footnote{Even these exceptions are 
disputed in Appendix~C.}.

For rings with superimposed electric and magnetic bending only two ``doubly magic'' cases are known, but there may be 
others involving higher Z nuclei.  The most promising of the known cases has doubly magic protons and He3 nuclei 
(of curiously close kinetic energies) both globally frozen; the doubled helion charge compensates for the
approximately doubled momentum.  Furthermore, counter-circulating in a ring of convenient and
inexpensive size, excellent polarimetry is also available.  A less promising doubly magic example has 
counter-circulating protons and positrons.  This case is less promising because the momenta are large enough to 
require an expensive ring, and because positron (and electron) polarimetry is not at all robust.

Doubly frozen pairings (with one or the other, but not both,  magic) of proton and deuteron beams, $pp$, $pd$, 
$dd$, and $dp$, are of special interest, both because their EDMs are of pre-eminent physical importance, and 
because these pairings can be investigated using well-established equipment and technology, most of which is 
currently available.  

Along with reproducible accelerating cavity frequency, doubly-frozen, the spin tunes (whose parameters other than 
$\beta$, $E_0$ and $B_0$  are constants of nature) can be stabilized to precision (inversely) proportional to run 
length.  \emph{This will permit precise electric and/or magnetic field reversal resettability to be performed.  The
reason this is important is that systematic errors can then be reduced by averaging over reversed configurations.}

In this way ``mutual co-magnetometry'' can exploit the ``perfect'' gyroscopic stabilization provided by the circulating 
MDMs themselves.  A conceptual design for a storage ring complex capable of exploiting this technology is presented 
below.  It is based on a reconfiguration of the COSY storage ring complex in Juelich, Germany, which has been the 
storage ring at which most of the required spin control technology has been developed. 

Electric/magnetic frozen spin storage rings will also be capable of measuring isotope MDM 
values with unprecedented precision. Though such rings can be built inexpensively, they are not very flexible.
For this reason they need to be developed using existing facilities that match their requirements\ \cite{Kolya-Snowmass}.

\begin{figure}[hbt]
\centering
\includegraphics[scale=0.28]{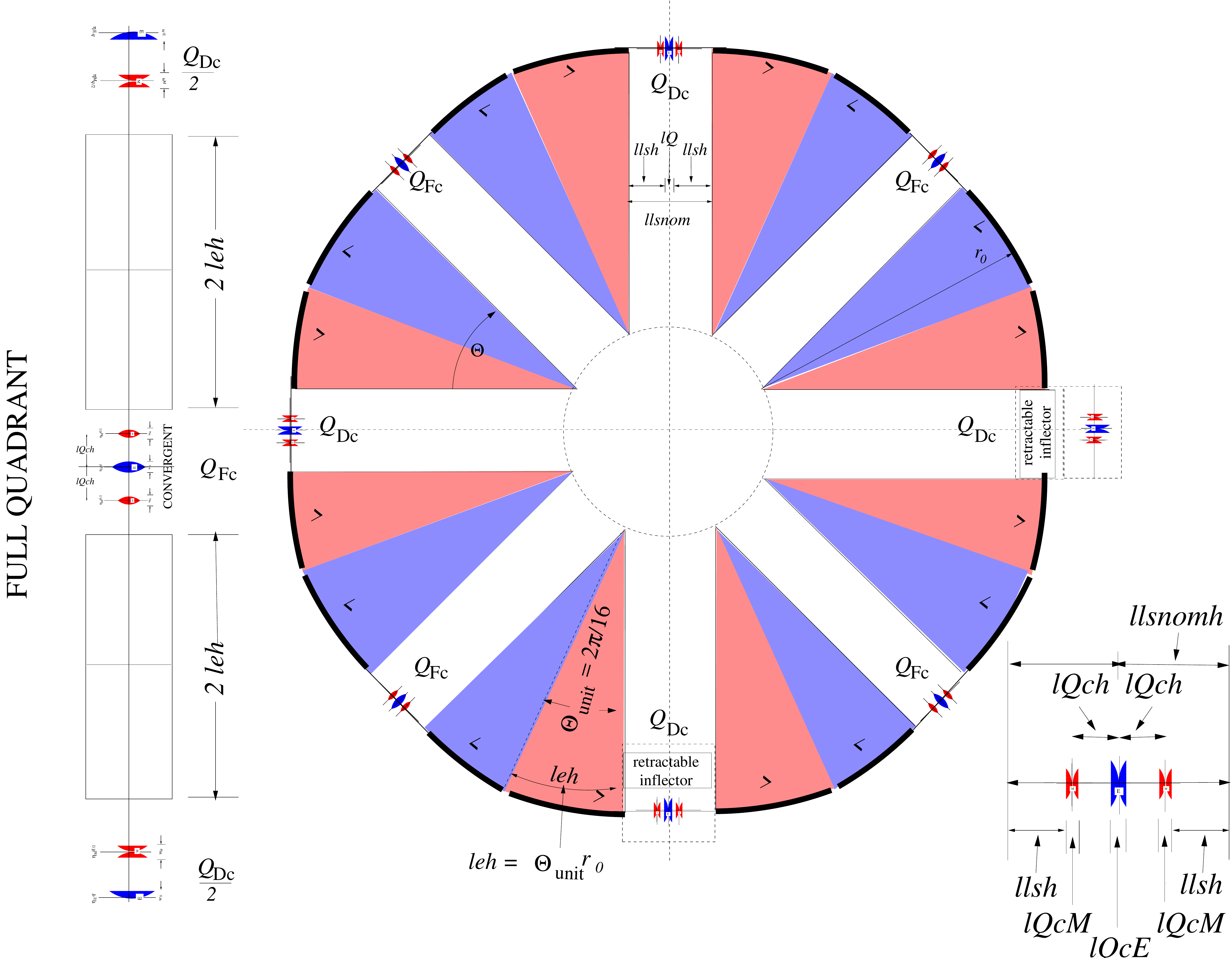}
\caption{\label{fig:PTR-layout-Toroidal8_102p2-mod}Lattice layouts for PTR, the proposed 
``EDM prototype'' storage ring.  The single quadrant ``model space'' figure on the left
can be conformally mapped to one quadrant of the ``physical space'' figure on the right.
Straight lines on the left map to perfect circles on the right.  Sector bends, distinuished
by blue and red color, though identical in the nominal design, are separately powered.
\emph{An extreme ``worst case'' example, investigated in Appendix~\ref{sec:Compromise-bending} has 
blue sectors representing pure electric bending and red sectors representing pure magnetic 
bending.  In this case, for the deuteron spin tune to cancel, the magnetic an electric arc lengths
would have to be adjusted.}  A ``compromise quadrupole
triplet'' is shown in the inset on the right.  The magnetic quad doublet and electric quad singlet
provide very weak ``optical trim correction''.  Their strengths are adjusted individually in 
the same $\eta_M,\eta_E$ ratio as the ring bend strengths.  Ideally these corrections would be
exactly superimposed, which would be mechanically challenging.  To thin element approximation 
this superposition is equivalent.  See Appendix~\ref{sec:Compromise}.}
\end{figure}

\begin{figure}[hbt]
\centering
\includegraphics[scale=0.4]{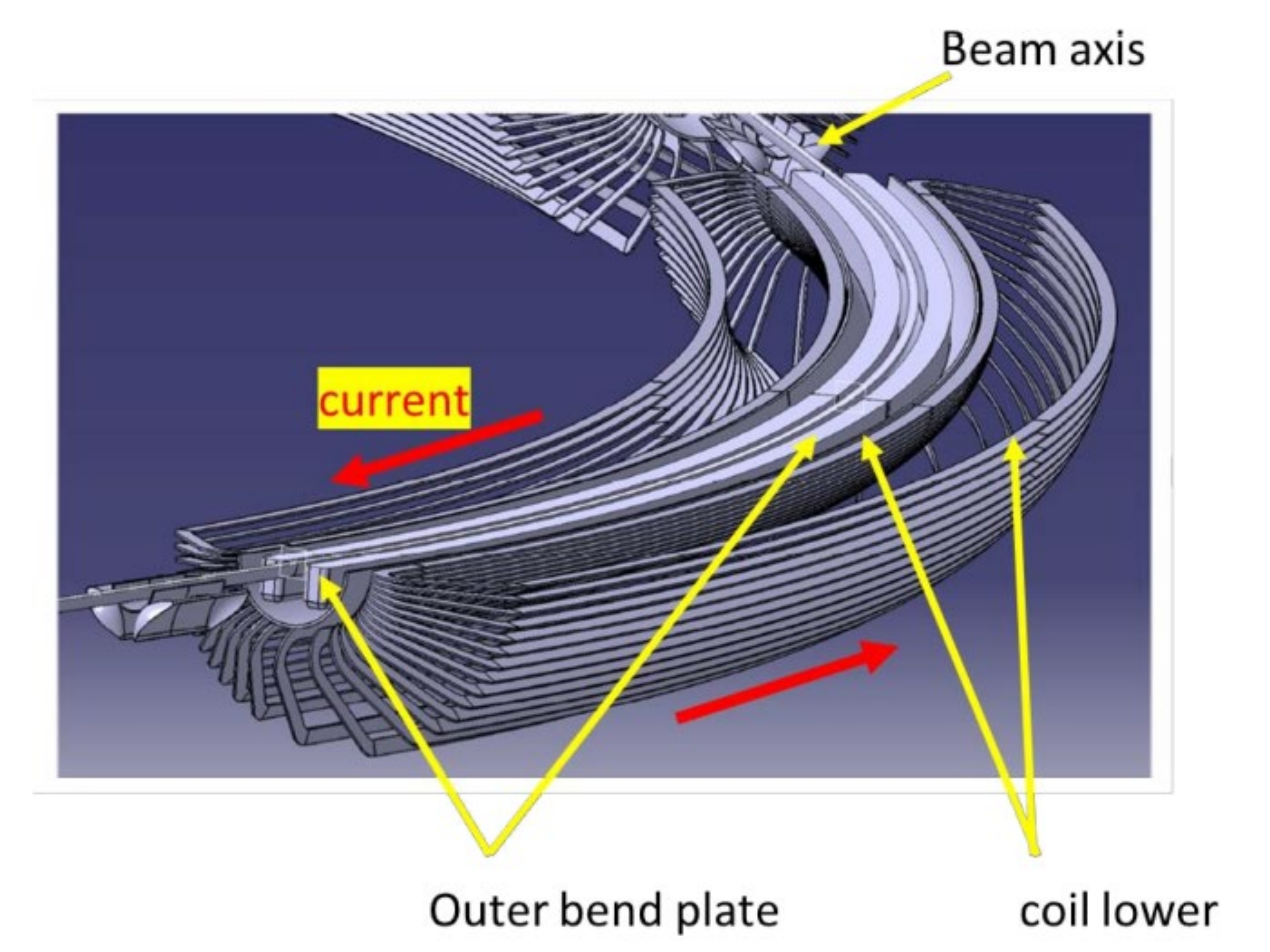}
\caption{\label{fig:SectorPerspective}Copied from reference~\ \cite{CYR}, a perspective mock-up of 
one sector of PTR, the superimposed E/M prototype ring.
Two ``short-circuited ends''cos\,$\theta$-dipoles surround the beam tube, within 
which the capacitor plates are accommodated.}
\end{figure}

Because the site for such an accelerator complex has not been formally selected, the ring circumference $\mathcal{C}$, 
nominally 100\,m, has been left left open in the computer codes producing the present report.  To maintain this 
flexibility the lattice design described can be expressed in ``scale invariant'' form\ \cite{CYR}.  But, to produce the 
numerical evaluations needed for concrete visualization, the PTR circumference is here taken to be 
$\mathcal{C}=102.2$\,m, as required to fit PTR, along with a bunch accumulator'' (BA) (required for rebunching 
the Juelich cyclotron proton or deuteron 25\,MHz cw-bunched cyclotron beams) into the COSY, Juelich beam hall.
A layout diagram for PTR is shown in Figure~\ref{fig:PTR-layout-Toroidal8_102p2-mod}.   A perspective view of one
octant is shown in Figure~\ref{fig:SectorPerspective}.   

\section{Testing time reversal symmetry}
\subsection{Electric dipole moment (EDM) measurement} 
The excess of matter over anti-matter in the present-day universe suggests 
the existence of violations of P and T symmetries substantially greater than are required by the 
``standard model'' of elementary particle physics. This provides the leading motivation for 
measuring EDMs, which must vanish if P and T symmetries are respected. 

Ultimate proton EDM measurement has been thought to require simultaneously counter-circulating, 
frozen-spin 233\,MeV protons in a ``large'' (depending on achievable electric field, and ranging 
`in size up to 800\,m circumference) nominal all-electric 
storage ring.  A step toward this goal is the ``small'' ($\sim$100\,m circumference) PTR ring  with 
superimposed electric and magnetic bending fields. Simultaneously counter-circulating frozen spin 
proton beams is one of the  ultimate features that can be demonstrated in PTR. 

The need for storage rings with predominantly electric bending is based on the
requirement to ``freeze the spins'' of stored particle beams. This is the first step
toward the measurement of electric dipole moments (EDMs) of the circulating
particles, as well as for other symmetry violation investigations.
Electrons (of either sign) along with the proton and triton baryons, are 
fundamental particles whose electric EDMs are potentially measurable in 
all-electric frozen spin storage rings. With superimposed magnetic bending
other particles, especially deuterons and helions (He3 nuclei), can have 
frozen spins.

In all cases the EDM values are so small that suppression of systematic errors
makes it essential for the storage ring to support simultaneous counter-circulating
beams.  With superimposed electric and magnetic bending, constructive bending superposition 
in one direction implies destructive superposition in the other direction. By itself, this 
does not preclude the possibility of simultaneously counter-circulating beams---but it does 
require major difference between the beams, either in particle type or momentum.

In contrast to a ``single particle table-top trap'', the essential feature of a storage ring 
``trap'' is that some $10^{10}$ particles can have their spins aligned in a polarized stored beam.
As demonstrated by Hempelmann et al. at the COSY lab\ \cite{Hempelmann},
this is a number of  particles large enough for the beam polarization to be detected externally, 
and fed back to permit external control of the beam polarization. 
Though the table large enough for any such a ``storage ring trap'' would be quite large, the 
level of achievable spin control, though classical, not quantum mechanical, can be comparable
to the control of one or a small number of polarized particles in a low energy trap.
Precision spin tune control  at COSY has been pioneered by  Eversmann et al.\ \cite{Eversmann}.

In true ``globally frozen spin'' operation the bunch polarizations remain globally frozen, for example parallel 
to the circulating beam momentum vector. For all-electric bending 
this fixes all kinematic quantities uniquely. But with superimposed electric and magnetic bending, 
the frozen spin condition can be met over bands of E/B ratios, specific to particle type. Furthermore, 
some experiments can be performed with ``pseudo-frozen spins'' that appear frozen locally, but 
not globally.

The leading observable effect of a static particle EDM would be an ``out-of-plane'' 
spin precession, orthogonal to MDM-induced ``in-plane'' horizontal spin precession caused by 
whatever magnetic and/or electric fields cause the particle orbit to follow a sequence
of horizontal circular arcs. With standard model EDM predictions being much smaller than 
current experimental sensitivities, detection of any particle's non-zero EDM would 
signal discovery of New Physics. 

A  ``nominal experimental proton EDM detectability target'' has conventionally been 
defined to be $10^{-29}\,e$\,cm.  An EDM of this magnitude could help to account for the 
observed matter/antimatter asymmetry of our universe while, at the same time being 
plausibly (only one or two orders of magnitude) larger than existing standard model 
calculations. 

Currently the proton EDM upper limit (as inferred indirectly by measuring the
Hg atom EDM) is roughly $2.1\times10^{-25}e\,$cm.\ \cite{CYR}
The original goal for PTR was to serve as prototype for the nominal, full-scale, 
all-electric proton EDM ring.  It has been possible, however to make the PTR design flexible
enough to support quite accurate EDM determinations and to achieve other goals, such
as precision measurement of anomalous magnetic moments of light nuclear isotopes.

\subsection{Spin tune comparator measurement of light isotope MDMs and EDMs}
(Except for synchrotron radiation) first generation electron/positron colliders such as the Cornell Electron Storage 
Ring (CESR) have demonstrated CPT-symmetry, and robust backward/forward (bf)-symmetry.
\emph{But the present paper considers only storage rings in which particle spins can be globally frozen}. 
This rules out rings with pure magnetic bending---there are no candidates. One turns, therefore, 
to \emph{electric} rings.   Of special interest are 
the 232.8\,MeV ``nominal all-electric'' all-electric proton EDM ring, which has been under discussion for more than a decade.
Also of interest (especially at Wilson Lab) could be a (small) all-electric 14.50\,MeV $e^+/e^-$- storage ring in which 
globally-frozen spin positrons and electrons can counter-circulate simultaneously. 

For particles at rest ``co-magnetometry'' in low energy ``table-top particle traps'' has been essential. 
For example, Gabrielse\ \cite{Gabrielse-eEDM} has (with excellent justification) described the measurement of the 
electron magnetic moment (with 13 decimal point accuracy) as ``the standard model's greatest triumph'', based 
on the combination of its measurement to such high accuracy and on its agreement with theory to almost the 
same accuracy.  

Especially for the \emph{direct} measurement of EDMs, storage ring technology with beam pairs that can counter-circulate 
simultaneously in a storage ring with superimposed electric and magnetic bending is required.  In this context the term 
``mutual co-magnetometry'' can be used to apply to ``beam type pairings'' for which both beams have frozen spins.
Usually only one of the beams can be ``globally'' frozen or ``magic'', however.  

Surprisingly, \emph{counter-circulating beams can  follow  accurately identical orbits, even when the bending 
is performed by superimposed electric and magnetic fields.} Here we show how this can enable the precise comparison of 
the spin tunes, and hence MDMs, of  different elementary particle types---hence the name 
``spin tune comparator''.  Both beams are assumed polarized and one of the beam type's MDM may be much 
better known than the other; for example $e^+$ better than $p$, or $e^-$ better than $e^+$. 

To the extent both MDMs are already known, the known ratio can be used to adjust a compensating element 
to match this ratio.  Though EDM measurement is scarcely mentioned in this spin tune comparator section, 
this remains the goal and \emph{spin tune comparator operation will permit Koop-wheel rotation reversal with 
almost arbitrarily high, frequency domain, accuracy. Within each run a precision EDM measurement (including systematic 
error contribution) can then be obtained.} Koop\ \cite{Koop-SpinWheel} argues (persuasively) that intentional roll of 
the Koop wheel can improve spin coherence duration.  Extraction of the EDM value then relies on linear
interpolation of data taken at negative and positive roll rates to cancel the torque applied to drive the ``roll''.

All of the most important measurement methods for ultimate EDM precision can be developed in the PTR ring. 
Furthermore, one can be more relaxed about parameter tolerance specifications in a small ``inexpensive'' 
prototype than in an eight times larger eventual ``ideal'' ring.  One of the tasks for PTR will be to 
investigate how tight the tolerances have to be, and how easily they can be met.
In this spirit, the present section derives quite tight tolerances to be aimed for (but not necessarily achieved) 
for measuring EDMs in a small ring with accuracy rivaling what can be achieved in the full scale ring.
This is work in progress. Unresolved technical issues are discussed in other sections.

Especially for the \emph{direct} measurement of EDMs, storage ring technology with beam pairs that can counter-circulate 
simultaneously in a storage ring with superimposed electric and magnetic bending is required.  In this context the term 
``mutual co-magnetometry'' can be used to apply to ``beam type pairings'' for which both beams have frozen spins.
Usually only one of the beams can be ``globally'' frozen or ``magic'', however.  

Compared to the electron, MDM accuracies of all other particles
are inferior by three orders of magnitude or more. 
A task for a spin-tune-comparator can be to
``transfer'' some of the electron's precision to the
measurement of other magnetic dipole moments. 
For example, the
positron EDM could perhaps be determined to almost the current 
accuracy of the electron's, and the (presumed) equality 
(in magnitude) of electron and positron MDM's could be improved
by several orders of magnitude.

Another motivation for developing spin tune comparison capability is to study 
beam-based self-compensation and self-calibration.  Accurately known counter-circulating 
MDM's can be used to cancel unknown differences between electric and
magnetic bending field shapes. ``Introspectron'' might be an apt task for such a 
ring---the purpose then being to investigate its own self-consistency. 
This is not without pure physics interest. With spins always a bit mysterious, 
confirmation (or not) to high accuracy, that CW and CCW spin tunes are identical could 
be informative. This possibility is pursued briefly in Appendix~E, which suggests 
investigation of the Aharonov-Anandan phase.  One supposes that any gravitational 
phase shift will be opposite for CW and CCW circulation.

The final section contains currently unresolved technical issues, the most important of which
concerns the achievability of strong electric fields surrounded by weak magnetic fields. 
This is only an issue for the small prototype ring PTR.  For an eight times larger ring, for
example situated in the AGS or Wilson lab tunnel, electric field strengths much in excess of 
a million volts per meter are not required (at least for the $p,h$ EDM difference measurement). 

\subsection{General relativistic influence on spin precession}
Various authors\ \cite{Obukhov:2016vvk}\cite{Silenko:2004ad}\cite{OrlovFlanaganSemertzidis}
have pointed out that general relativity (GR) introduces
effects that could be measurable at the precision level promised by proposed EDM rings. A
paper by L\' aszl\'o and Zimbor\' as\ \cite{Laszlo-Zimboras} has 
calculated the GR influence on storage rings designed for
MDM measurement, such as the PTR ring under discussion here.

The nominal EDM value can be compared to a general relativistic (GR) 
out-of-plane precession effect, mimicking an EDM of the order of  $10^{-28}\,e$\,cm,
associated with the storage ring's orientation in the earth's magnetic field. 
Depending on storage
ring details, this reliably calculable ``background precession'' will provide a 
``standard candle of convenient magnitude'' calibration of any EDM 
measurement.  

Other GR references include
\ \cite{Silenko:2006er}\cite{Silenko:2015jqa}\cite{Khriplovich:1997ni}\cite{Pomeransky:2000pb}.
These authors have calculated the GR effect of the earth's gravitational 
field on the experiment referred to above as the nominal all-electric version of EDM
measurement. They show that the GR effect mimics the EDM effect with a
magnitude such that, if mistakenly ascribed to the proton EDM, would 
produce a spurious proton EDM value of $3\times10^{-28}\,e$\,cm. 
This is about thirty times greater than the precision anticipated for
the ring and not inconsistent with the Orlov, Flanagan,
and Semertzidis estimate\ \cite{OrlovFlanaganSemertzidis}.

Interpretation of gravitational effect as Aharonov-Anandan spin tune shift is discussed in 
Appendix~\ref{sec:Aharonov-Anandan}.

\section{Mutual co-magnetometry provided by common orbits in the same ring}
\subsection{Fractional bending coefficients}
To aid the discussion of rings with superimposed electric and magnetic bending, (to be abbreviated as 
``EM-superposition'') fractional bending coefficients $\eta_E$ and $\eta_m$ can be defined by
\begin{equation}
\eta_E = \frac{q}{pc/e}\,\frac{E_0r_0}{\beta},\ 
\eta_M = \frac{q}{pc/e}\,cB_0r_0,
\label{eq:BendFrac.2}
\end{equation}
neither of which is necessarily positive.  These bending fractions satisfy
\begin{equation}
\eta_E + \eta_M = 1\quad\hbox{and}\quad
\frac{\eta_E}{\eta_M} = \frac{E_0/\beta}{cB_0}.
\label{eq:BendFrac.2p}
\end{equation}
The ``potencies'' of magnetic and electric bending are in the ratio $cB_0/(E_0/\beta)$ because  the electric
centripetal force in electric field $E_0$ is stronger than the centripetal force produced by magnetic field 
$cB_0=E_0$ by the factor $1/\beta$, as regards bending charge $q$ onto an orbit with the given radius of curvature 
$r_0$. Also, when expressed in term of spin tunes, the  ``potencies'' of magnetic and electrically induced MDM 
precessions are in the same ratio as the bending potencies.  Defining a ``normalized momentum'' 
\begin{equation}
\widetilde{p} = \frac{pc/e}{q},
\label{eq:BrhoErhro.1}
\end{equation}
the fractional bending coefficients can be expressed in terms of ``Brho'' and ``Erho'' factors;
\begin{equation}
\eta_E = \frac{1}{\widetilde{p}}\,E_0r_0/\beta
       = \frac{E_0r_0/\beta}{E_0r_0/\beta + cB_0r_0}, \quad
\eta_M = \frac{1}{\widetilde{p}}\,cB_0r_0       
       = \frac{cB_0r_0}{E_0r_0/\beta + cB_0r_0}.
\label{eq:BrhoErhro.2M}
\end{equation}
With superimposed electric and magnetic fields, the combination $E_0r_0/\beta + cB_0r_0$ plays the role
that $cBr_0$ plays for purely magnetic lattices.  This imports into EM-superposition the ``geometrization
of transverse optics''---which is to say ``Newtonian optics with lens strengths quantified by focal lengths''. 
\emph{An essential inference from Eqs.~(\ref{eq:BendFrac.2}) through (\ref{eq:BrhoErhro.2M}) is that, if
forward and backward focusing strengths are in the same ratio as E/M bending strengths then 
the forward and backward focal lengths are in the same ratio as the normalized $\tilde p$ values}.
This is strictly true for vertical focusing (irrespective of electric and magnetic bending fractions) but
true only to a first approximation (i.e. vanishing in the $\eta_M=0$ limit) for horizontal focusing.
With this limitation, forward and backward focal lengths will be the same, element by element and, 
for a lattice that is forward/backward symmetric the forward and backward $\beta_y$-functions will be
identical and the $\beta_x$-functions approximately the same.  

This issue is discussed in numerical detail in Section~\ref{sec:PTRLatticeOptics}.
It is shown that forward/backward symmetry can be satisfied closely but not quite exactly in the 
PTR ring.  Though it is possible to superimpose electric and magnetic bending with high precision, it is much
more difficult to superimpose electric and magnetic lumped quadrupole fields.  For this reason, a significant 
fraction of the present paper is devoted to designing a PTR ring that is stable and robust, using
toroidal field shaping electrodes and magnet coils along with ``triplet, compromise quadrupoles'' (described in detail
in Appendix~\ref{sec:Compromise}) that approximately 
superimpose electric and magnetic lumped quadrupoles in the same ratio as the bending fields. To linearized
 order, the lattice properties will then be almost identical for forward and backward traveling beams. 

\subsection{Frozen spin mutual co-magnetometry}
In the idealized storage ring to be discussed, the electromagnetic
fields are ``cylindrical'' electric ${\bf E}=E_0{\bf\hat x}r_0/r$ (with $E_0<0$) and, 
superimposed, uniform magnetic field ${\bf B}=B_0{\bf\hat y}$ (with $B_0>0$).
The bend radius is $r_0>0$. Terminology is useful to specify the relative
polarities of electric and magnetic bending:
Cases in which both forces cause bending in the same sense will be called
``constructive'' or ``frugal'';  Cases in which the electric and magnetic
forces subtract will be referred to as ``destructive'' or ``extravagant''.

For a particle with spin circulating in a (horizontal) planar magnetic storage ring,
its spin precesses around a vertical axis at a rate proportional to the particle's 
anomalous magnetic dipole moment, $G$.  For an ``ideal Dirac particle'' (meaning $G=0$) 
\emph{in a purely magnetic field} the spin precesses at the same rate as the momentum---pointing always 
forward, for example.
Conventionally the spin vector's orientation is specified by the angle $\alpha$ between 
the spin vector ${\bf S}$ and the particle's momentum vector ${\bf p}$ (which is tangential, by definition). 
For such a ``not-anomalous'' particle the spin-tune $Q_M$ 
(defined to be the number of $2\pi$ spin revolutions per particle revolution) 
therefore vanishes, in spite of the fact that, in the laboratory, the spin has actually precessed 
by close to $2\pi$ each turn.  

In general, spin 1/2 particles are not ideal Dirac particles; the directions of their spin axes deviate 
at a rate proportional to their anomalous magnetic moments, $G$, and their spin tunes differ from 
zero even in a purely magnetic field.  Note also, that a laboratory electric field produces a magnetic 
field in the particle rest frame, so a particle in an all-electric storage ring also has, in general, a 
non-vanishing spin tune $Q_E$. Along with $G$ and $Q$, all of these comments apply equally to the 
polarization vector of an entire bunch of polarized circulating particles.  

By convention, in the BMT-formalism\ \cite{BMT}\cite{Wolski}, the orientation of the spin vector 
${\bf S'}$ is defined and tracked in the rest frame of the circulating particle, while the 
electric and magnetic field vectors are expressed in the lab. The spin equation of motion 
with angular velocity $\omega$ is 
\begin{equation}
\frac{d{\bf S'}}{dt} = \omega\times{\bf S'},
\label{eq:BMT.1} 
\end{equation}
with orbit in the horizontal $(x,z)$ plane assumed, where
\begin{align}
\omega
 &=
-\frac{qe}{\gamma mc}\,
\bigg(\Big(G\gamma\Big)cB_0 + \Big(\big(G - \frac{1}{\gamma^2-1}\big)\gamma\beta^2\Big) \frac{E_0}{\beta}\bigg)\,{\bf\hat y} \notag\\
 &\equiv
-\frac{qe}{\gamma mc}\,\bigg((Q_{M})cB_0 + (Q_E)\,E_0/\beta\bigg)\,{\bf\hat y},
\label{eq:BMT.2} 
\end{align}
With $\Delta\alpha$ being the in-plane spin precession advance relative to the beam direction each turn, the ``spin tune'' is defined to
be $Q_s = \Delta\alpha/(2\pi)$.
Spin tunes in purely electric and purely magnetic rings are given by
\begin{equation}
Q_E = G\gamma - \frac{G+1}{\gamma},
\quad
Q_M = G\gamma,
\label{eq:BendFrac.7}
\end{equation} 
where $\gamma$ is the usual relativistic factor.
Note that the sign of $Q_M$ is the same as the sign of $G$, which is positive for
protons---proton spins precess more rapidly than their momenta in magnetic fields. 
Deuteron spins, with $G$ negative, lag their momenta in 
magnetic fields.  With $G$ positive, $Q_E$ increases from -1 at zero velocity, eventually switching sign
at the ``magic'' velocity where the spins in an all-electric ring are ``frozen'' (relative 
to the beam direction).  When a particle spin
has precessed through 2$\pi$ in the rest frame it has also completed one full revolution
cycle from a laboratory point of view; so the spin-tune is a frame invariant quantity. 

\subsection{Forward-backward beam pairings}
For brevity one can discuss just electrons (of either sign), 
protons($p$), deuterons($d$), tritons($t$), and helions($h$); or even just $p$,
$d$, and $h$, based on the consideration that (with the exception of
helion beam) most of the apparatus, and all of the 
technology, needed for their EDM  measurement is presently available at COSY
laboratory in Juelich, Germany.  Polarized helion beams of appropriate
energy will soon be available at BNL.

The circulation direction of a so-called 
``master beam'' (of whatever charge $q_1$) is assumed to be CW or, equivalently,
$p_1>0$. A secondary beam charge $q_2$ is allowed to have either 
sign, and either CW or CCW circulation direction.

Ideally both beam polarizations would be frozen ``globally'' (meaning spin tune $Q_S$ is zero and the angle $\alpha$ 
between polarization vector and momentum is constant everywhere around the ring).  (Somewhat weaker) ``doubly-frozen'' 
can (and will) be taken to mean that a ``primary beam'' locked to $Q_S=0$, circulates concurrently with a ``secondary'' 
beam that is ``pseudo-frozen'', meaning the spin tune is locked to an unambiguous, exact, rational fraction (other 
than zero).  Only if the secondary beam rational fraction is exactly zero, would the terminology ``doubly-magic'' 
be legitimate.

Such particle pairings are expected to make direct MDM or EDM difference measurements of unprecedented 
precision possible.  For any  arbitrary pairing of particle types $(p,p),\ (p,d),\ (p,e+)$, etc., continua of such 
doubly-frozen pairings are guaranteed.  This is taken to imply that both beam polarizations can be phase-locked,
which is one of the requirements for mutual co-magnetometry.

A design particle has mass $m>0$ and charge $qe$, with electron charge 
$e>0$ and $q=\pm 1$ (or some other integer). These values produce circular 
motion with radius $r_0>0$, and velocity ${\bf v}=v{\bf\hat z}$, where the motion
is CW (clockwise) for $v>0$ or CCW for $v<0$. With $0<\theta<2\pi$ being 
the cylindrical particle position coordinate around the ring, the angular 
velocity is $d\theta/dt=v/r_0$. 

(In MKS units) $qeE_0$ and $qe\beta c B_0$ are commensurate forces, 
with the magnetic force relatively weakened by a factor $\beta=v/c$ 
because the magnetic Lorentz force is $qe{\bf v}\times{\bf B}$. 
By convention $e$ is the absolute value of the electron charge.  When $e$
appears explicitly, usually as a denominator factor, its purpose in 
MKS formulas is to allow energy factors to be expressed in electron volts (eV)
in formulas for which the MKS unit of energy is the joule. 
Newton's formula for radius $r_0$ circular motion, expressed in terms of 
momentum and velocity (rather than just velocity, in order to be relativistically valid)
can be expressed using the total force per unit charge in the form
\begin{equation}
\beta\frac{pc}{e} = \Big(E_0 + c\beta B_0\Big)\,qr_0,
\label{eq:CounterCirc.1} 
\end{equation}
Coming from the cross-product Lorentz magnetic force, the factor $q\beta cB_0$
is negative for backward-traveling orbits because the $\beta$ factor 
is negative.

A ``master'' or primary beam travels in the ``forward'', CW direction. 
For the secondary beam, the $\beta$ factor can have either sign.
For $q=1$ and $E_0=0$, formula~(\ref{eq:CounterCirc.1}) reduces to a standard 
accelerator physics ``cB-rho=pc/e'' formula.  For $E_0\ne 0$ the formula 
incorporates the relative ``bending effectiveness'' of $E_0/\beta$ 
compared to $cB_0$.  As well as fixing the bend radius $r_0$,
this fixes the magnitudes of the electric and magnetic bend field values 
$E_0$ and $B_0$. To begin, we assume the parameters of a frozen spin ``master'',
charge $qe$, particle beam have already been established, including the signs
of the electric and magnetic fields consistent with $\beta_1>0$ and $p_1>0$.  
In general, beams can be traveling either CW or CCW.  For a CCW beam both $p$ and 
$\beta$ have reversed signs, with the effect that the electric force is unchanged, but the 
magnetic force is reversed. The $\beta$ velocity factor can be expressed as
\begin{equation}
\beta = \frac{pc/e}{\sqrt{(pc/e)^2 + (mc^2/e)^2}}.
\label{eq:CounterCirc.2} 
\end{equation}
Introducing the abbreviations, 
\begin{equation}
\mathcal{E} = qE_0r_0,
\quad\hbox{and}\quad
\mathcal{B} = qcB_0r_0, 
\label{eq:Alterations.2}
\end{equation}
it has been shown in the companion paper\ \cite{RT-ICFA} that all circular orbits
satisfy the equation 
\begin{equation}
p_m^4 - 2cB_0(qr_0)p_m^3 + (\mathcal{B}^2-\mathcal{E}^2)p_m^2 - \mathcal{E}^2m^2 = 0,
\label{eq:AbbrevFieldStrengths.3-rev} 
\end{equation}
where $p_m$ is the momentum of a particle of mass $m$.

Curiously, expressed as a quartic equation for $p_m$, the absence of a term linear in $p_m$ causes this equation 
to be ``independent of $r_0$'' (or rather of the product ``$qr_0$'') in the sense that the product factor $(qr_0)$ 
can be chosen arbitrarily without influencing any essential implications derivable from the equation.  This and other 
properties can be confirmed by pure algebraic reasoning, based on the missing linear in $p_m$ factor, or by explicit 
partially-numerical factorization of the left hand side into the product of four factors linear in $p_m$. 

Independence of the $qr_0$ has huge importance for planning any EDM ring prototype.  In the present paper,
Tables~  \ref{tbl:p-d2}, \ref{tbl:p-He3}, \ref{tbl:full-scale-p-He3}, and \ref{tbl:full-scale-CESR} 
give parameters for rings of bending radius, respectively, 11.0\,m, 50.0\,m, 95.5\,m, and 85.0\,m.  In every case, 
in spite of the seeming complexity of these tables, $r_0$ can be changed arbitrarily without difficulty. 
The electric and magnetic fields need only to be scaled inversely to match the kinematic parameters to the bending 
radius.  In this sense the 11.0\,m PTR ring is a faithful prototype for the eight times larger ``nominal full scale'' rings. 

These considerations have removed some, but not all of the sign ambiguities introduced by the quadratic 
substitutions used in the derivation of Eq.~(\ref{eq:AbbrevFieldStrengths.3-rev}). 
The electric field can still be reversed, thereby doubling the set of solutions of the equation. 
Note that this change cannot be compensated by switching the sign of $q$, which also reverses the 
magnetic bending.

It is important also to realize, in cases where efficient polarimetry is available, that \emph{the sign of 
$B$ (or $E)$ can be reversed or reset to high accuracy without the need even to measure $B$ (or $E$).} 
This feature, exploiting the constancy of electron MDMs, has enabled the masses of all resonances discovered 
in electron/positron colliders to be determined with such high accuracy.

\subsection{Electric and magnetic contributions to spin tune}
The combined field spin tune $Q_s$ can be expressed in terms of the fractional bending coefficients;
\begin{equation}
Q_S = \eta_EQ_E + \eta_MQ_M.
\label{eq:BendFrac.6} 
\end{equation}
Superimposed electric and magnetic bending permits beam spins to be frozen 
``frugally''; i.e. with a ring smaller than would be required for all-electric 
bending; for spin tune $Q_S$ to vanish requires
\begin{equation}
Q_S = \eta_{_E}Q_E + (1-\eta_{_E})Q_M = 0.
\label{eq:SpinPrecess.5m}
\end{equation}
Solving for $\eta_{_E}$ and  $\eta_{_M}$, 
\begin{equation}
\eta_{_E} = \frac{G\gamma^2}{G+1}, \quad
\eta_{_M} = \frac{1+G(1-\gamma^2)}{G+1} = \frac{1-G\beta^2\gamma^2}{G+1}.
\label{eq:SpinPrecess.5}
\end{equation}
For example, with proton anomalous magnetic moment $G_p=1.7928474$, trying $\gamma=1.25$, 
we obtain $\eta_{_E} = 1.000$ which agrees with the known proton 
233\,Mev kinetic energy value in an all-electric ring. 
For protons in the non-relativistic limit, 
$\gamma\approx1$ and $\eta^{\rm NR}_E \approx2/3$.

The electric/magnetic field ratio for the primary beam (with anomalous moment $G_1$) to be magic is
\begin{equation}
\frac{\eta_E}{\eta_M} = 
\frac{E_0/\beta}{cB_0} = \frac{G_1\gamma_1^2}{1-G_1\beta_1^2\gamma_1^2}.
\label{eq:SpinPrecess.5pp}
\end{equation}
For given $\beta_1$, along with this equation and the required bend radius $r_0$, this fixes the 
electric and magnetic fields to the unique values that globally freeze the primary beam spins.
With $1\rightarrow2$ subscript replacement, the same frozen beam formulas apply to the secondary 
beam; note, though, that the $\beta$ factor has opposite sign. To be ``doubly-magic'' both beams must 
satisfy this relation.  At this time the only known dual solutions are $(p,p)$, $(p,h)$, and $(p,e^+)$.

\section{Prototype ring (PTR) EDM measurement capability}
\subsection{Toroidal focusing PTR lattice design}
\emph{Responsible planning} for an eventual ``{nominal all-electric''} all-electric EDM storage ring with simultaneously 
counter-circulating $\mathcal{E}=232.8$\,MeV frozen spin proton beams \emph{requires the construction of a prototype 
ring (PTR).} To maintain flexibility the PTR lattice design initially is ``scale-invariant'' and site independent.
But, for numerical examples: $\mathcal{C}$=102.5\,m, roughly 1/8 scale relative to the nominal all-electric ring. Except for 
construction itself, PTR needs only present day technology and apparatus available in the COSY facility.

Though planned initially as a prototype, and constrained by cost minimization compromised performance, PTR
will, itself, be capable of the performance promised in the title of the present paper.  \emph{This is largely
the result of improved understanding of the {\rm EDM} measurement capability improvement produced by spin
tune phase-locking, understanding the {\rm BSM} power of measuring the difference of proton and 
helion {\rm EDM}s, and recognizing that ``quadrupole free'' toroidal focusing will assure long spin coherence 
time, especially with betatron phase space ``mixing'' provided by running on the ``difference resonance''.}

There is a substantial literature describing the physical importance of measuring EDMs of 
elementary particles.  The most recent comprehensive report is due to the CPEDM collaboration, based mainly on 
experience of the JEDI group using the COSY ring in Juelich, Germany, and described in a CERN Yellow Report\ 
\cite{CYR} (referred to here as ``CYR'') feasibility study for a storage ring to search for EDMs of charged particles.

Measurement of the proton EDM relies on the development of a storage ring with predominantly electric bending, 
capable of storing simultaneously counter-circulating frozen spin proton beams.  A conclusion of the CYR study
is that a low energy prototype EDM ring, here referred to as ``PTR'', as economical as practical, is needed 
to investigate problematical issues.

The Phase~I goals for PTR are, 
first, to store an intense (not necessarily polarized) proton beam and, second, to store a simultaneous 
counter-circulating beam. Phase~II will superimpose magnetic bending in order to store and freeze an intense 
frozen-spin 45\,MeV proton beam.  

There is considerable overlap between Chapter~7 of the CYR report and the present paper.  
Figure~\ref{fig:SectorPerspective} is a perspective mock-up of one sector of PTR copied
from that report.  Though the PTR superperiodicity is $N_{\rm super}=4$, it is sufficient 
for most purposes to display just one octant. The ring is also forward/backward symmetric. 
There have been substantial lattice design changes subsequent to the CYR manuscript commit  date.  
The currently revised design is exhibited in Figure~\ref{fig:PTR-layout-Toroidal8_102p2-mod}.

Quadrupoles are shown but, as mentioned previously, they are intended to be as weak as possible,
consistent with their main roles, which are to enable operational procedures requiring large
changes in ring optics, and to compensate for any unexpected systematic optical deficiencies
associated with the fabrication of the toroidal electrodes.

Based on much the same optics that has enabled scanning transmission electron microscopy (TEM and STEM) of
such spectacularly acute spatial resolution, toroidal focusing can be expected (for the same reason) to provide
exceptionally long spin coherence times (SCT) for polarized beams.  This can be expected to enable experimentation 
into the role played by particle spin in time-reversed situations.

The decision to rely entirely on toroidal focusing, carries with it a significant operational 
penalty resulting from the fact that, once fabricated, no smooth way of tuning the toroidal 
focusing $m$ value has been established. Theoretical discussion of these issues in detail is
contained in companion paper\ \cite{RT-Positron-q}.  Cancellation of spin polarization decoherence
is greatly enhanced by toroidal focusing.  As mentioned at the start of this paper there is
considerable overlap of that paper with this paper. Here we list only essential references from 
that paper:\  
\cite{ConformalHandbook},
\cite{AlbrechtToroidal},
\cite{ToffolettoLeckeyRiley},
\cite{CourantSnyder},
\cite{CourantDirichlet},
\cite{Erni-TEM},
\cite{Good},
\cite{Magiera},
\cite{Nehari}.
In combination, along with other technical details, these papers explain how the requirement 
of conformal invariance is satisfied by the PTR design in the present paper. 

The dominant bending and focusing is performed by the red and blue toroidal sectors shown
in Figure~\ref{fig:PTR-layout-Toroidal8_102p2-mod}.  The $m$-values are displaced slightly, with 
$m_{\rm blue}\leqq m_{\rm red}$. The purpose for this is to preserve 
flexibility while the focusing design is being refined.  At the time of final construction it 
will be preferable for red and blue sectors to be identical.  Otherwise, every sector bend has 
identical superimposed electric and magnetic bending.  As fabricated, all magnet bends 
will probably be powered in series (with possible small shunt compensation) and all electrode
voltages only slightly adjustable (for closed orbit steering). 

In spite of the small number of adjustable parameters, there is sufficient flexibility to tune PTR
to some of the operational MODES mentioned in previous reports.  Computationally the optical design
has been performed using a MAPLE-BSM code, where ``BSM'' is an acronym for ``Belt and Suspenders, 
Modifiable''.  The lumped quadrupoles can provide the somewhat stronger (separated function) focusing 
needed for robust ring commissioning operations.  However, a COLLIDING BEAM mode described in previous 
reports cannot be achieved.  This is because backward/forward symmetric ``low beta squeeze''
(for increased luminosity) is impractical with superimposed electric and magnetic bending.  

As well as violating forward/backward lattice function symmetry, lumped quadrupoles inevitably
introduce chromaticity into ring optics.  Customarily in storage rings this chromaticity is
compensated by one, or typically, at least two, families of sextupoles.  Such sextupoles prove to
be acceptable for unpolarized beams, but their presence impairs the ring performance for polarized
beams. Their presence is the leading cause of spin precession decoherence, which limits the
spin coherence time (SCT).  For ultimate EDM measurement precision, it will therefore be essential
for the lumped quadrupoles to be as weak as possible.

As mentioned previously, the various lattice optical designs are scale invariant and site-independent.
But, to make lattice parameters and plots more intuitive, the ring circumference is taken to be 
102.5\,m, matched exactly to the ring circumferences of the proposed bunch accumulator (BA) described 
below.  As fixed by this circumference, in all subsequent lattice function plots, the horizontal axis 
coordinate is the longitudinal position coordinate $s$. 

\begin{figure}[hbt]
\centering
\includegraphics[scale=0.5]{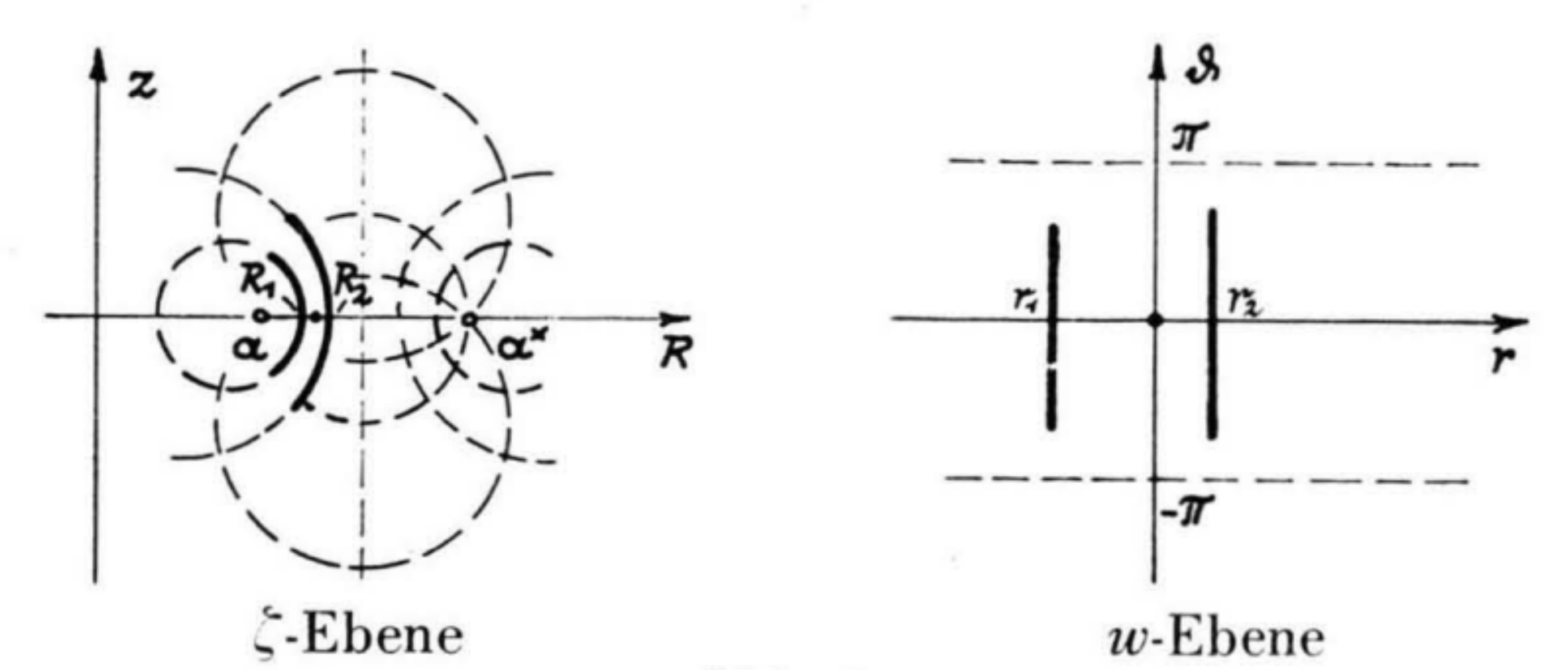}
\caption{\label{fig:ToroidalShape}Photocopied, for emphasis, from a historically important paper of 
Albrecht\ \cite{AlbrechtToroidal} who provided the original toroidal focusing formulation. 
``Physical space'' analytic complex function on the left is conformally mapped to ``model space'' 
analytic complex function on the right.  
This figure shows field lines and equipotential contours in the central
horizontal ring plane for the toroidal coordinate system used in developing the Wollnik transfer matrices 
used in the present paper. 
Electrodes cut from the full toroid surface are shown by bolder curve sections.
Electrically isolated, they are held at equal but opposite electric potentials, with electric field lines
joining exactly normal at both electrodes.  Electrodes, and normal entry particle trajectories 
are perfect circles.  The Wollnik linear transfer matrices used in the present paper correspond to 
propagation through the good field regions between these electrodes.  }
\end{figure}

\subsection{EDM measurement strategy\label{sec:MeasurementStrategy}}
As regards the orientation of the beam polarization, it is essential to distinguish between ``in-plane'' and 
``out-of-plane'' orientations, where ``the plane'' refers to the ring beam plane, which is presumed to be 
horizontal.  \emph{In-plane precession}, is routinely induced by ideal (rest frame) magnetic fields acting 
on beam particle magnetic dipole
moments (MDMs).  Ideally, \emph{out-of-plane precession} can be induced only by P- and/or T-violating symmetry violating
torques due to electric fields acting on particle EDMs.  But, in practice, the inevitable existence of unintentional
magnetic fields acting on particle MDMs can also induce out-of-plane precession.  The average radial magnetic field
$\langle B_r\rangle$ is expected to be the dominant source of systematic error in the determination of particle EDMs.

The leading strategy for reducing this source of systematic error is to produce simultaneously counter-circulating 
frozen spin beams on ``identical'' orbits.  This condition is to be achieved by adjusting local vertical beam deflection
corrections to cause the counter-circulting orbits to be identical.  This cancels the dominant systematic error source 
$\langle B_r\rangle$ by canceling the average out-of-plane (vertical) orbit separation. We refer to this capability as  
``self-magnetometry''. The precision with which the orbits can be matched vertically depends on the precisions of the 
beam position monitors (BPMs) that measure the vertical beam positions, and on the ring lattice sensitivity to the 
magnetic field errors causing the orbits to be vertically imperfect. 

Assuming both beam spins are frozen, at least the ``primary'' beam$-1$ will, by convention,  
be globally frozen; spin tune $Q_{s1}=0$.  Ideally 
both beams would have $Q_s=0$ but, ordinarily, the ``secondary'' beam-2 will be only locally frozen; $Q_{s2}$ exactly equal 
to a rational fraction other than $0/1$.  In this condition both beam polarizations can be phase-locked,  allowing both
beam spin tunes to be set and re-set with frequency domain precision, meaning that synchronism can be maintained for runs 
of arbitrary duration.  Since the RF frequency can also be restored to arbitrarily high precision, conditions 
can be set 
and  re-set repeatedly, without depending upon high precision measurement to the electric and magnetic bend fields.
This allows, for example, the magnetic bending field to be reversed with high precision, as would be required to 
interchange CW and CCW beams. 
This capability can be referred to as \emph{stabilizing all fields  by phase locking both revolution frequencies
and both beam polarizations, using their own MDMs as ``magnetometric gyroscopes''}.

\noindent
\subsubsection{Encapsulation of the EDM measurement strategy\label{sec:StrategyEncapsulation}}
\begin{itemize}
\item 
 Use threefold phase locking:
     \begin{enumerate}
     \item Primary beam~1, CW,  \emph{globally frozen}
     \item Secondary beam~2, CCW, in general, locally ($Q_s=$ non-zero rational fraction) frozen
     \item Synchronous bunch capture onto almost identical orbits with different ``best matched'' 
           beam~1 and beam~2 harmonic numbers in a common RF cavity,
     \end{enumerate}
\item
which accurately matches beam~1 and beam~2 orbits.
\item
and allows conditions to be reset precisely, including magnetic or electric field reversal, 
without the need for unachievably high precision direct electric and magnetic field measurement.
\item
This supports averaging over interchange of conditions to reduce systematic error.
\footnote{Recent ``table-top'' measurements of molecular EDMs have been surveyed by
Alarcon et al.\cite{Alarcron-EDM} Due to the physical complexity of these laboratory experiments, it is 
impractical to reverse the direction of molecular beam transversal.  The fact that the storage ring 
EDM measurement has no such limitation enables some systematic errors to be cancelled by averaging
over forward and backward transversals.}
\end{itemize}

\subsection{High intensity beam injection from CLIP to replace bunch accumulation using BA}
\begin{figure}[hbt]
\centering
\includegraphics[scale=0.55]{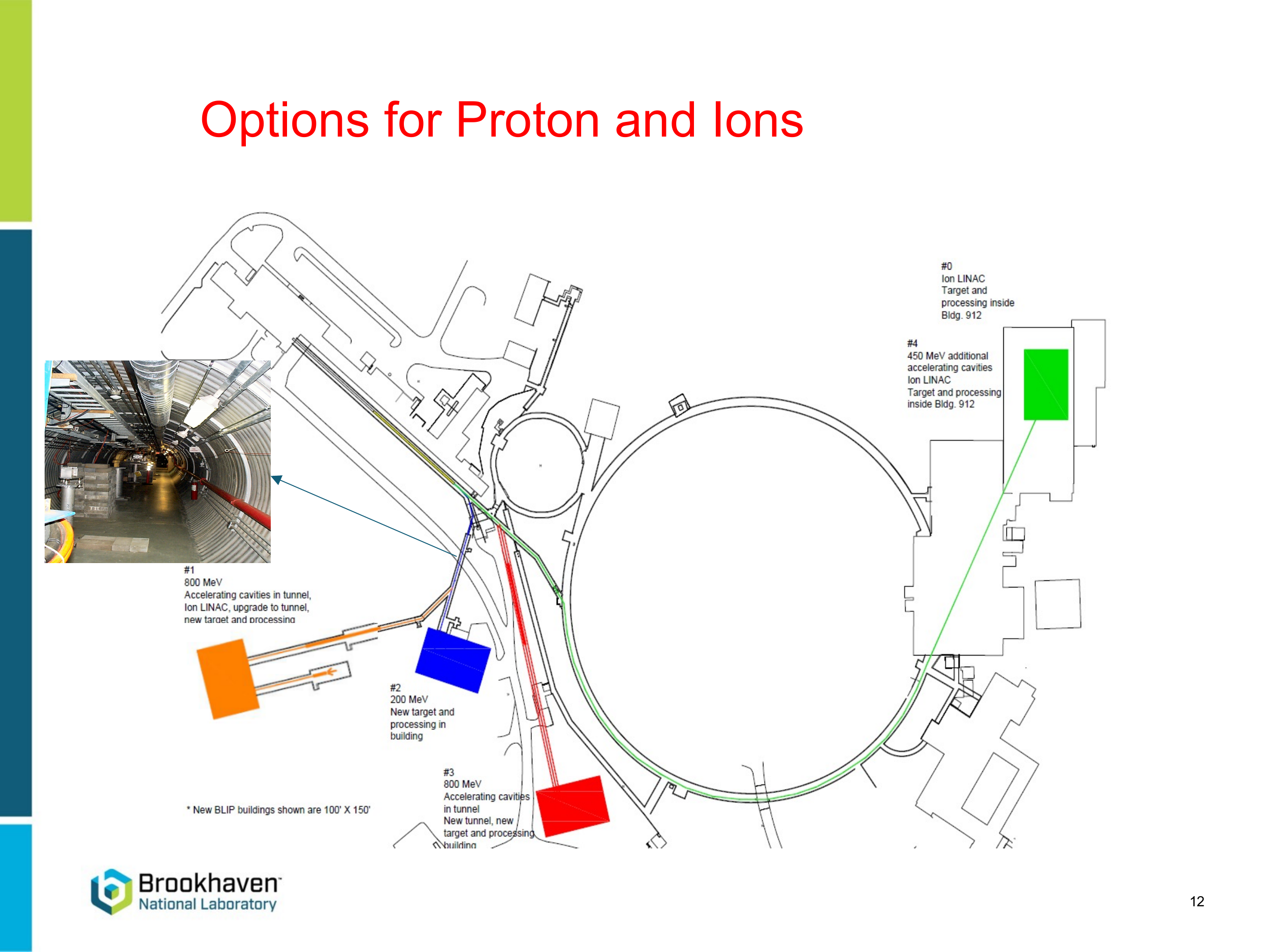}
\includegraphics[scale=0.165]{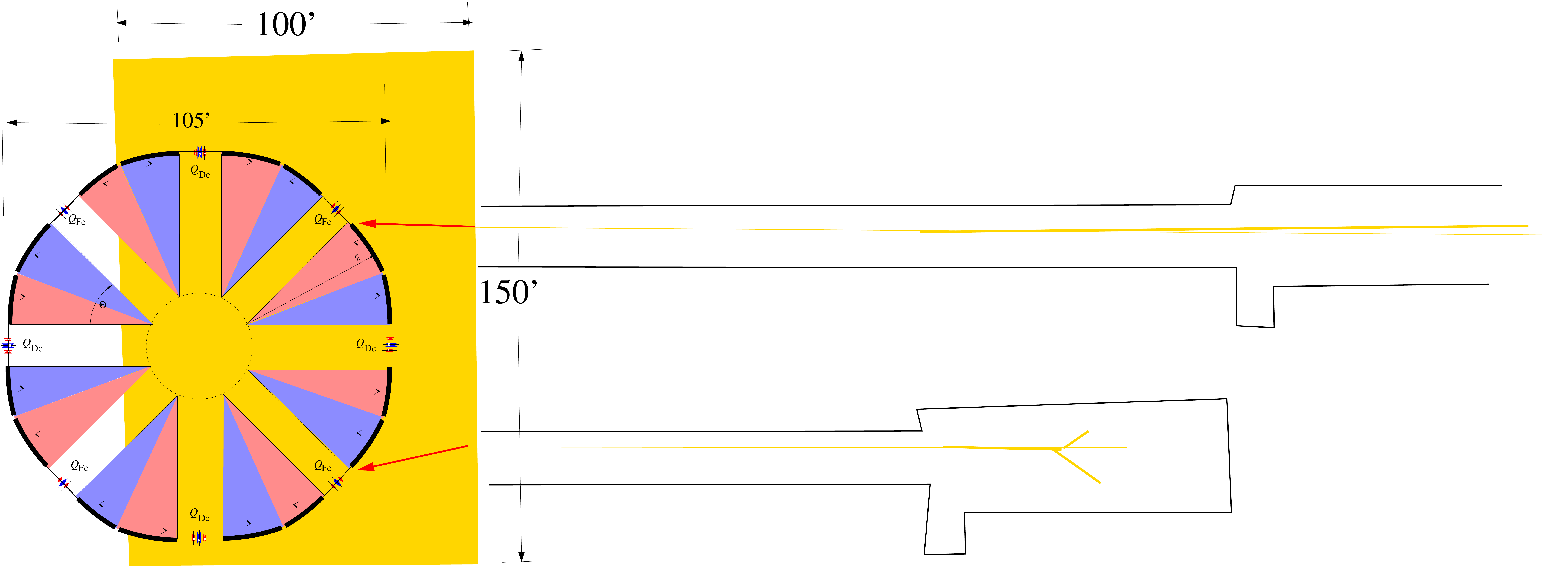}
\caption{\label{fig:CLIP-p-accelerator-pic}Cut and paste mock-up envisaging PTR within the CLIP complex.
Above, shown in outline only, the small ring is the existing booster, and the larger structure is 
the AGS.  Proposed new structures are shown in color.
Below is PTR, shown as if situated in the vicinity of (slightly canted and blown-up) copy
of (orange) CLIP Building 1, with beam injected from CLIP.}
\end{figure}

Figure~\ref{fig:CLIP-p-accelerator-pic} displays a conjectured placement of PTR in the 
polarized beam injection line of the (currently under review) BNL Center for Linac Isotope Production (CLIP),
including the proposed light Ion linac (LIL). The anticipated polarized injected beam currents from LIL 
are a factor of roughly 6000 times larger than from the COSY cyclotron described previously. This 
is comparable with the 10,000 times increase in bunch current promised by the BA bunch accumulator 
shown in Figure~\ref{fig:COSY-hall-mod7}. 

Furthermore CLIP promises a gourmet menu of polarized nuclear isotopes, including most, if not all, shown
on Koop's menu, Figure~\ref{fig:KoopParameters}.  Most exciting of all the choices is the combination
of protons, $p$, and helions, $h$.  It is this pairing that can provide the most precise test of BSM physics
that is likely to be achievable in the quite near future. Furthermore the range of available beam energies
is luxurious enough to require no detailed discussion.

As mentioned previously, mutual co-magnetometry with PTR can be expected to permit convenient and precise MDM 
measurements for any nuclear isotope having lifetime in excess of a minute or so.  Because of its
concentration on the proton/helion EDM difference measurement, this capability is not
discussed further in the present paper.
%
%
%

\section{Lattice design and performance}
\subsection{Ideal thick lens toroidal focusing\label{sec:ThickToroidalFocusing}}
Transfer matrices from one symmetry point to the next of a storage ring with superperiodicity $N_s$ 
have the form
\begin{equation}
{\bf M_1} = 
\begin{pmatrix} \cos\mu        & \beta\sin\mu     \\
         -\frac{\sin\mu}{\beta} &  \cos\mu   
\end{pmatrix}
 \equiv
\begin{pmatrix} C         & \beta S     \\
         -\frac{S}{\beta} &       C 
\end{pmatrix},    
\label{eq:ThickToroidal.1}
\end{equation}
where $\mu$ can stand for $\mu_x$ or $\mu_y$, $\beta$ can stand for $\beta_x$ or $\beta_y$
and similarly for $C\equiv\cos\mu$ and $S\equiv\sin\mu$. Squaring ${\bf M_1}$ and
applying trigonometric identities produces
\begin{equation}
{\bf M_2}\equiv {\bf M_1^2}  
 =\begin{pmatrix} \cos2\mu      & \beta\sin2\mu \\
         -\frac{\sin2\mu}{\beta} &  \cos2\mu   
\end{pmatrix}.
\label{eq:ThickToroidal.2}
\end{equation}
Squaring again and applying trigonometric identities produces
\begin{equation}
{\bf M_4}\equiv {\bf M_1^4}  
 =\begin{pmatrix} \cos4\mu      & \beta\sin4\mu \\
         -\frac{\sin4\mu}{\beta} &  \cos4\mu   
\end{pmatrix}.
\label{eq:ThickToroidal.3}
\end{equation}
By applying these formulas separately to horizontal, $x$, and vertical, $y$ transverse coordinates produces
\begin{equation}
{\bf M^x_4}\equiv {{\bf M^x_1}^4}  
 =\begin{pmatrix} \cos4\mu_x      & \beta_x\sin4\mu_x \\
          \frac{\sin4\mu_x}{\beta_x} &  \cos4\mu_x   
\end{pmatrix}
\quad\hbox{and}\quad
{\bf M^y_4}\equiv {{\bf M^y_1}^4}  
 =\begin{pmatrix} \cos4\mu_y      & \beta_y\sin4\mu_y \\
         -\frac{\sin4\mu_y}{\beta_y} &  \cos4\mu_y   
\end{pmatrix}.
\label{eq:ThickToroidal.4}
\end{equation}
\emph{What makes these formulas fascinating is that the parameters $\beta_x$ and $\beta_y$ can be 
chosen arbitrarily, and not necessarily equal.}  This is important for maintaining conformal mapping validity with 
a substantial fraction of the horizontal focusing being ``geometric'' which does not contribute to vertical focusing.
Then, by constraining the betatron phase advances $\mu_x$ and $\mu_y$, for example as
\begin{equation}
\mu_x = \mu_y + \frac{\pi}{2},
\label{eq:ThickToroidal.5}
\end{equation}
one obtains simpler relations.  In this case 
\begin{equation}
4\mu_x = 4\mu_y + 2\pi  \overset{\rm eff.}{\ =\ } 4\mu_y,
\label{eq:ThickToroidal.6}
\end{equation}
since increasing an angle by $2\pi$ has no detectable effect.
In this paper lattice design meeting this constraint is referred to as
``ideal, thick lens toroidal focusing''.   
Even simpler for tuning the PTR ring would be the choice $\mu_x\approx\pi/2$, in which case
\begin{equation}
{\bf M^x_4}
 \approx\begin{pmatrix} 1      & \beta_x\sin4\mu_x \\
       -\frac{\sin4\mu_x}{\beta_x} &  1   
\end{pmatrix}
\quad\hbox{and}\quad
{\bf M^y_4}
 =\begin{pmatrix} 1      & \beta_y\sin4\mu_y \\
       -\frac{\sin4\mu_y}{\beta_y} &  1   
\end{pmatrix},
\label{eq:ThickToroidal.7}
\end{equation}
which are close to identity matrices which, in linear approximation, describe drifts of length 
$\beta_x\sin4\mu_x$ and $\beta_y\sin4\mu_y$ with (small) 2,1 elements representing weak net focusing).
$\beta_x$ and $\beta_y$ are independently adjustable, and all matrix elements are simply 
expressible in terms of the two parameters, $C$ and $S$, defined in Eq.~(\ref{eq:ThickToroidal.1})
and satisfying the condition $S^2+C^2=1$.  

Currently the PTR lattice design shown in Figure~\ref{fig:PTR-layout-Toroidal8_102p2-mod} 
applies these formulas only up to ${\bf M^x_2}$ and ${\bf M^y_2}$.  For operational flexibility, the ring
optics is then controllable by the triplets (located at symmetry points) and masquerading as superimposed 
electric/magnetic quadrupoles.  In this sense the ring behavior will resemble a separated function
FODO lattice with free, arbitrarily strong,  $F$ and $D$ focusing strength parameters. (All of this has 
to be consistent with $F$ and $D$ being turned off as nearly as possible during actual EDM measurement.)

\subsection{Transfer matrix for beam evolution through toroidal bending elements\label{sec:Wollnik}}
The content of this paper has been limited to this point, mainly to resolving issues identified in the
(pre-Covid) CERN Yellow Report.  ``Covid-enforced leisure'' has resulted in the revolutionary promotion
of PTR from ``EDM prototype ring'' to ``ambitious research project'' as explained so far and in the rest of
the paper.

All-electric bending fields exist in the tall slender gaps 
between inner and outer, almost-plane-vertically, curved-horizontally, 
electrodes. The radial electric field dependence is 
\begin{equation}
E= E_r\sim\frac{-1}{r^{1+m}},
\label{eq:WW-AG-CF.1}
\end{equation}
Our convention has the electric field $E_0$ being negative since the nominal beam charge
is positive and the force is centripetal. 
The corresponding electric potential parameterization specializes to inverse power law 
dependence on the bend radius $r$ with 
the electric potential defined to vanish on the design orbit;
expressed as power series in $\xi=x/r$, it is given by
\begin{align}
V(r)
 &=
\frac{E_0r_0}{ m }\,
\bigg(
(1-\xi)^m-1
\bigg)  \label{eq:WeakFoc-mod.2}\\
 &=
-E_0r_0
\bigg(
\xi + \frac{1-m}{2}\xi^2 + \frac{(1-m)(2-m)}{6}\xi^3\dots
\bigg).
\notag
\end{align}
This formula simplifies spectacularly for the Kepler $m$=1 case.  As it happens, for relativistic particles, 
$m=1$ also marks the edge of a region to which the term ``toroidal optics'' applies. Beyond this edge 
orbits spiral in toward the center.  The other edge is marked by the $m=0$ ``cylindrical edge''; beyond this edge 
vertical orbits diverge to infinity.  These edges are defined more accurately below by Eqs.~(\ref{eq:Wollnik-TM-params}).

For focusing index in the (approximate) range $0 < m < 1$ linearized horizontal evolution of initial displacement 
vector ${\bf V_{x0}}$ through a thick bending and focusing element element of length $w$ is given by 
\begin{equation}
 {\bf V_{x1}} = {\bf M_4}{\bf V_{x0}} = 
\begin{pmatrix} c_x     &  s_x    & 0 &  d_xN_t    \\
              -k_x^2s_x &  c_x    & 0 & s_xN_t/r_0 \\
            -s_xN_t/r_0 & -d_xN_t & 1 &  -N_t^2    \\
                0      &    0    &   0   &    1    
\end{pmatrix}{\bf V_{x0}},
\quad\hbox{where}\quad 
{\bf V_{x0}} = \begin{pmatrix} x_0 \\ \theta_{x0} \\ s \\ \delta p /p \end{pmatrix}
\label{eq:Wollnik-TM-x}
\end{equation}
\begin{equation}
{\bf V_{y1}} = {\bf M_2}{\bf V_{y0}} = 
\begin{pmatrix}   c_y    &  s_y       \\
               -k_y^2s_y &  c_y       
\end{pmatrix}{\bf V_{y0}},
\quad\hbox{where}\quad 
{\bf V_{y0}} = \begin{pmatrix} y_0 \\ \theta_{y0} \end{pmatrix}.
 \label{eq:Wollnik-TM-y}
\end{equation}
The parameters are given by
\begin{equation}
k_x = \frac{1}{r_0}\sqrt{1 - m + \frac{\eta_E}{\gamma^2}},\quad
k_y = \frac{1}{r_0}\sqrt{m},\quad
N_t = 1 + \frac{\eta_E}{\gamma^2}, 
\label{eq:Wollnik-TM-params}
\end{equation}
where $N_t$ accounts for time dilation.
The transfer matrix components for elements of length $L$ are expressed in terms of expressions
\begin{equation}
c_x = \cos(k_xL),\ s_x = \sin(k_xL),\ d_x = \frac{1-c_x}{r_0k_x^2},\ t_d=\frac{k_xL-s_x}{r_0^2k_x^2},\\
c_y = \cos(k_yL),\ s_y = \sin(k_yL).
\label{eq:MissingEqName}
\end{equation}
Due originally to Wollnik\ \cite{Wollnik}, but here expressed in more conventional modern notation, these
formulas have been copied from Lebedev\ \cite{Lebedev}, except for having been further generalized to allow superimposed
electric and magnetic bending.
\footnote{Subtle detail: ring tunes estimated from transfer matrices based on accepting 
large values of $L$ in advance in Eqs.~(\ref{eq:MissingEqName}) may be seriously incorrect due to incorrect aliasing with 
thick element focusing. Rearranging Eqs.4.6 from reference\ \cite{RT-ICFA}, and solving for $\tan\psi$ 
produces an expression for phase $\psi(s)$ in terms of matrix elements evaluated locally, for example at arbitrary origin;
\begin{equation}
\psi(s)
 = 
\tan^{-1}\,
\frac{{\bf M}_{1,2}(0, s)}
     {\beta_0{\bf M}_{1,1}(0, s) - \alpha_0{\bf M}_{1,2}(0, s)}.
\label{TM.17}
\end{equation}
Like all inverse trigonometric formulas, this equation has multiple solutions.
But, with the {\tt .sxf} granularity being required to be fine enough, one
can (in principle) sequentially obtain unique phases. With
$\psi(s)$ starting from $\psi(0)=0$, as $s$ increases
from $s_i$ to $s_{i+1}$ there is a unique solution of Eq.~(\ref{TM.17}),
$\psi(s_i)\ge\psi(s_{i-1})$ such that the function $\psi(s)$ increases 
monotonically, as required.
Because the interval from $s_{i=1}$ to $s_i$ is non-zero, $\psi$ can,
superficially, advance discontinuously; the correct solution is the least
discontinuous. The same calculation has to be done for both the $x$ 
and $y$ betatron sectors. 
Even though, theoretically, the phase advances 
monotonically, numerical errors can cause Eq.~(\ref{TM.17}) to
give local phase decrease. Requiring betatron phase $\psi(s)$ to increase monotonically
(as shown most satisfyingly in Figure~\ref{fig:FDDF-ToroidalQuadrant-betas}) can cause $\beta(s)$ shape
variation to seem curious.  This behavior was encountered especially while producing 
Figure~\ref{fig:0p32349-0p99990-0p00909}.  This issue is pursued further in 
Appendix~\ref{sec:Aharonov-Anandan}.}

This accounts for the appearance of $\eta_E$ in
Eqs.~(\ref{eq:Wollnik-TM-params}). (Recall that $\eta_E+\eta_M=1$, noting that one or the other of $\eta_E$ and 
$\eta_M$ can be negative, in which case the centripetal forces combine ``destructively''. 
But the fractions must sum to 1 to produce the correct $E/B$ bend ratio.)

Operating on orbit displacement vector ${\bf V}$, ${\bf MV}(L)$ describes, to linear approximation, 
how ${\bf V}$ has evolved after propogation through bend element of length $L$.
For pure magnetic bending $\eta_E=0$.  For partial electric bending the $1+\eta_E/\gamma_0^2$ combination, 
which approaches 1+$\eta_E$ in the nonrelativistic limit, is a velocity dependent time dilation correction 
factor which modifies dilated time $t_d$ as well as the ``geometric focusing'' supplied by the curvature.  

Notice that there is dependence, in Eqns.~(\ref{eq:Wollnik-TM-params}), 
on bending radius $R_0$ in both $k_x$ and $k_y$ and dependence on $\eta_E$ in $k_x$, though not in $k_y$.  
In the present paper the absence of dependence of $k_y$  on $\eta_E$ is referred to as ``vertical only, geometricization''. 
This means that vertical focusing, and therefore $\beta_y$, is independent of beam momentum (provided the
bending force matches the design ring radius). 
\emph{But horizontal focusing depends on beam momentum (or, rather,  momentum/charge) through its 
dependence of $k_x$ on $\eta_E$}.  

In the case of doubly-frozen or doubly-magic 
counter-circulating beams, the vertical focusing is the same for both beams, independent of beam 
circulation direction.  But the horizontal lattice optics depends on beam momentum and is different for
counter-rotating beams.  Fortunately, for motion in a purely horizontal plane, both spin tunes 
$Q_x$ and $Q_y$, as defined in Eqs.~(\ref{eq:BendFrac.7}), are ``constants of the motion'',
independent of horizontal betatron oscillation amplitude to all orders.  By symmetry, the spin tunes are 
independent of vertical betatron oscillation amplitude, but only to linear order.  This makes it 
important, therefore, for $\beta_y(s)$ to be exactly independent of beam circulation direction, as it is. 
(This comment is pursued in Appendix~\ref{sec:Hybrid} which discusses the so-called
``hybrid option'' for ring focusing.) 

The matrix ${\bf M_2}$ has determinant 1, which is the only requirement for a 2x2 matrix to be 
symplectic.   Block diagonalized into 2x2 matrices, the blocks 
on the main diagonal of  ${\bf M_4}$ are individually symplectic, and the determinant of the blocks on the 
other diagonal vanish. Taken together, these conditions show that ${\bf M_4}$ is also ``symplectic''.

The quotation marks here are intended to serve as a warning that, iterated indefinitely, 
 ${\bf M_4}$ is actually divergent.  The lower two rows of ${\bf M_4}$ include ``chromatic'' 
i.e. momentum-dependent effects---primarily dispersion and time-of-flight. Another idiosyncrasy of 
accelerator formalism is that (based on a fast-slow approximation) these chromatic effects are counted 
as ``second'' rather than ``first'' order effects.  This is related to the fact that it is only the orbit 
shape that can be said to be stable for all time.  (No meaning whatsoever is attached to orbit position
as a function of time.) 

The important feature of conformally-mapped toroidal optics is that ``everything of importance'' electrodes, 
equipotentials, on and off-momentum closed orbits remain perfect circles.  The divergent behavior of iterating 
${\bf M_4}$, represents the longitudinal beam spreading associated with the spread of particle velocities.
So a bunched beam will gradually ``de-bunch'', eventually becoming a uniform ``coasting beam.  As established by 
McMillen and Veksler, this debunching can be prevented by using an RF cavity to capture all (or most) of the
particles into stable longitudinal ``buckets''.

\section{PTR lattice optics\label{sec:PTRLatticeOptics}}
\subsection{Parameters for PTR situated in the COSY beam hall}
\begin{figure}[hbt]
\centering
\includegraphics[scale=0.45]{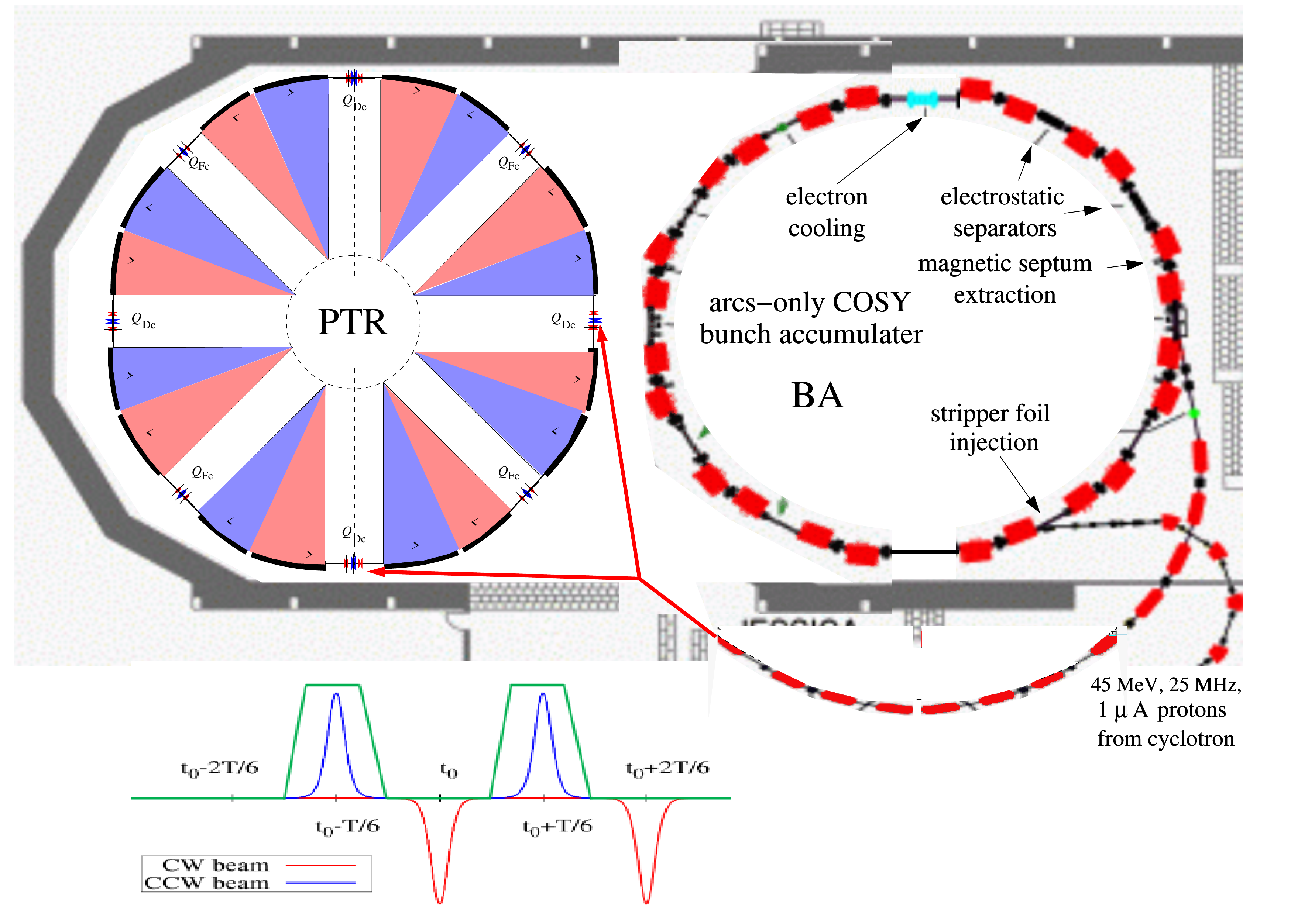}
\caption{\label{fig:COSY-hall-mod7}Pre-CLIP implementation of PTR in the COSY Hall. The bunch accumulator, BA,
consists of an  ``arcs-only COSY'', rebuilt with existing electron cooling, 
stripper-foil injection and electrostatic-magnetic extraction.  Both of these requirements would be met
by simply retaining existing equipment.  The inset figure shows pairs of polarized bunches, transferred 
on successive injection cycles, from BA to counter-circulate in PTR.
}
\end{figure}

Parameters and lattice function values for the lattice design illustrated in 
Figure~\ref{fig:PTR-layout-Toroidal8_102p2-mod} are given in Table~\ref{tbl:MODEparameters2}. Included in this
table are parameters appropriate for Figure~\ref{fig:COSY-hall-mod7}, which shows the
positioning of PTR and ``BA'' which is a ``bunch accumulator ring'' needed to prepare polarized beam
bunches for injection into PTR.  

\setlength{\tabcolsep}{2pt}
\begin{table}[h]\small
\caption{PTR and COSY-arcs-only bunch accumulator (BA) parameters. \label{tbl:MODEparameters2}}
\centering
\begin{tabular}{ccccc}             \hline
file name             & variable         & unit &    BA             &     PTR        \\ 
                      &  name            &      &  COSY-arcs-only   &               \\ \hline
circumference         & {\tt circum}     &  m   &   102.250         &    102.250     \\ \hline
bend radius           &  {\tt r0}        &  m   &                   &     11.0       \\
E field., 30 MeV proton.  &  E               & MV/m &                   &    5.370       \\ \hline 
long straight length     &  {\tt llsnom}    &  m   &                   &    4.142        \\
totol available straight & $16\times${\tt llsh} & m &                 &     32          \\
electrodes/quadrant   &                  &      &                   &      4         \\
bend/electrode        & {\tt Thetah}     & radian &                 &  $2\pi/16$     \\
electrode length      & {\tt Leh}        &  m   &                   &   4.32       \\ \hline 
\rowcolor[gray]{.9}
PTR stored p's no BA  &                  &      &                   & $0.6\times10^7$ \\ 
\rowcolor[gray]{.9}
COSY-arcs-only BA     &                  &      & $0.6\times10^{11}$ & $0.6\times10^{11}$ \\ \hline
min/max horizontal beta    &   $\beta_x$      &  m   &                   &  9.60/10.83   \\
min/max vertical beta    &   $\beta_y$      &  m   &                   &  16.6/24.8         \\ \hline
horizontal tune       &    $Q_x$         &      &                   &  1.726        \\
vertical tune         &    $Q_y$         &      &                   &  0.673       \\ \hline 
\end{tabular}
\end{table}

\subsection{``No lumped quadrupole'' toroidal optics lattice functions}
\begin{figure}[hbt]
\centering
\includegraphics[scale=0.60]{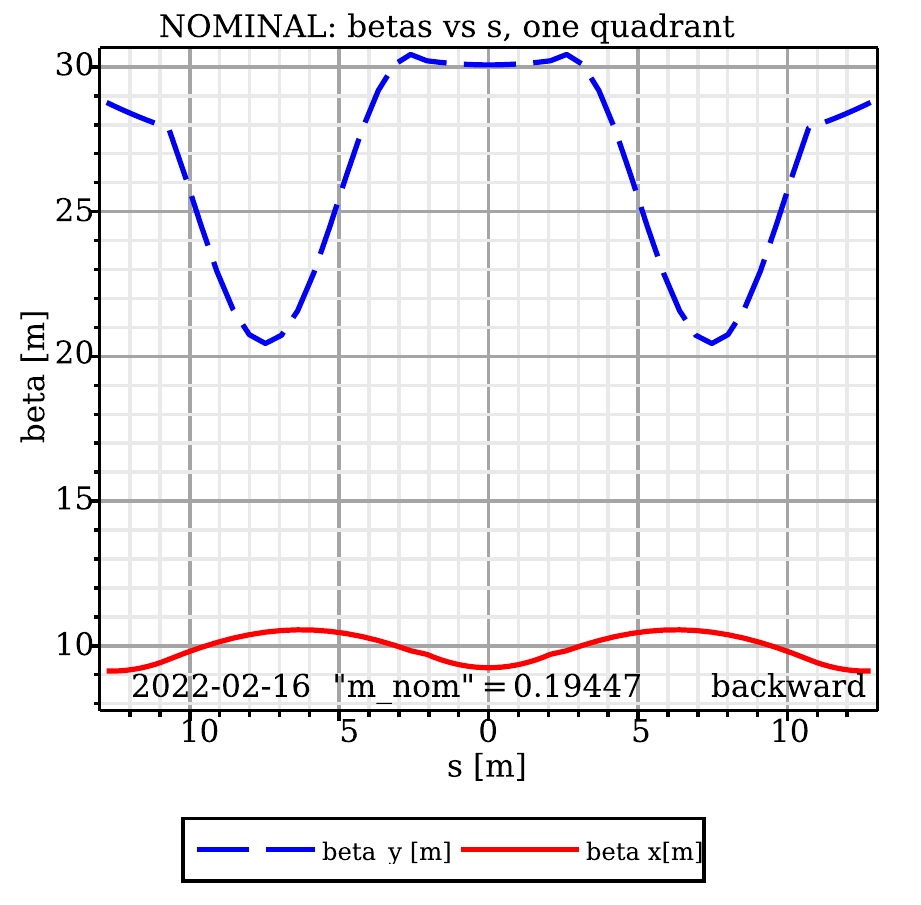}
\includegraphics[scale=0.60]{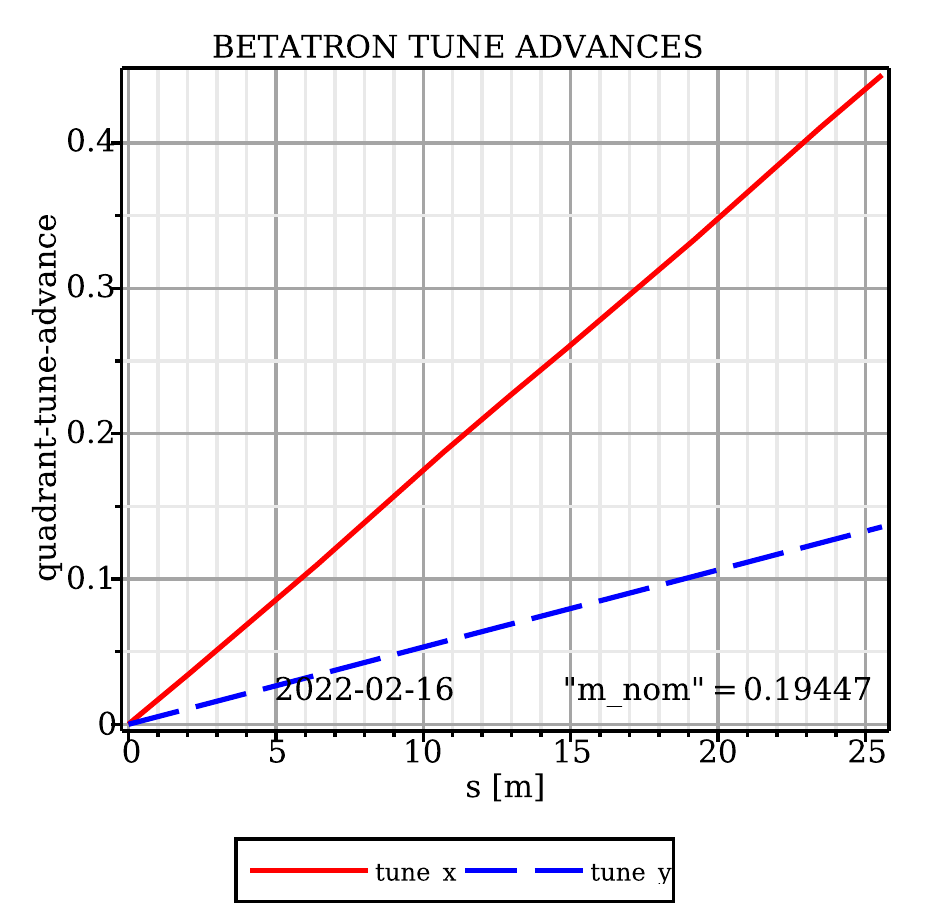}
\includegraphics[scale=0.60]{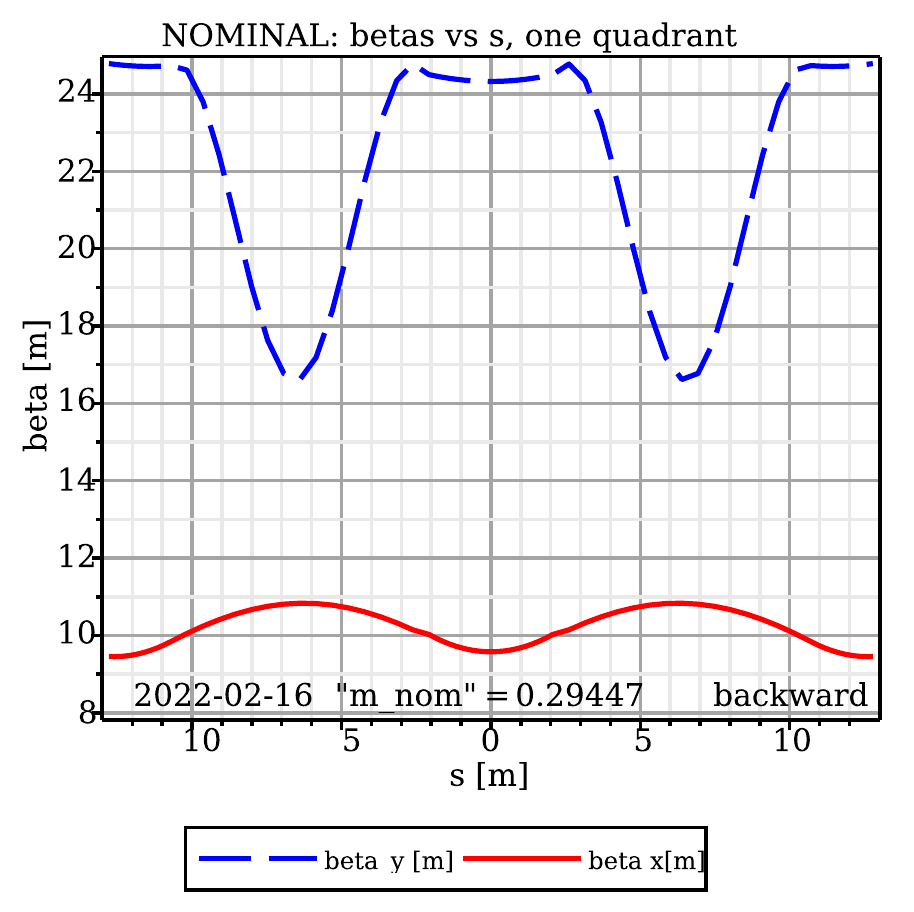}
\includegraphics[scale=0.60]{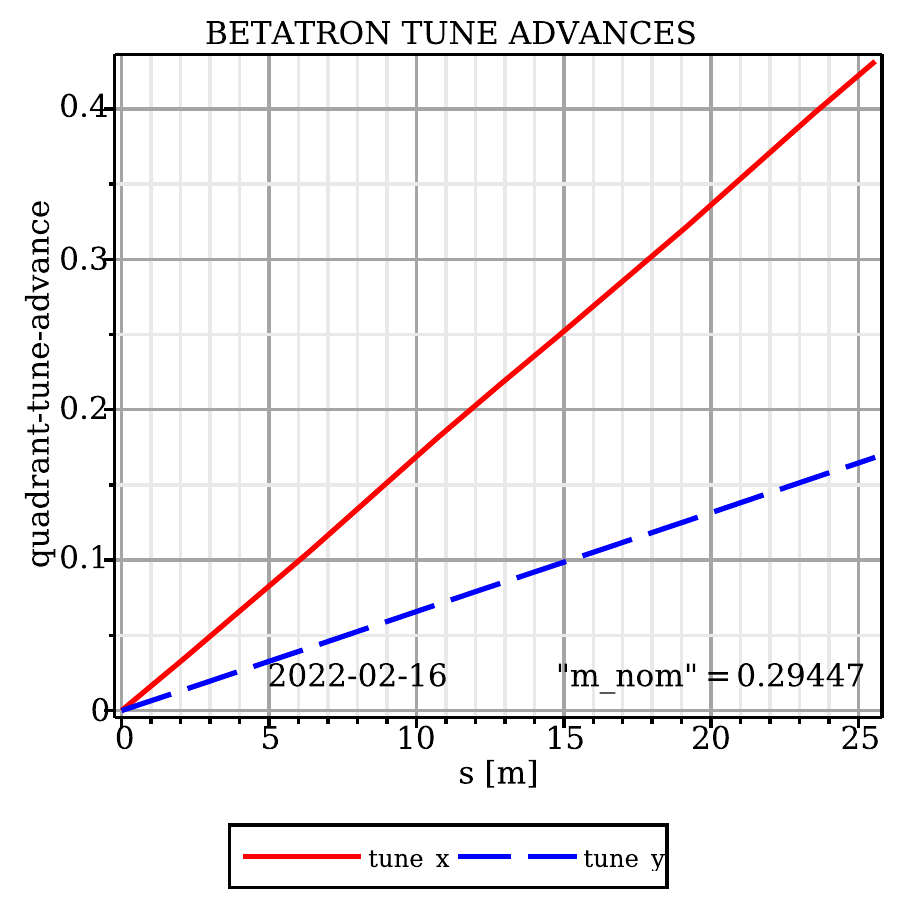}
\includegraphics[scale=0.60]{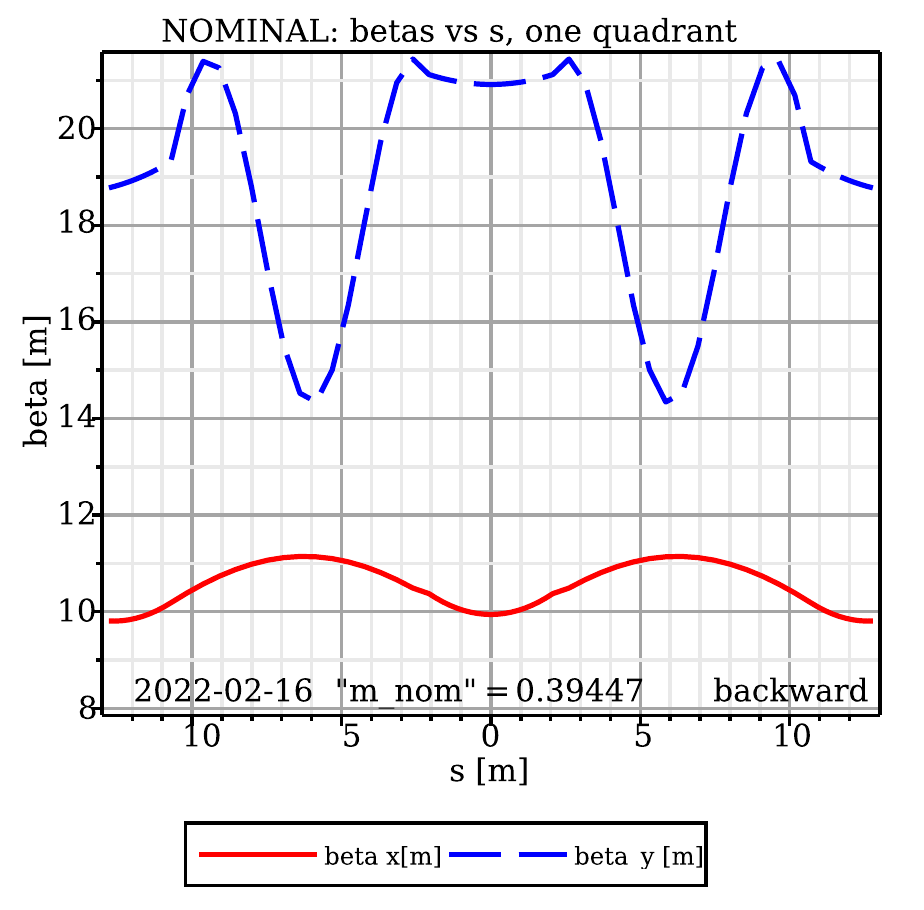}
\includegraphics[scale=0.60]{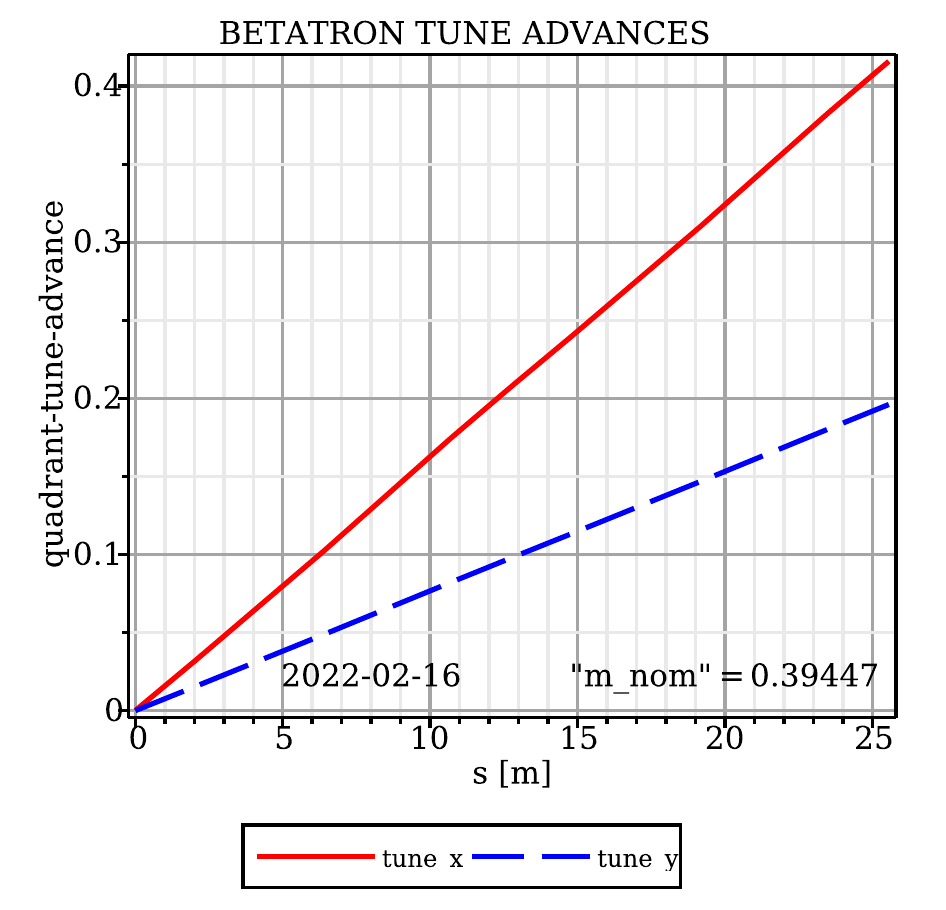}
\caption{\label{fig:FDDF-ToroidalQuadrant-betas} PTR optical functions for one quadrant centered
at a D-focusing quadrupole straight section, with all quadrupole strengths set to zero. 
{\bf Left:\ }beta functions 
plotted aginst longitudinal coordinate $s$; {\bf Right:\ }betatron tune advances plotted against 
longitudinal coordinate $s$; full ring tunes are four times the right hand intercepts. Regarded 
as a spectrometer, the upper case quadrant is ``under-focused'' the bottom case ``over-focused''.  
The central row is approximately matched to $\alpha_{\rm D}=0$ at the crossing point with all 
quadrupoles turned off. From this condition the full ring would be stabilized with very weak 
quadrupole strengths.}
\end{figure}

Figure~\ref{fig:FDDF-ToroidalQuadrant-betas} illustrates the result of lattice tuning in a storage ring with
``no lumped quadrupoles''.  The quotation marks here acknowledge that PTR will, in fact, have lumped quadrupoles.
But their strengths will be adjustable such that (under unrealistically ideal conditions) their strengths
will be tuned to exactly zero. The tuned-up situation is shown in the middle row of the figure; 
the upper/lower rows represent varying $m_{\rm nom}$ (where $m_{\rm nom}$ is another name for the $m$ parameter 
appearing in Eqs.~(\ref{eq:Wollnik-TM-params})) below/above the tuned-up condition.  For fixed ring geometry, 
quadrupole free, toroidal focusing required for long term stability in both horizontal and vertical planes 
depends on the value of this single parameter $m_{\rm nom}$.  

From the point of view of operational practicality, this behavior is problematical in two ways; once constructed,
neither the ring geometry nor the value of $m_{\rm nom}$ can be tuned conveniently.  One can contemplate altering  
$m_{\rm nom}$ by using a smooth mechanical way of adjusting the vertical curvature of the electrodes producing
the electric field---the horizontal curvature is fixed by the constant bending radius constraint---but no satisfactory
design exists for adjusting the electrode shape. This is because an electrode supple enough to be easily adjustable, 
would almost certainly be insufficiently rigid to guarantee the long term mechanical and thermal shape 
constancy required for precision mutual co-magnetometry.  The absence of lumped quadrupoles would mean that the entire ring 
design has to be ``dead-reckoned''.
%


For this reason the quadrupoles shown in Figure~\ref{fig:FDDF-ToroidalQuadrant-betas} are critical to the design
but need to be accurately tuneable to small values, consistent with achieving adequate ring stability.  
One might reasonably expect at least one iteration in which, having first established the quadrupole settings 
needed for stability, the electrodes could be mechanically deformed during a period of ring shutdown to enable 
the quadrupoles to be weaker.

This loss of operational flexibility has to be regarded as a serious impediment to achieving pure toroidal
focusing.  Nevertheless, the (eventual) benefits of toroidal focusing are so great, especially long SCT, that this 
``inconvenience'' has to be accepted.   Further investigation will be required to make sure the quadrupoles can
be strong enough to overcome realistically large initial design, fabrication, and positioning errors.  

\section{Forward and backward traveling ring optical functions\label{sec:BackwardForward}}
\begin{figure}[htb]
\centering
\includegraphics[scale=0.57]{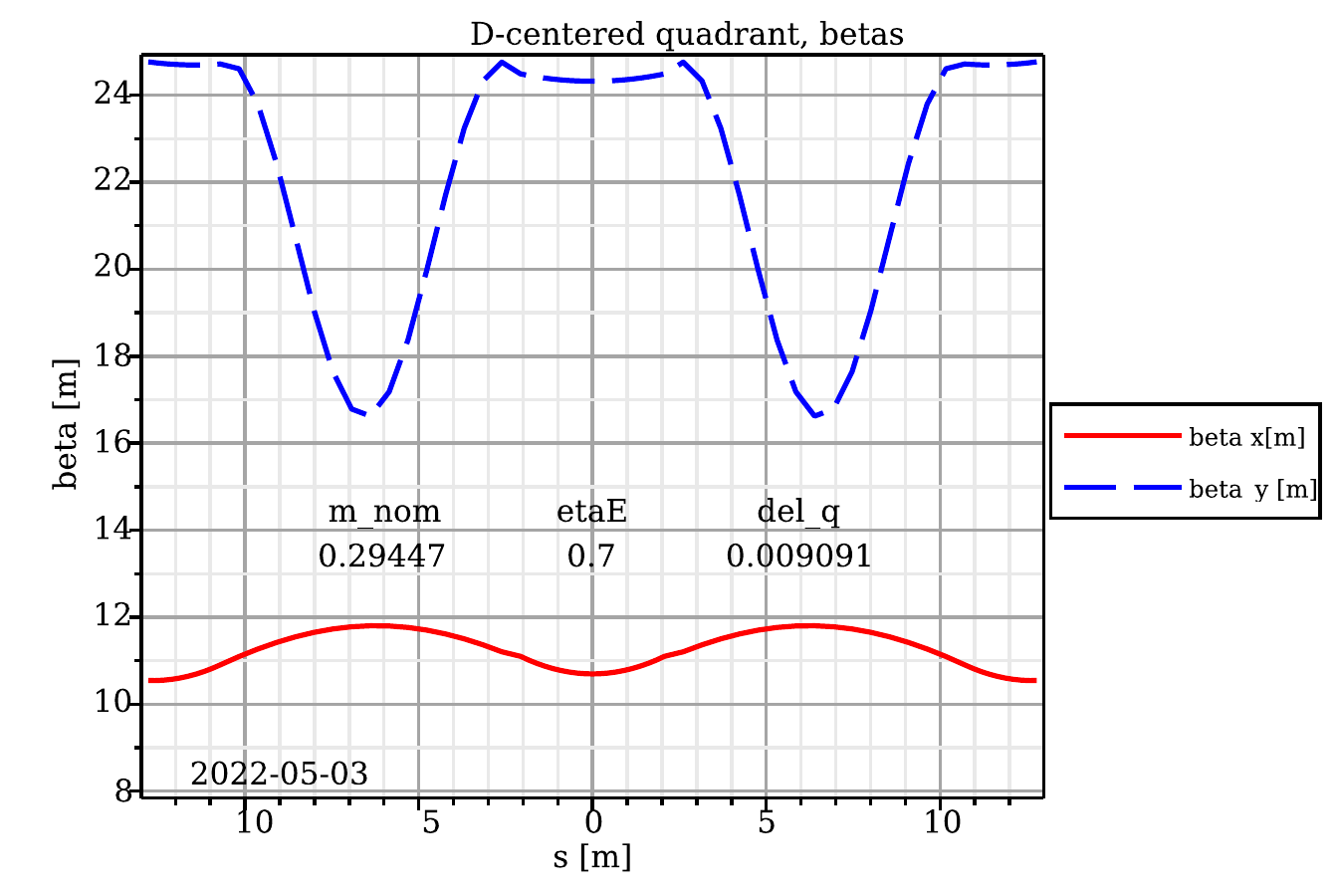}
\includegraphics[scale=0.57]{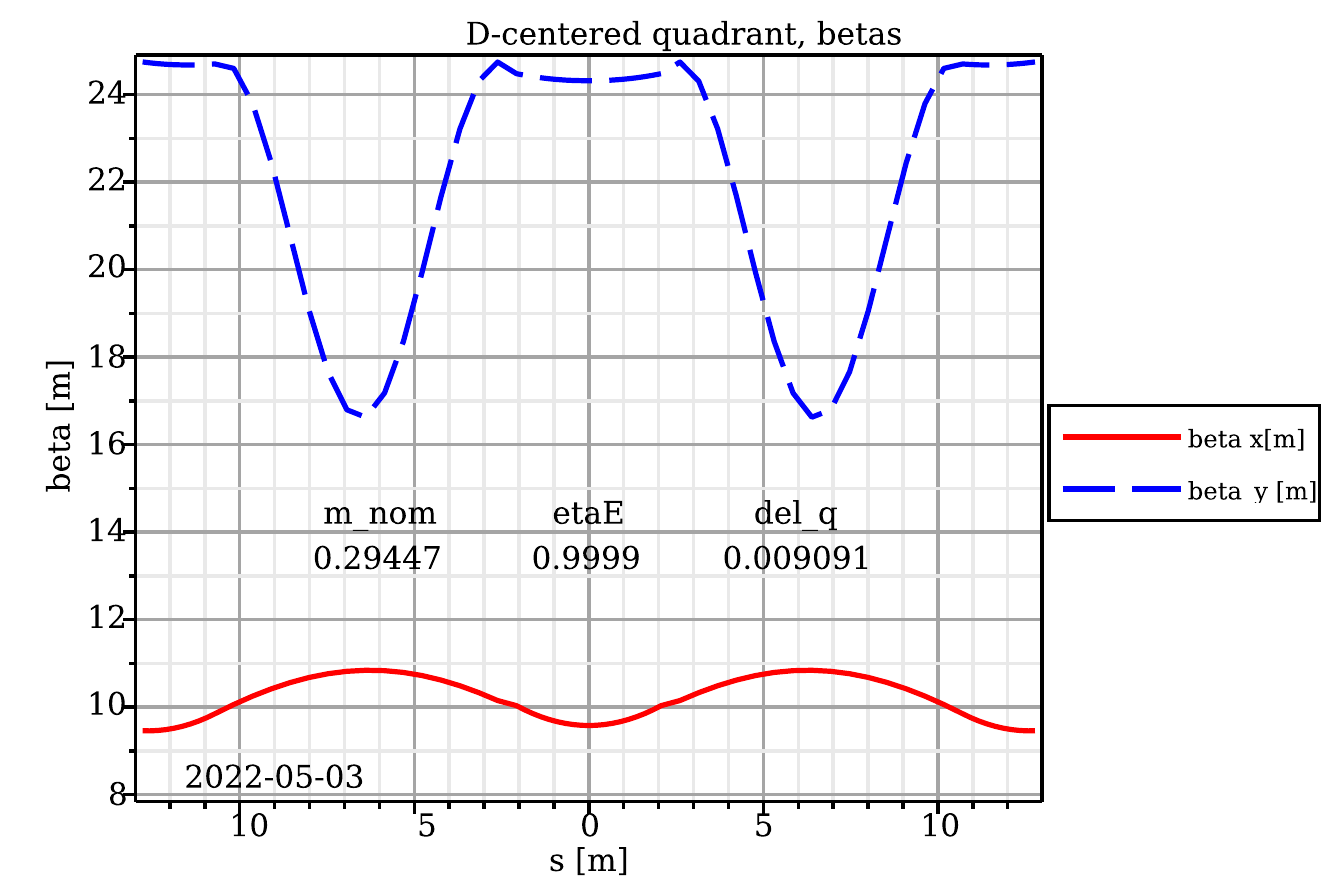}
\includegraphics[scale=0.57]{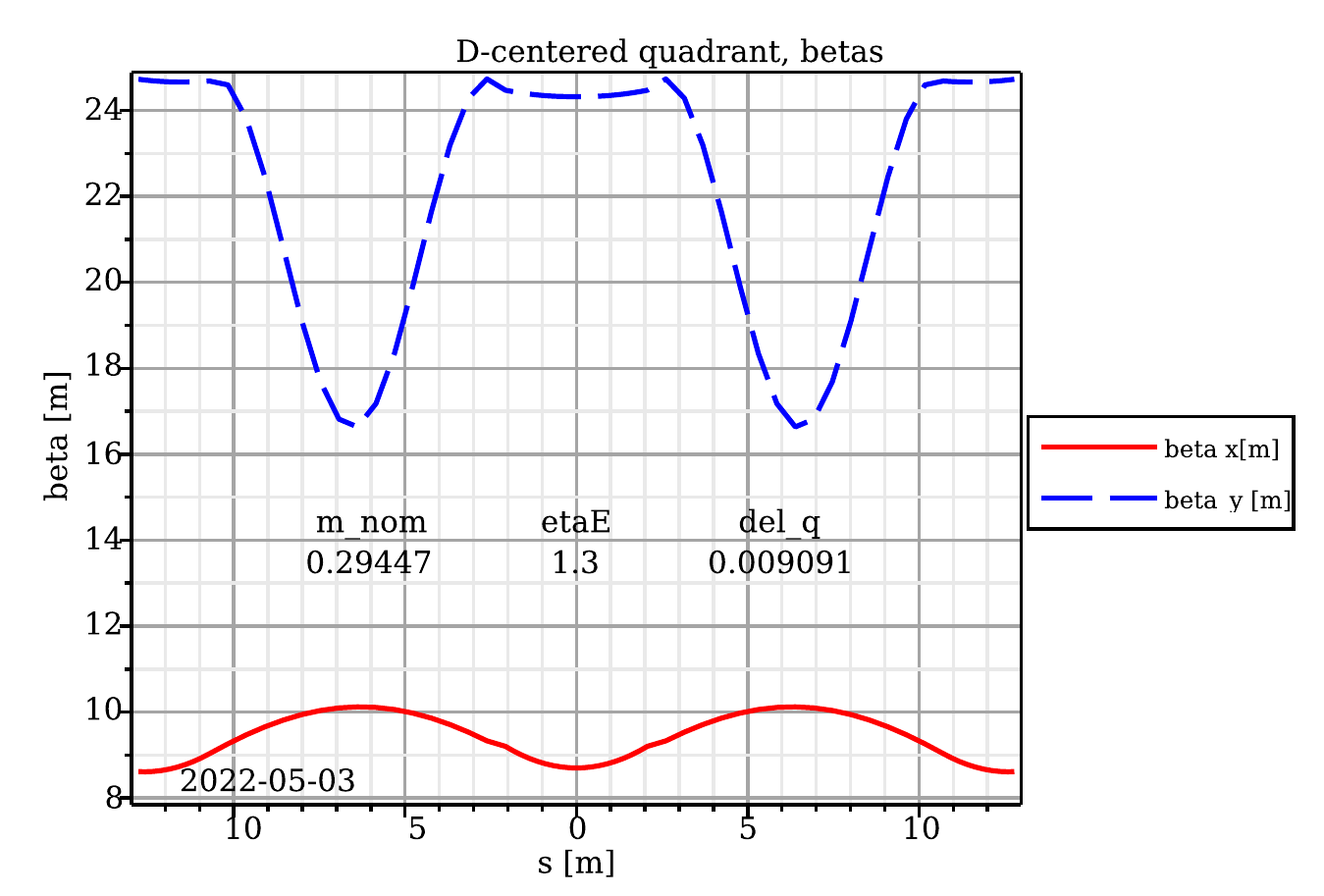}
\caption{\label{fig:etaE-dependence}PTR beta functions with $m_{\rm nom}=0.29447$, 
corresponding to the ``tuned-up'' value for the central case in Figure~\ref{fig:FDDF-ToroidalQuadrant-betas}.
In this configuration, in spite of being ``thick lenses'' the intervening bends act like drift sections
(of length not equal to the arc length in general).
Optically the full ring then acts like a separated function FODO lattice with (invisible) alternating gradient
lumped quadrupole with focusing strength $del_q=0.1/r_0=0.009091$/m.  Fractional electric field
bending fraction $\eta_E$ varies from 0.7 in the top graph to 1.3 in the bottom.  This variation leaves
$\beta_y$ unchanged, but causes $\beta_x$ to vary (approximately proportional to $1/\sqrt{\eta_E})$.
 }
\end{figure}

\begin{figure}[h]
\centering
\includegraphics[scale=0.52]{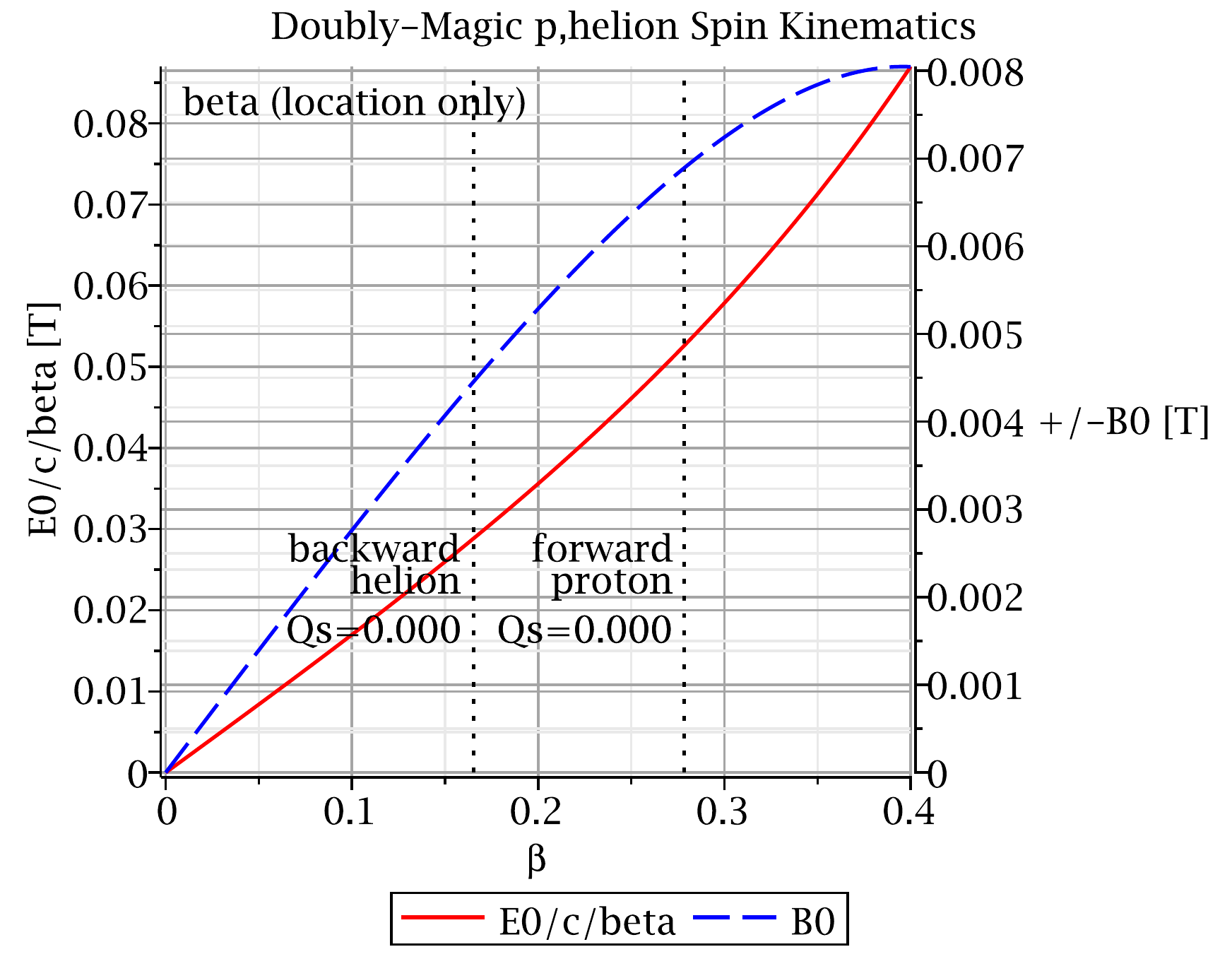}
\caption{\label{fig:DoublyMagic-p-helion}Doubly magic, simultaneously-circulating frozen spin  proton
and helion (He3 nucleus) beams. The axes register electric ``field'' $E0/v$ and magnetic field $B0$, both
measured in Tesla units. Both axes apply to the ``forward'', $\beta_p\approx0.28$ frozen spin proton beam
indicated by vertical broken line. The $\pm$ magnetic field sign indicates the magnetic force is 
constructive for protons, destructive for helions. The vertical broken line at $\beta\approx0.17$ 
indicates that a simultaneously backward traveling helion beam with that velocity would circulate stably,
also with zero spin tune.
}
\end{figure}

\begin{figure}[t]
\centering
\includegraphics[scale=0.42]{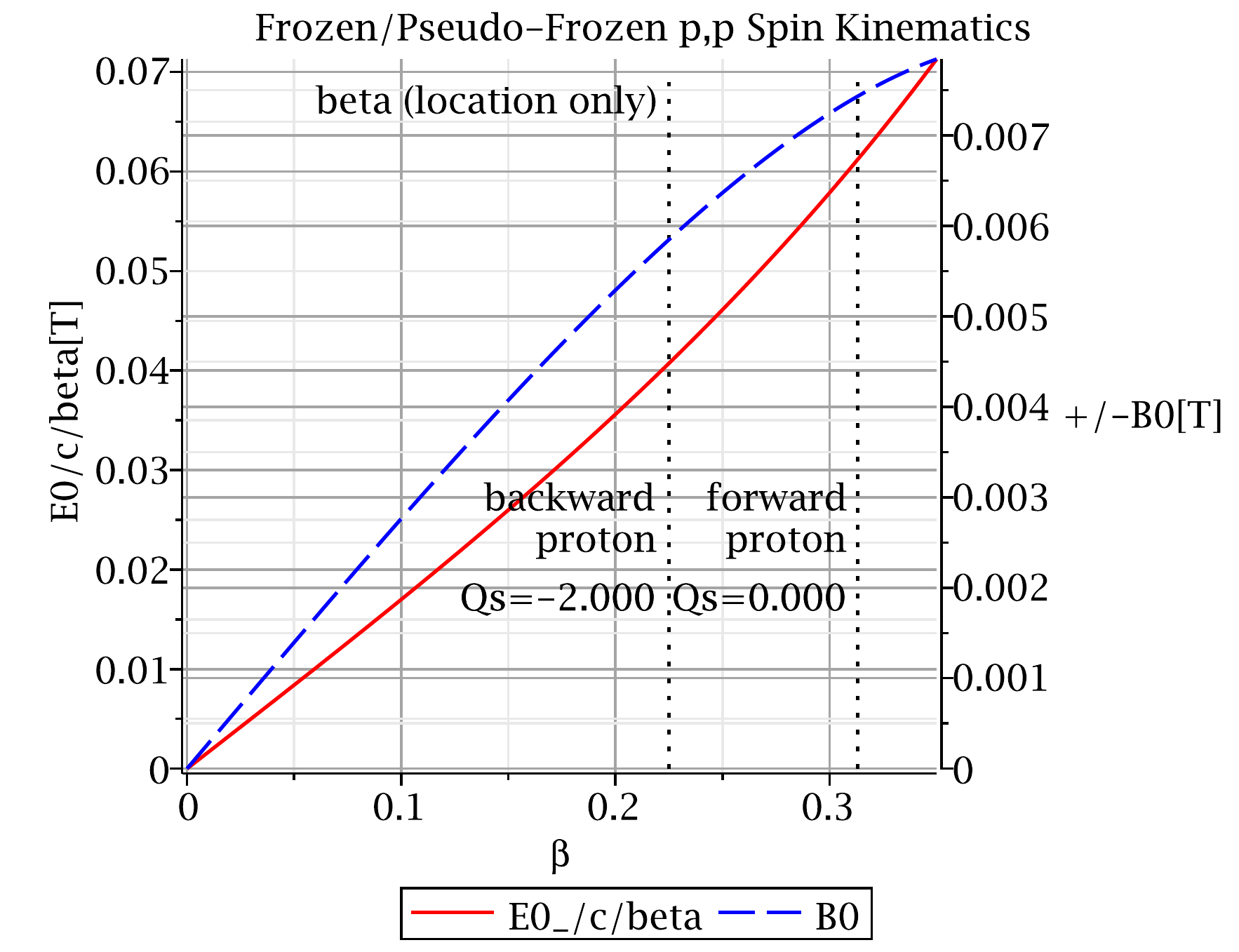}
\includegraphics[scale=0.42]{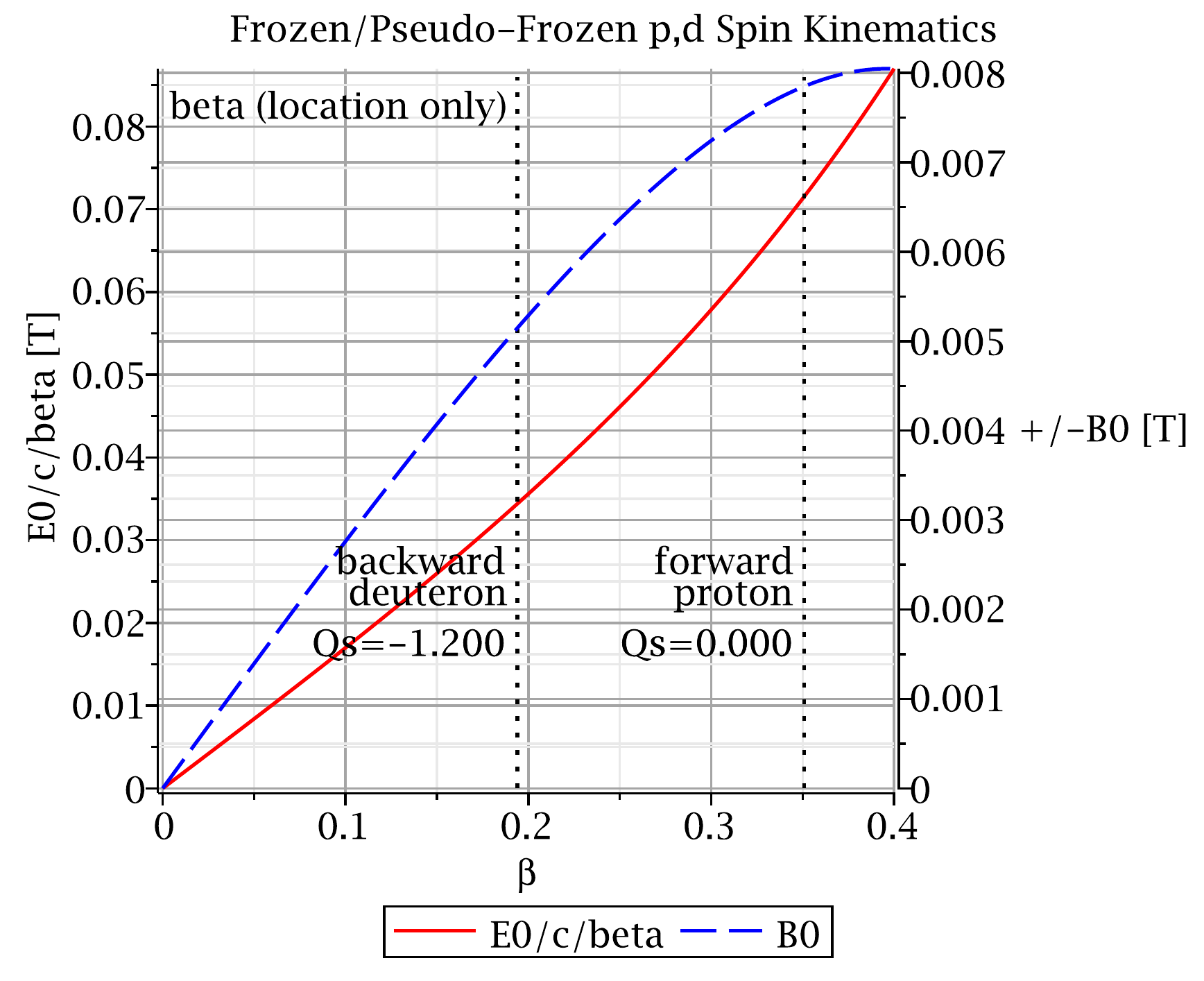}
\caption{\label{fig:MagicPseudoMagic-p-p}
Simultaneously counter-circulating beam kinematics in PTR. Left-hand (coarse) scale is $E/v$, 
right-hand (fine) scale is $\pm B$. The $\pm$ factor distinguishes forward/backward. The horizontal scale 
is $\beta=v/c$.  A forward traveling, 49.65\,MeV polarized proton beam has momentum such that spin tune 
$Q_s$ vanishes. A backward traveling (slightly adjustable) 24.73\,MeV proton beam can be tuned, e.g. 
to be pseudo-frozen.}
\end{figure}

Forward and backward traveling beam optics dependence is illustrated in Figure~\ref{fig:etaE-dependence}.
If the focusing is frugal in the forward direction, for example, $\eta_E=0.7$, $\eta_M=0.3$, then it
is extravagant, approximately $\eta_E=1.3$, $\eta_M=-0.3$ in the backward direction.
Bend parameters for ``doubly magic'' (both spin tunes equal zero) forward proton beam and backward 
traveling helion beams are plotted in Figure~\ref{fig:DoublyMagic-p-helion}. Similarly, 
Figure~\ref{fig:MagicPseudoMagic-p-p} shows parameters for a ``doubly frozen'' proton/proton combination, 
on the left, and a  ``doubly frozen''proton/deuteron combination on the right. In these graphs a 
forward traveling, polarized proton beam has momentum such that, with the electric and magnetic fields shown, 
the forward traveling proton spin tune $Q_s$ vanishes; in other words the ``magic'' condition for frozen spin 
is satisfied for the forward traveling protons.  Because the ring elements are backward/forward symmetric the 
lattice optics is satisfactory for both forward and backward beam propagation.  


On further reflection, forward/backward compatibility  may seem 
to be impossible; with superimposed electric and magnetic bending, if the bending is ``frugal'' 
(i.e. constructive) for the forward traveling beam, it is ``extravagant'' (i.e. destructive) for 
the backward traveling beam.  Does this not make it impossible for same particle type, but opposite 
direction beams to counter-circulate simultaneously; one or the other beams would surely be lost. 

No!  There is a ``loop-hole'' in this argument. If both beams are protons, then the particle 
momenta in the two beams can be sufficiently different for their ``$B\rho$ and $E\rho$ values'' 
(confusing accelerator jargon for beam momentum matched to the required
bending) to be appropriate for both beams; i.e. for both beams to have the same radius of curvature.  
The motivation for the $B\rho$, $E\rho$ usage is to ``geometricize'' the transverse optics, effectively 
converting accelerator focusing design into Newtonian optics. In the language of geometric optics, quadrupole 
strengths are expressed as inverse focal lengths.  It is this feature that allows the choice of 
PTR length scale to be deferred\footnote{There is no equivalent trick for ``geometricizing'' 
longitudinal dynamics---parameters such as RF frequency depend on times-of-flight along
particle orbits, a concept not present in geometric optics.}.

For counter-circulating beams having the same particle type, for example protons, one may
have $\beta_p= 0.31304$ for the forward beam, and $\beta_p=-0.31304\times41/47$ for the backward beam.
Figure~\ref{fig:MagicPseudoMagic-p-p} contains graphical descriptions by which the parameters of
simultaneously counter-circulating beams can be established. This requires reduced momentum for
a CCW proton beam that can counter-circulate in a ring with the same accelerator settings. These 
settings could, for example, freeze the spins of the CW beam. This would still leave some freedom
concerning the spin tune of the CCW beam.  Though the CCW beam could not be arranged to have 
globally frozen spins it could, in some cases, be adjusted to be 
pseudo-frozen.  With counter-circulating beams of different particle type, PTR could
serve as an ``MDM comparator''. This could be used, for example, to make a highly precise measurement 
of the ratio of spin-tunes (and hence MDMs) of two simultaneously circulating beams.

Though both beams on the left graph in Figure~\ref{fig:MagicPseudoMagic-p-p} are protons, the spins of only
one can be frozen---with constructive bending there is only one magic combination of electric and magnetic 
bending.  There remains one freedom, however---for example to pseudo-freeze the counter-circulating beam.
The graph on the right of Figure~\ref{fig:MagicPseudoMagic-p-p} shows the kinematics of counter-circulating
protons and deuterons.

\section{Canceling decoherence by intentional operation ``on the coupling resonance''}
A powerful strategy for increasing the spin coherence time (SCT) (which is equivalent to 
``canceling spin decoherence'') is to intentionally mix degrees of freedom. Some such cancellation 
occurs automatically in longitudinal synchrotron oscillation, as particle energies systematically cycle
between minimum and maximum values.  What is proposed in this section 
is to cancel decoherence associated with transverse (i.e betatron) oscillations, by intentionally
mixing horizontal and vertical degrees of freedom.  

For the $\eta_E$=1 optics shown in Section~\ref{sec:BackwardForward} the \emph{fractional} betatron tunes
$Q_x/Q_y=0.726/0.673$ are quite close, meaning close to a difference resonance.  The effect of any skew quadrupole, 
present in the ring either intentionally or unintentionally, is to excite ``out of plane'' betatron oscillations.
As the tunes are brought closer together the amplitude of cross plane excitation increases until, exactly on 
resonance, the beams become round. 

Though $x,y$ coupled sum resonance is fatal, coupled difference resonances are especially stable for dual beam 
stability (though disappointingly not helpful for high luminosity beam-beam operation, as first investigated 
and established at the Orsay storage e+/e- storage ring ACO\ \cite{Haissinski} as well as at 
CESR\ \cite{ChromaticitySharing}\cite{RoundBeams}.)  The optimal ACO tune ratio was $Q_x/Q_y=1.86/2.86$.

\begin{figure}[hbt]
\centering
\includegraphics[scale=1.0]{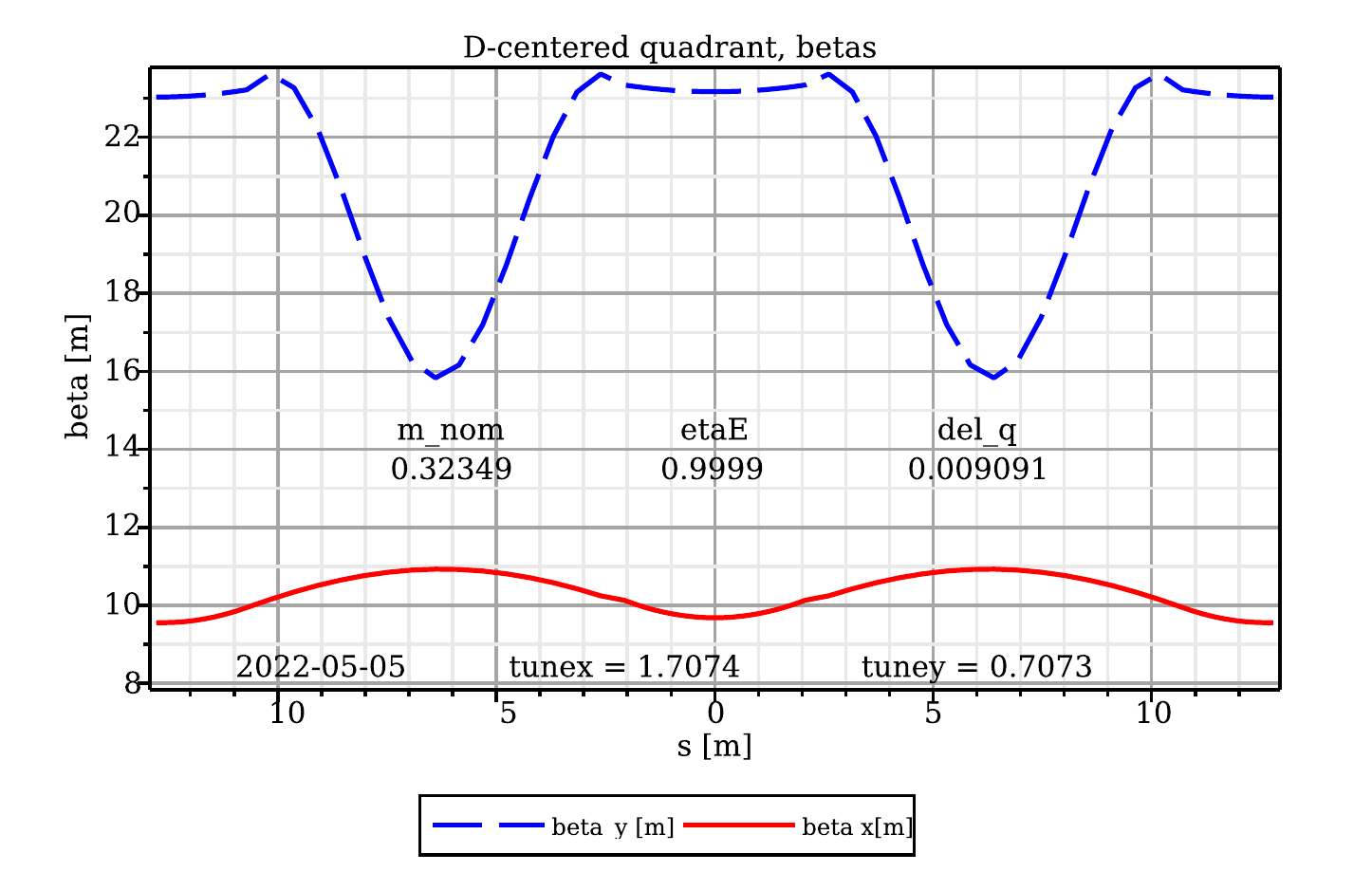}
\caption{\label{fig:0p32349-0p99990-0p00909}
Lattice functions for essentially all electric bending ($\eta_E=0.9999$), and quite weak lumped quadrupole tuning
with  ($del\_q=0.009091\,{\rm m}^{-1}$).
The horizontal and vertical fractional tunes have been made equal (0.7073). In this condition, by itself, the presence of 
betatron motion causes no beam depolarization.  
}
\end{figure}

With careful tuning the fractional PTR tune equality can be made exact (as well as shifted away from third 
integer resonance, to which $Q_y$ has so far been unacceptably close.  The result of doing this tuning is shown in 
Figure~\ref{fig:0p32349-0p99990-0p00909}.  Relative to the central case in Figure~\ref{fig:etaE-dependence} this 
mainly entailed the change: focusing index $m_{\rm nom}= 0.29447\longrightarrow0.32349.$ The result of this change is shown in
Figure~\ref{fig:0p32349-0p99990-0p00909}.  The altered fractional tunes have become equal, as shown at the bottom 
of the plot.  Otherwise lattice focusing functions have been little changed.\footnote{Raw lattice description files 
for PTR tuned to the conditions of Figure~\ref{fig:0p32349-0p99990-0p00909} are available at 
\href{https://github.com/jtalman/ual1/tree/master/examples/lattices}{Click Here} ,
in four different formats, in four different sub-directories, con-xml, nocon-xml, adxf, and sxf,
always with filename having matching suffix, such as PTR\_m\_nomEQ0p0.32349\_sl4.sxf. }

Figure~\ref{fig:0p32349-0p99990-0p00909} represents ``fully mixed'' long coherence time operation in the
all-electric case but, for realism, with a small, one part in 10,000 magnetic bending component.
Replacing $\eta_E$ by 1.0001 and/or reversing the sign of $del_Q$ in any combination, produces 
essentially the same graph. Replacing $\eta_E$ by 1, however, produces no solution.  This is
of no physical significance---the code failure is caused by zero denominators in the calculation 
of magnetic field influence on the kinematics.

On the coupling resonance any ``cross-plane coupling'' causes the beam to become round (or rather
to have equal horizontal and vertical emittances, because their sum $\epsilon_x+\epsilon_y$ 
is conserved)\ \cite{CourantSnyder}).  For any particular particle $\epsilon_x$ and $\epsilon_y$ can be 
interpreted as ``Courant-Snyder invariants'' (whose sum is a constant of the motion).   
By virtue of cycling regularly between horizontal and vertical betatron motion, in this condition, betatron 
motion induces the same spin precession for every particle having the same $\epsilon_x+\epsilon_y$ sum.
As a result (ignoring rare stochastic events) there is no spin decoherence among all particles lying on any 
centered ellipse in betatron phase space---which is to say, every particle.  In this condition, by itself, 
the presence of betatron motion causes no beam depolarization. This has been confirmed 
by simulation, indirectly, in the past.

The true situation is not quite as simple as has just been described.  With betatron oscillations ignored, 
the presence of dispersion causes the beam to be dispersed radially, with off-momentum bending radius proportional 
to momentum, which is different for every particle.  At the same time there is a similar mixing in the synchrotron 
phase space.  This complicates the previous discussion.  The longitudinal mixing itself is purely beneficial
from the point of view of maximizing SCT.  But the chaotic mingling of transverse and longitudinal mixing
has the potential for introducing unpredictable resonance effects. This possibility will have to be addressed
by simulation or by more sophisticated analysis than has been attempted here.

To the extent that the relativistic $\gamma$ factor is the same for every particle and the resulting motion 
remains purely horizontal, synchrotron oscillation by itself would also not induce depolarization via decoherence.  
Since neither of these conditions is exactly true there will be a systematic precession proportional
to $\gamma$ offset. Since both electric spin tune $Q_E$ and magnetic spin tune $Q_M$ depend
(only) on $\gamma$, there is a strong tendency for spin decoherence to be closely correlated with
energy offset.  To leading order in synchrotron oscillation amplitude, the decoherence associated with 
synchrotron oscillations would therefore still cancel. Regrettably, without care, synchrotron oscillations tend to 
be quite nonlinear. However, by superimposing weak third harmonic RF cavities, the spin decoherence from
this source can also be canceled to high accuracy. These effects can also be studied by simulation. 

Optimistically one concludes that decoherence will be sufficiently canceled to have become insignificant,
even with no powered sextupoles employed to increase polarization SCT.
This does not yet mean that day-long runs can be anticipated.  The phase-locking on which everything in this paper 
is predicated relies on efficient polarimetry.  Yet, at this time, non-destructive polarimetry does not exist. 
One can assume, therefore, that run durations will be dominated by particle loss associated with 
polarimetry. 

It should be understood that most of the tasks PTR has to perform in its role as EDM prototype do not depend on
subtle effects such as have been discussed in this section, nor on the no-quadrupole requirement for optimizing 
EDM measurement precision that led to the lattice design shown in Figure~\ref{fig:PTR-layout-Toroidal8_102p2-mod}.
As already anticipated in the CYR, the commissioning is expected to proceed in successive 
phases each lasting as long as a year.  It is only in later phases, for optimal MDM and EDM measurements to proceed, 
that the delicate toroidal focusing requirements will need to be met.  

\section{``Quadrupole free'' PTR transfer matrix lattice design}
\subsection{Transfer matrix algebra}
For a $2\times2$ transfer matrix ${\bf M}$ to be symplectic it must satisfy the algebraic relation
\begin{equation}
{\bf M}^{-1} = -{\bf S}{\bf M}^T{\bf S},
\quad\hbox{where}\quad {\bf S} = \begin{pmatrix} 0 & -1 \\ 1 & 0  \end{pmatrix},
\label{eq:M01p}
\end{equation}
and the only imposed constraint for $2\times2$ transfer matrices requires them to have unit determinant. For any higher 
(necessarily even) dimensions the same equation is satisfied with ${\bf S}$ being replicated along the diagonal.
This introduces symplecticity conditions going beyond the unity determinant requirement.

Transfer matrices ${\bf M_x}$ and ${\bf M_y}$ for one PTR superperiod as well as powers up to ${\bf M_x^8}$ 
and ${\bf M_y^8}$, which describe two consecutive complete revolutions follow: 
\begin{math}
\\
{\bf Mx} = 
\left[\begin{array}{cccc}
 0.33015 &  9.04118 &  0.0 &  8.68780 
\\
 - 0.09855 &  0.33015 &  0.0 &  1.27816 
\\
 - 1.27816 & - 8.68780 &  1.0 &  204.00230 
\\
  0.0 &  0.0 &  0.0 &  1.0 
\end{array}\right]
\end{math}

\begin{math}\label{(14)}
{\bf My} = \left[\begin{array}{cc}
 0.95513 &  12.95238 
\\
 - 0.02167 &  0.95513 
\end{array}\right]
\end{math}

\begin{math}
{\bf Mx^2} = \left[\begin{array}{cccc}
- 0.78201 &  5.96980 &  0.0 &  23.11206 
\\
 - 0.06507 & - 0.78201 &  0.0 &  0.84395 
\\
 - 0.84395 & - 23.11206 &  1.0 &  385.79588 
\\
  0.0 &  0.0 &  0.0 &  1.0 
\end{array}\right]
\end{math}

\begin{math}
{\bf My^2} = \left[\begin{array}{cc}
 0.63162 &  24.74233 
\\
 - 0.04139 &  0.63162 
\end{array}\right]
\end{math}

\begin{math}
{\bf Mx4} =  \left[\begin{array}{cccc}
 0.22307 & - 9.33687 &  0.0 &  10.07647 
\\
  0.10177 &  0.22307 &  0.0 & - 1.31996 
\\
  1.31996 & - 10.07647 &  1.0 &  732.58076 
\\
  0.0 &  0.0 &  0.0 &  1.0 
\end{array}\right]
\end{math}

\begin{math}
{\bf My^4} = \left[\begin{array}{cc}
- 0.62514 &  31.25562 
\\
 - 0.05229 & - 0.62514 
\end{array}\right]
\end{math}

\begin{math}
{\bf Mx^8} = \left[\begin{array}{cccc}
- 0.90048 & - 4.16564 &  0.0 &  24.64854 
\\
  0.04541 & - 0.90048 &  0.0 & - 0.58890 
\\
  0.58890 & - 24.64854 &  1.0 &  1491.76252 
\\
  0.0 &  0.0 &  0.0 &  1.0 
\end{array}\right]
\end{math}

\begin{math}
{\bf My^8} = \left[\begin{array}{cc}
- 1.24343 & - 39.07858 
\\
  0.06537 & - 1.24343 
\end{array}\right]
\end{math}

\subsection{Interpretation of the transfer matrices}
Even for the $4\times4$ matrices (mentally-partitioned $2\times2$) one can check mentally (at least at the 10\% level) that 
all determinant values are close to 1.  Displayed only to 6 place accuracy, each of these matrices satisfies 
Eq.~(\ref{eq:M01p}) to machine precision. In fact, this too is an understatement; these matrices are, in fact, 
analytically exact.  They have been obtained purely algebraically. 

This is all that is required for the transverse propagation to be Hamiltonian, and hence loss-free.
No paraxial approximation, nor small angle, nor small amplitude approximations have been made; no truncated 
power series have been employed, no linear, quadrupole, sextupole and higher orders introduced. 
Admittedly these statements are true primarily because the starting transfer matrix was assumed to be exact.
This is an excellent (but not perfect) assumption for the toroidal focusing design, which has been
adopted in modern day (Scanning) Transmission Electron Microscopy (STEM), as described most recently
in a textbook by P. Erni\ \cite{Erni-TEM}, with clear elementary physical discussion, especially in Section~5.

\subsection{Longitudinal optics; beam capture into stable buckets}
In spite of this splendid transverse evolution description one cannot fail to notice that the horizontal 4x4 
matrices are, in some sense, divergent. Concentrating on the $M_{34}$ elements, referring to 
Eq.~(\ref{eq:Wollnik-TM-x}), and 
visualizing $\delta$p/p=$10^{-5}$ as a possible fractional momentum offset, one finds that, after a single turn, the 
longitudinal arc length parameter $s$ will have advanced by 0.75\,cm, relative to a central particle. 

For a few turns in a ring of circumference 100\,m this advance would be scarcely noticeable, but circulating at
roughly a MHz rate for, say, two minutes, the beam spread will be of order $10^6\,$m. In other words the beam will
have by then spread around the ring into a uniformly distributed coasting beam.  
To keep the beam bunched, which is required for phase-locking, the PTR beam has to be quickly captured into stable 
RF buckets.  This should be straightforward.

\section{Practical simultaneous frozen spin beam combinations\label{sec:PracticalCombinations}}
A selection of practical frozen spin combinations are given in Tables~\ref{tbl:p-d2} and \ref{tbl:p-He3}.
Phase-locked parameters are given as rational numbers (i.e. ratios of integers). 
Table~\ref{tbl:p-d2} shows ``doubly frozen'' proton, deuteron combinations, for which the primary beam spin 
tune vanishes, while the secondary beam spin tune, though a rational fraction, is not zero.  
Table~\ref{tbl:p-He3} shows the ``doubly magic'' proton, helion case in which both spin tunes vanish.  
In all of these cases the RF cavity frequencies are also given as rational fractions.  In this condition the 
RF frequency can also be phase-locked, but it is only the primary beam for which the EDM-induced, out of 
plane precession accumulates monotonically. 

In all cases the frequencies are ``over-determined'' in the sense that the circumferences of primary and 
secondary beam, though close, are not quite identical.  The defects are spelled out in the figure captions.  
With orbits not quite identical, the precession-inducing error fields to which the beams are subject are not 
quite equal.  To leading order, the EDM measurement error will be canceled by averaging over runs with primary 
and secondary beams interchanged.

Initially this defect can be analyzed and compensated for by theoretical calculation of corrections to the 
measured EDM values.  The issue can also be faced experimentally be introducing a \emph{very weak} secondary 
RF cavity, synchronized to the secondary beam and strong enough to alter the secondary beam radius to match 
the primary beam radius.  To avoid affecting the primary beam, this longitudinal RF signal, synchronized to 
the secondary beam  would have to be gated to affect only the secondary beam. This would require gate opening 
and closing switching times at the 10\,ns level. 
\setlength{\tabcolsep}{1pt}
\begin{table}[htb]
\centering
\begin{tabular}{|c|ccccc|cc|cccccc|c|} \hline 
 bm  &     m1 & G1      &q &$\beta_1$&  K1  & E0*  &  B0* & m2     &  G2   & q  &$-\beta_2/\beta_1$&  KE2 & Qs2  & bm \\ 
  1  &    GeV &         &  &         & MeV  & MV/m & mT  & GeV    &       &    &  $*\dagger$      &  MeV &      &  2 \\ \hline
     & $r_0*=$ & 11.0\,m &  & Qs1=0  &&  {\bf PRESENT}    & {\bf DAY, 2022}  &    &       &    &     PTR         &      &      &    \\ \hline
   p & 0.9383 & 1.7928  & 1& 0.31304 & 49.7 & 6.26 & 8.11  & 0.9383 & 1.79  &  1 & 41/57           & 24.7 &-2/1  &  p \\ 
   p & 0.9383 & 1.7928  & 1& 0.29175 & 42.7 & 5.33 & 7.76  & 1.8756 &-0.57  &  1 & 53/100          & 22.8 &-9/8  &  d \\ 
   d & 1.8756 & -0.5713 & 1& 0.18000 & 31.1 &-7.73 & 74.12 & 1.8756 &-0.57  &  1 & 83/172          & 7.12 &-25/31&  d \\ 
   d & 1.8756 & -0.5713 & 1& 0.17760 & 30.3 &-7.52 & 73.11 & 0.9383 & 1.79  &  1 & 35/67           & 4.06 &43/127&  p \\  \hline
\end{tabular}
\caption{\label{tbl:p-d2}
Quantities expressed as rational fractions are exact and phase locked to arbitrarily high accuracy.
Except for PTR construction itself, these pairings use only \emph{present day technology and apparatus}.
Beam~1: is globally frozen in every case ; $Qs1=0/1$; so \emph{the beam~1 EDM signal accumulates monotonically}.
Except for $r_0*$, $*\dagger$, E0* and B0* columns, all entries are \emph{EXACT}, either \emph{integers}, or
(truncated) \emph{physical constants} or \emph{calculable,} kinematic quantities.
``Closed orbit mismatches'', from the  $\dagger$ column, are
0.71927 = 41/57   - 0.0000282, 
0.52997 = 53/100  - 0.0000299,
0.48261 = 83/172  + 0.0000518,
0.52235 = 35/67   - 0.0000380.
}
\end{table}
\begin{table}[htb]
\centering
\begin{tabular}{|c|ccccc|cc|cccccc|c|} \hline 
 bm  &     m1 & G1      &q &$\beta_1$&  K1  & E0*  &  B0* & m2         &  G2   & q    &$-\beta_2/\beta_1$& KE2 & Qs2 & bm \\ 
  1  &    GeV &         &  &         & MeV  & MV/m & mT   & GeV        &       &      &  $*\dagger$     &  MeV &     &  2 \\ \hline
     & $r_0*=$ & 11.0\,m &  &        &      & {\bf NEAR} & {\bf FUTURE} &       &      &    &     PTR    &      &      &    \\ \hline
   p & 0.9383 & 1.7928  & 1& 0.27831 & 38.6 & 3.90 & 6.13  & 2.8084     &-4.18  &  2   & 107/180         & 39.2 & 0/1  &  h \\ 
   h & 2.8084 & -4.1842 & 2& 0.16544 & 39.2 & 3.90 & -6.13 & 0.9383     & 1.79  &  1   & 180/107         & 38.6 & 0/1  &  p \\  \hline
\end{tabular}
\caption{\label{tbl:p-He3}{\bf Doubly magic, proton/He3 combination case: }Quantities expressed as rational fractions are exact and 
can be phase locked to arbitrarily high accuracy, Rational fraction entries in the $\dagger$ column of the table are fixed by RF 
synchronism; the orbit circumference ratios are close, but not exact.  Precise magnetic field reversal is assured by three-fold 
phase locking as spelled out in Section~\ref{sec:MeasurementStrategy}.}
\end{table}

\clearpage

\section{Error estimation}
\subsection{Insignificance of horizontal- relative to vertical- orbit dipole moment control}
One notes, from Eqs~(\ref{eq:BendFrac.7}), that electric spin tune $Q_E$ and 
magnetic spin tune $Q_s$ depend kinematically only on the relativistic factor $\gamma$, which is
phase locked and globally constant for stored beams.  Though not stated previously, this simple behavior 
assumes the central orbit lies in a single (typically horizontal) plane.  For constant $\gamma$ MDMs 
are therefore unaffected by horizontal bend errors to all orders. By up-down symmetry, MDMs
are constant only to linear order in vertical displacement error.  As a consequence, it is mainly
vertical positioning that needs to be tightly specified; here to $\pm 0.1\,$mm, or less,
as required.

\subsection{Insignificance statistical- relative to systematic- error in EDM measurement}
An essential difference between neutron and proton EDM measurements comes from the fact that, though dominant
for the neutron EDM measurement, the statistical proton EDM measurement error is negligible by comparison.  A convenient way
to exploit this is to note that the statistical error for a ``nominal week long'' ($2\times10^5$\,s) storage ring measurement 
will be about $\pm10^{-29}\,e$\,cm.  This is an accuracy level which (conventional) thinking has suggested to be a plausible BSM 
proton EDM.  For all practical purposes, the implication for storage ring EDM measurement is that we can ignore counting statistics
and concentrate entirely on systematic errors. 

\subsection{Reduction of systematic error by averaging over reversed beam conditions}
\begin{itemize}
\item
CW/CCW beam reversal requires reversing the magnetic field.
\item
Setting, reversing, and resetting magnetic field with frequency domain accuracy relies 
only on MDMs acting as ``magneto-sensitive gyroscopes'' for phase locking both betatron tunes.
\item
This avoids the need for (unachievably precise) magnetic field measurement, while allowing systematic error 
reduction by averaging over CW/CCW reversal.
\end{itemize}

\subsection{Error estimation for the doubly magic proton-He3 combination\label{sec:ErrorEstimation}}
The doubly magic proton-He3 combination has been especially emphasized in the present paper. This is because, 
with equipment currently in hand, or at least described in careful design proposals, the detection of BSM
physics, in the form of non-vanishing EDMs of fundamental particles is possible inexpensively in the near future, 
using proven operational techniques.  Parameters for the (essentially unique) proton/He3 \emph{doubly-magic} pairing 
are given in Table~\ref{tbl:p-He3}.

The EDM signature is out-of-plane MDM precession.  The dominant systematic error is due to unintentional
(and unknown) radial magnetic field error acting on particle MDMs, which mimic the effect of intentional electric field 
acting on particle EDMs.  Such unknown magnetic sources are discussed in Chapter 3 of reference\cite{EDM-Challenge}.
In particular, Section~3.3.4 of that book argues that the systematic storage ring proton EDM error can be estimated to
be comparable to the systematic Ramsey neutron EDM error (coming from the same source). Copying from the first
sentence of the abstract to the present paper, the current day neutron EDM systematic error is
$\pm0.2_{\rm sys}\times10^{-26}\,e$\,cm (five times smaller than the neutron EDM statistical error
in the same experiment).

Another way of estimating the storage ring proton EDM systematic error is based on the precision with which the vertical
separation of countercirculating beams can be minimized---zero average vertical beam separation would imply 
vanishing radial magnetic field average, $\langle B_r\rangle$ and, therefore, zero spurious out-of-plane precession.
Uncertainties in this estimate come from the estimated beam position monitor (BPM) precision, and the extent to which 
the vertical BPM sensitivity can be increased by reducing the vertical focusing strengths.  As regards order of magnitude, 
estimates by the author\cite{EDM-Challenge}, calculations by Tahar and Carli\ \cite{TaharCarli} and by Lebedev\ \cite{Lebedev} 
are consistent with the $\pm0.2_{\rm sys}\times10^{-26}\,e$\,cm systematic error estimate given in the preceeding paragraph. 

\subsection{Systematic error cancellation in $\Delta$ = EDM[proton]-EDM[He3] measurement}
Treated as composite, the proton consists of quarks and gluons, but treated as a fundamental particle, the proton
EDM should vanish, or at least be very small.  Consisting of two protons and a neutron, the He3 nucleus
is unambiguously composite.  To the extent that T-symmetry is respected, as in the standard model, He3 would
still be required to have vanishing EDM.  Recognizing that it is quite challenging for the electric center and
the mass center of two protons and one neutron to be identical, in an experiment looking for T-violation, one can 
conjecture that the magnitude of any He3 EDM is significantly greater than the proton EDM.  

For the mutual co-magnetometry emphasized in the present paper, this suggests treating any out-of-plane proton
precession as spurious (i.e. caused by spurious radial magnetic field acting on the proton EDM).  Why not, therefore,
utilize the proton for mutual co-magnetometry by phase locking the proton EDM into the horizontal plane, which amounts to 
cancelling $\langle B_r\rangle$.  In this circumstance, any measureable out-of-plane precession of the He3 beam 
polarization can be interpreted as a measure of the He3 EDM.  This is a picturesque way of visualizing our experiment
designed to measure $\Delta$ = EDM[proton]-EDM[He3].  (Among the many other things that could go wrong) there
remains the possibility that, by chance, proton and He3 EDMs are equal, but opposite in sign---this would be both 
surprising and unfortunate. In any case non-zero $\Delta$ would still imply BSM physics.

\subsection{Anticipated systematic errors for various beam configurations} 
Based on the previous discussion, systematic errors anticipated in various configurations are listed in Table~\ref{tbl:SystError}.
In all cases one will reduce systematic error by averaging over CW and CCW circulation.
The ``consecutive'' column systematic error assumes that simultaneous CW and CCW beam circulation has been established.
Successive runs alternate between CW and CCW.  ``Concurrent'' columns assume the simultaneous phase-locked circulation of 
CW and CCW beams, with at least one beam globally frozen in the ``singly magic'' case and both beams globally frozen 
in the ``doubly magic'' case.  The final column pulls out all the stops, taking advantage of the cancellation of
systematic error for $\Delta$ in the proton-He3 case. 
\begin{table}[htb]
\centering
\begin{tabular}{ccccc} \hline 
run sequencing        & consecutive           & concurrent             & concurrent            & $\Delta$    \\ \hline
                      & singly magic          & singly magic           & doubly magic          &  p-He3      \\
                      & $\pm\sigma_{\rm syst.}$ & $\pm\sigma_{\rm syst.}$ & $\pm\sigma_{\rm syst.}$ &             \\
                      & $e$\,cm               & $e$\,cm                & $e$\,cm                &  $e$\,cm   \\ \hline
single run            & $10^{-26}$             & $10^{-27}$             & $10^{-28}$              &            \\
polarization reversal & $0.5\times10^{-26}$    & $0.5\times10^{-27}$    & $0.5\times10^{-28}$     &            \\
CW/CW reversal        & $10^{-27}$             & $10^{-28}$             & $10^{-29}$              & $<10^{-29}$ \\ \hline
\end{tabular}     
\caption{\label{tbl:SystError} Systematic error estimates for various EDM measurements using the PTR storage ring.}
\end{table}

\section{Practical limits, uncertainties, issues and ackowledgements}
The content of this paper has been largely theoretical--- not theoretical particle physics, but theoretical
experimental physics.  What has been missing is practical experimental physics including much of
which is essentially electrical engineering.  Since mutual co-magnetometry depends on \emph{simultaneous} 
counter-circulating beams, accurate electric/magnetic superposition is essential. There are grounds for both 
optimism and pessimism in this area concerning the PTR design.

Experience with the AGS-Analogue electron ring, as described in considerable detail in references
\ \cite{RT-Instrumentation-paper}, \cite{EDM-Challenge}, and \cite{RT-JDT-AGS-Analogue}, provides grounds
for optimism concerning purely electric storage rings.  With a circumference of 43\,m, less than the
proposed PTR circumference in the ratio 43/102, the electron kinetic energy was 10\,MeV (which 
corresponds approximately with 20\,MeV proton kinetic energy, as regards achievable electric field value), 
the AGS-Analogue ring was built quickly, inexpensively and without much bother.  The cost of the ring 
itself was undoubtedly much smaller than the cost of the building housing it.  

Based only on the AGS-Analogue ring experience, the proposed PTR ring could be expected to meet
the requirements of the $p$, $d$, and $h$ examples spelled out in Section~\ref{sec:PracticalCombinations}.  

This is likely to remain true for PTR, especially if the voltage and current power supply circuits
remain simple.  For design purposes, in a ring with eight-fold super-periodicity, one could power each
octant or quadrant individually. This would make beam steering easy.  All electrodes would be powered in parallel,
all magnet coils in series.  Even simpler, if construction tolerances allow it, would be to power 
all ring elements from just two supplies.  Individual element powering would probably 
lead to serious and useless complication.

Less carefully documented is the evidence concerning electrical breakdown and, especially,
sparking ``catalyzed'' by the presence of weak local magnetic fields.  For PTR, \emph{for which superimposed electric 
and magnetic fields are required for best systematic EDM error reduction}, this is undoubtedly the area 
of greatest concern for simultaneous beam frozen spin functionality in PTR.

Experience seems to have shown that occasional sparking in high electric field bending elements
is unavoidable.  On the other hand, at least with short enough electrodes (i.e. of small capacitance) 
the breakdown causes no permanent damage. Ignoring possible damage to sensitive electronics, sparking rates 
less than one per day would only be a nuisance, but greater than one per hour would make quality data collection 
impossible. 

To reduce capacitance the length per construction unit will have to be quite short, perhaps one meter.
By butting construction lengths quite closely (perhaps a few mm separation, with resistive separation
spacers for registration) the electrode field uniformity degradation at transitions should be 
acceptably small.  

Further details concerning PTR and mutual co-magnetometry are contained in 
references\ 
\cite{COSY-stripper-explained},
\cite{Doubly-frozen},
\cite{RT-RAST}.
\cite{Septier},
\cite{Germain},
\cite{Germain-Rohrbach},
\cite{Germain-Tingley},
\cite{RohrbachElectricPrisms}.

At some cost in EDM systematic error coming from out-of plane spin precession 
(the so-called ``geometric phase problem'') counter-circulating orbit matching could be sequential, 
rather than simultaneous. Electric and magnetic bending could alternate on a quite short distance scale 
to preserve much the same frozen spin combinations as are possible with simultaneous beams. This would, 
however, require magnetic field reversal between runs in order to reverse the circulation direction..  
Spin tune phase locking would make this possible, as has been explained above, without the need for 
impractically precise magnetic field measurement.  This would also aggravate the problem of time 
dependent magnetic field error bursts---an important source of error also in the neutron EDM determination.

\noindent
{\bf Acknowledgments:\ }
All acknowledgments in the companion paper\ \cite{RT-ICFA} to this one apply to the present paper.  Special
acknowledgment to Hans Str\"oher, Frank Rathmann, Alex Nass, Kolya Nikolaev and my son John 
can be repeated; along with continuous support, all have made substantial contributions, suggestions and 
corrections to this paper.  Andras Laszlo provided helpful comments.  More significant yet have been the 
contributions of Christian Carli who, in a continuing series of communications, has found mistakes and 
formulated insightful questions that have invariably resulted in major correction, reorganization and 
clarification.  

\appendix

\section{Magnetic/electric field strength ratio constraints\label{sec:RatioConstraint}}
\subsection{Backward/forward (bf) ring symmetry}
A moving picture of particle orbits in a \emph{magnetic} storage ring, with a clock visible, can produce a record of
``forward'' evolution. By running the moving picture backwards one can produce a predicted time-reversed record. 
To test time-reversal symmetry one needs to establish initial orbit conditions that match the previous
final orbit conditions, but with reversed momenta (P-reversal), clock jiggered to run backwards (T-reversal), in
preparation for photographing the time-reversed evolution.  Traveling backward in the same magnetic field, the same 
particles would be deflected in the wrong way to retrace their paths. This could be remedied by reversing the particle 
charge, for example by switching (C-reversal) from protons to anti-protons. 
Because of the close connection between damping and T-conservation, the near absence of synchrotron radiation causes
protons to be favored over electrons in the present context.  The Fermilab Tevatron $p\bar p$ collider can therefore be 
said to provide the best  ``direct'' laboratory demonstration of CPT-symmetry---in this case, in the electromagnetic 
domain.  (Here ``direct'' is meant to exclude inferences reliant on elementary particle theory.) 

Because particles and anti-particles have opposite charge, a storage ring containing only magnetic elements
is automatically ``(C-bf)-symmetric'' (``C'' meaning ``charge'' and ``bf'' meaning ``backward-forward''), 
for particle-antiparticle colliding beams.  This is a very high level of symmetry in the sense that forward and 
backward orbits are identical even with off-axis orbits and arbitrarily-displaced, arbitrarily-imperfect, elements 
(provided, of course, that the imperfections are, themselves, purely magnetic). 

\subsection{Magnetic/electric element strength ratio constraint}
Like thoroughbred race horses, storage rings are ``high-strung''---meaning, tempermental, recalcitrant, etc.
We use instead the term ``finely-tuned'', with much the same meaning as intended when applied by particle 
theorists when discussing ``unnaturally accurate'' parameter equality, but with better 
etymological justification for the usage\ \cite{RT-scaling}. 

The CW ring functions properly, for example, only after its ``quadrupole knobs'' 
have been finely tuned. But having adjusted a quadrupole knob to stabilize the CCW beam might have 
destabilized the CW beam. This will complicate two-beam tune-up. 

The implications for the electric/magnetic storage ring proposed in this paper are therefore serious. 
For a start, to make forward and backward orbits identical, the ratio of electric to magnetic bending 
strengths have to be everywhere as identical as possible. Even then, as already explained, because ``constructive'' 
bending in one direction is ``destructive'' in the other direction, the beam particles of the counter-circulating 
beams have to differ, either in particle type or in (absolute value of) momentum. 

Another implication of the previous discussion is, ideally, that the ring quadrupoles should 
have superimposed magnetic and electric field strengths in the same ratio as for the bending elements. 
For the same reason, other correction elements, for example for beam steering, must have 
magnetic and electric field strengths in the same ratio.  This would include sextupoles as well
(should they be required) but their opposite spatial symmetry, along with their minuscule absolute 
strength, means that purely electric sextupoles should be adequate for tuning PTR chromaticities. 

Yet another implication is that, to the extent the particle velocities are relativistic,
effects, such as Lorentz contraction and time dilation, need to be accounted for. Fortunately, for the
EDM prototype ring, the betas are only about 0.3, so the effects are fairly small.  
The situation resembles an all-electric EDM ring viewed from a rotating frame of reference. In a rotating
frame an ``all-electric'' ring would have significant magnetic bending.
Until now such effects have been unimportant; they have not (nor need be) accounted for in particle tracking 
codes. But for counter-circulating beams of different velocity, the beams have to be treated individually. 

The purpose for the present section has been to explain how bf-symmetry can be restored by meeting one
(highly restrictive) condition: the ratio of electric to magnetic field strengths would ideally be the same 
everywhere---this means, especially, that the electric and magnetic bending field shapes have to be identical 
and be superimposed as exactly as possible. \emph{This applies also to steering corrections and focusing corrections.}
Deviations from meeting these constraints will impose the need for applying theoretically calculated 
(probably via simulation) corrections to measured EDM values.

\section{``Triplet compromise'' magnetic/electric quadrupole design\label{sec:Compromise}}
Ideally, the discrete lattice quadrupoles would have electric and magnetic fields superimposed in the
same ratio $\eta_E/\eta_M$ as in the bend regions.  In frozen-spin proton PTR application the magnetic 
bending is hardly ``perturbatively small'' compared to the electric bending; $\eta_M/\eta_E\approx1/3$.
This presents a serious design problem. 

\begin{figure}[hbt]
\centering
\includegraphics[scale=0.45]{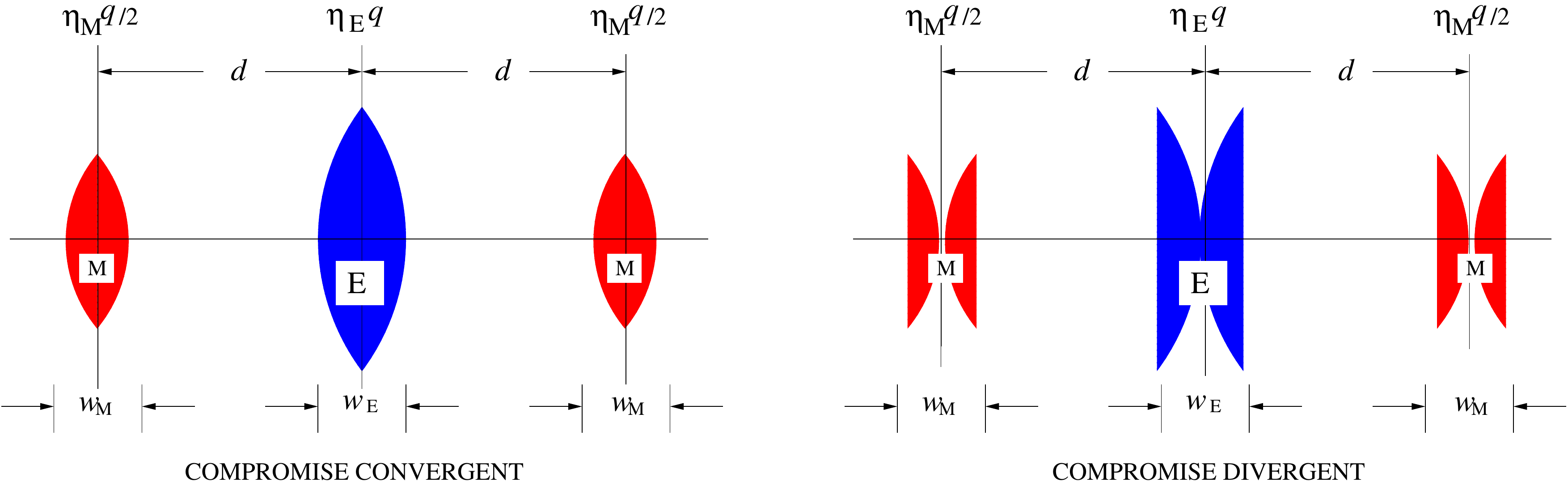}
\caption{\label{fig:BF-ReversibleQuad}''Compromise triplet'' representation of an electric quadrupole with (weak)
magnetic quadrupole superimposed. Blue quads are electric, red magnetic. For the primary beam all quads combine 
constructively; for the secondary (backward-traveling) beam the red signs reverse.  In thin element 
approximation, with the quad strengths shown, the triplet approximates a thin lens of strength $q=1/f$, with
(perturbatively small but) unequal forward and backward focal lengths $f$.  The purpose for this complication is 
to make backward and forward $\beta$-functions as nearly identical as possible.  For sufficiently small values of 
both $d$ and $\eta_M/\eta_E$ the triplet acts like a singlet quadrupole with weak magnetic field superimposed 
on strong electric field.}
\end{figure}

Clearly the coils producing the magnetic field need to be outside the electrodes producing the electric field.
Any scheme with (insulated) current carrying conductors situated between electrodes
is doomed to fail, since the electric and magnetic quads would be skew relative to one another.
Yet, the electric and magnetic quadrupole field reference radii would preferably be equal, for the field ratio 
to be constant over the full ring aperture. As a compromise an approximate match can be obtained by replacing 
each quadrupole by the symmetric quadrupole triplet arrangement shown in Figure~\ref{fig:BF-ReversibleQuad}.  
This provides the required match in thin-element approximation, though not in higher order approximation.
This provides the primary motivation for reducing lumped quadrupole strengths to the extent possible.

With each of the three quadrupoles forming the ``triplet'' on the left side of Figure~\ref{fig:BF-ReversibleQuad}
being ``focusing'', treated as a single quadrupole, its symbol would be ``F''.  Similarly, the ``triplet''
on the right, treated as a single quad,  symbolized as ``D''.

Because the thin element approximation is valid only perturbatively, and the $\eta_M/\eta_E\approx1/3$  ratio is not
negligible, the forward and backward optical functions can never be exactly equal.  This does not seriously impair the EDM 
determination, provided that both forward and backward optics are stable.  With quads situated at crossing points,
as in the design shown in Figure~\ref{fig:PTR-layout-Toroidal8_102p2-mod}, there should be comfortably broad bands 
of joint stability.

\section{``Hybrid'' EDM ring design issues\label{sec:Hybrid}}
\subsection{Inconvenient hybrid ring operational characteristics }
Ring designs with electric bending and magnetic focusing have come to be referred to 
as ``hybrid''\ \cite{HybridEDM}.  Assuming the central orbit for both CW and CCW orbits pass through 
the centers of every quadrupole (which is always at least approximately true) then CW and CCW central 
orbits are identical, irrespective of whether the quadrupoles are electric, or magnetic, or some
combination of both.

Horizontally focusing/defocusing quads are conventionally denoted by ``F''/''D''.
In a ring with electric bending and separated function magnetic FODO focusing, the transverse CW and 
CCW beam profiles are forward/backward symmetric but with horizontal maxima coincident with
vertical minima, and vice versa.  This does not mean, however, that horizontal 
and vertical beam widths are the same.  This is because there is ``geometric focusing'' which ``focuses'' 
horizontally but not vertically; $\beta_y$ maxima will coincide with $\beta_x$ minima 
alternating with                 $\beta_y$ minima coinciding with $\beta_x$ maxima. 
As a consequence the transverse shapes of backward and forward stored beams superimpose approximately 
only after relative longitudinal displacement by one half-cell. 

Once one has accepted the systematic EDM error coming from imbalanced lattice functions for CW 
and CCW beam shapes being different, some difficult operational problems remain.  Several 
modes of operation will be needed in a storage ring capable of EDM measurement. Ideally
there is a lattice matched to each of these modes, which include injection, initial 
commissioning, possible optimization for polarimetry, possible bunching and rebunching
and optimization for EDM measurement.

Here we consider only the most basic of these, which, during commissioning, is the
establishment of ``identical'' CW and CCW central orbits and, during data collection
is the exact preservation of central orbits.  Smoothing the CW and 
CCW design orbits individually can be performed using routine procedures.

Ideally (though often unnecessarily extravagant) for a ring with $N_c$ cells there would
be $N_c$ vertical BPMs situated at the locations of $N_c$ $\beta_y$ maxima, and    
$N_c$ horizontal BPMs situated at the locations of $N_c$ $\beta_x$ 
maxima\ \cite{RT-Universal}; these are points of maximum BPM sensitivity in each case.  
Typically there are fewer steering elements than BPMs, but here we assume there are the 
same numbers $N_c$ vertical and $N_c$ horizontal, optimally situated around the ring. 

For establishing the design orbit there are $2N_c$ measured BPM positions that need to be 
best fit, in a Gaussian least square sense, using the settings of the $2N_c$
steering adjusters.  For a single beam this is routinely achieved by satisfying $2N_c$
linear equations in $2N_c$ unknowns.  The coefficients in these equations are calculable
``influence functions'' $T_{ij}$ which, multiplied by adjuster ``$i$'' strength, produces
the displacement at detector ``$j$'' due to adjuster ``$i$''.  For a ring lattice 
(such as PTR) that is bf-symmetric, having solved these equations to find the smoothest 
possible CW orbit, the CCW orbit will be the same to good accuracy.

But, for a ring with magnetic quadrupoles, more conditions need to be met.  For a start 
one might assign $N_c$ of the adjusters to the CW beam and the remaining $N_c$ adjusters 
to the CCW beam.  With $N_c$ being a significantly large number, the CW closed orbit 
established by the CW adjuster set would produce a tolerably good closed CW orbit.  
Applied separately, the same procedure could be applied to produce a good CCW closed orbit.

Unfortunately, the influences of the CW and CCW adjuster sets are not independent.
Changing the CW adjusters will foul up the CCW orbit and vice versa.  This complicates 
the optimization terribly.  One can only seek the most favorable setting for every 
adjuster. Again a least squares fit can be sought. Two-way influence functions $T_{ij}^{CW}$
and $T_{ij}^{CCW}$ can be calculated.  Each coefficients of the least squares sum
would be the coherent sum of the $T_{ij}^{CW}$s and the $T_{ij}^{CCW}$s, each
multiplied by its corresponding adjuster ``$i$'' and summed over all ``$j$'' BPMs.

This complication is brought about by the hybrid choice having electric bends and magnetic
quadrupoles.  The PTR ring lattice optics favored for EDM measurement as described in the
present paper is backward/forward symmetric.  Provided that the closed orbit steering adjusters 
respect the $\eta_E/\eta_M$ field ratio as all other elements in the ring, the backward and forward 
beam steering will behave almost identically, and the forward and backward beta functions will be close 
in value. This has been demonstrated for PTR in Figure~\ref{fig:etaE-dependence}.

\subsection{Splitting the degenerate closed orbit fixed point in ``all-electric'' 
$p$ or $e$ EDM rings\label{sec:FixedPoint}}
``Appendix A: Existence of Equilibrium Orbits'' in the famous Courant and Snyder 
paper\ \cite{CourantSnyder}, that introduced alternating gradient (AG) accelerator optics, contains mathematics 
that is curiously rigorous for an accelerator physics paper.  This reflects a long heredity in
Mathematical Physics:  Ernest Courant, son of Richard Courant, student of Hilbert, colleague of Riemann, 
student of Gauss, all in G\"ottingen, Germany.  Apart from the well known Courant and Hilbert volumes,
Richard Courant is perhaps best known for his Hilbert space refinement of the 
``Dirichlet Principal''\ \cite{CourantDirichlet} which Courant and Snyder exploit as the mathematical basis 
for the long term stability of three dimensional closed orbits in magnetic accelerators. 
Their Appendix A proves that magnetic rings have stable 3D closed orbits.  Then, specializing
in Appendix B to a more detailed Hamiltonian description of orbits in rings with up/down symmetry, 
they demonstrate the stability of 2D orbits lying in a horizontal plane. 

By now it has been adopted as ``intuitively obvious'' that all storage rings have stable 3D orbits. 
For rings with both electric and magnetic bending this may or may not be true; generalization to
the assumption that equal-energy, globally-frozen spin, proton beams can counter-circulate simultaneously in
a 233\,MeV hybrid storage ring is a large and questionable step to take. Because of their opposite 
sign $\eta_M$ values, for the beams to simultaneously have vanishing spin tunes requires their $\gamma$
offsets to have compensating magnitudes.  Though challenging, one supposes that this condition can be 
met, perhaps by shifting their equilibrium synchrotron phases in a common RF cavity.  

The situation is somewhat different in the nominal, truly all-electric, ``nominal all-electric'' 233\,MeV proton 
beam (with electric quadrupoles) described in the CERN Yellow Report\ \cite{CYR} and in the present paper. 
Instead of quoting the proton spin tune at 233\,MeV as $Q_s=0$ one could as well express the spin tune as 
degenerate, $Q_s=\pm0$.  Because of imperfect magnetic shielding the $\eta_M$ values, as well as having 
opposite sign, will have (slightly) different magnitudes. This will split the degeneracy, producing a 
situation with one beam having slightly positive spin tune along with energy slightly in excess of the 
nominal magic 233\,MeV value.  For the other beam both of these shifts will be reversed. 

As explained in the CYR, because the zero spin tune contour passes through zero at the exact symmetry point, 
the sign reversals of both $Q_s$ and magnetic field offsets needed to preserve vanishing spin tune
occur at precisely the same point.  As a result, counter-circulating $p$ beams 
will satisfy the conditions for being doubly-magic in the presence of either intentional or unintentional 
magnetic bending.  In this way the $p,p$ combination, can meet the doubly-magic requirement of opposite sign
spin tunes.  Inevitably, though, the energies of CW and CCW beams will be slightly different.

This has established conditions under which counter-circulating $p,p$ beams can both have vanishing 
spin tunes, irrespective of whether the focusing is electric or magnetic, and of the fact that their
beam energies will have to be different.  A natural procedure, in any case, would seem to be to 
intentionally allow for the inclusion of significant intentional, adjustable magnetic bending.
This has always been contemplated as a possible ingredient of the ``nominal all-electric'' program.

\section{Parameters for a mutual co-magnetometer EDM ring in AGS or Wilson tunnel}
Table~\ref{tbl:full-scale-p-He3} contains parameters for a full-scale mutual co-magnetometry EDM ring
that would fit into the AGS tunnel with bend radius $r_0$=95.5\,m, matching 
the Omarov et al. reference\ \cite{HybridEDM} proposal. Such a ring could have basically the same 
``quadrupole-free'' design as PTR, but probably with super-periodicity of 16, and bending radius of 
approximately 95\,m.  This table contains some of the same parameters as are given in Table~\ref{tbl:p-He3}. 
As before, all entries on the left side of the table are ``magic'' with globally frozen, spin 
tune zero parameters but spin tunes on the right have not been converted to closest rational 
fraction required for (pseudo-) freezing the spins.  Repeating a previous statement, for phase-locking
beam spin tunes, as required for setting and resetting and stabilizing beam conditions, only pseudo-frozen spin 
tunes are required. 

The top row represents, for example, the doubly magic pairing of a forward $p$ beam and a backward $h$ beam.
The second row then provides the parameters of an $h$ beam that would travel in the forward direction in the
ring with the same electric and magnetic field values.  Precise measurement of this spin tune 
would provide a significant consistency check of the apparatus. The next two rows describe essentially the 
same situation as the first two rows, but with the helion beam treated as primary and the proton beam as secondary. 

The top two four-line blocks represent ``doubly-magic'' cases.  The following lines
represent generic situations in which the primary beam is ``magic''; though the other beam may be pseudo-frozen, its spin tune
cannot vanish.  The final two deuteron rows are provided both to acknowledge the special importance of measuring the deuteron EDM 
measurement, and indicate the disadvantageous kinematic conditions that result from the deuteron's small anomalous MDM. 
The deuteron case is ironic in the sense that, though the deuteron's quite small anomalous MDM makes it the baryon
closest to being an ideal Dirac fermion, result in kinematic parameters that make the simultaneous storage of counter-circulating 
deuteron kinematically difficult.  These comments are continued in Section~\ref{sec:Compromise-bending}.

Compared to PTR, the nominal all-electric full-scale ring bending radius is an order of magnitude greater.  The most important 
consequence of this is that the electric electric field is roughly an order of magnitude less.  One can
confidently expect that an electric field of this magnitude can have superimposed magnetic field
without electrical breakdown.
\setlength{\tabcolsep}{1pt}
\begin{table}[htb]\footnotesize 
\centering
\begin{tabular}{|c|ccccc|cc|ccccccc|c|} \hline 
    bm &     m1 & G1 &     q1 &  beta1 &    KE1 &     E0 &    B0 & m2 &     G2 &     q2  & beta2 &  KE2 & bratio & Qs2 &  bm \\ 
      1 &    GeV &        &    &        &    MeV &   MV/m & mT   & GeV &        &        &        &    MeV &        &        &      2 \\ \hline\hline
       & $r_0*=$ & 95\,m &  &        &      &       &       &                    &        &        &        &        &        & &        \\ \hline
     p & 0.9383 & 1.7928 &  1 & 0.27831 & 38.5941 & 0.5513 & 0.8659 & 2.8084 & -4.1842 &  2 & -0.16544 & 39.2396 & -0.59444 & 7.48827e-05\P\  &      h\\ 
     p & 0.9383 & 1.7928 &  1 & 0.27831 & 38.5941 & 0.5513 & 0.8659 & 2.8084 & -4.1842 &  2 & 0.22206 & 71.9095 &  0.79788 & -1.98943e+00 &      h\\ \hline
     h & 2.8084 & -4.1842 &  2 & 0.16544 & 39.2393 & 0.5513 & -0.8658 & 0.9383 & 1.7928 &  1 & -0.27830 & 38.5924 &  -1.6822 & -4.62057e-05\P\  &      p\\ 
     h & 2.8084 & -4.1842 &  2 & 0.16544 & 39.2393 & 0.5513 & -0.8658 & 0.9383 & 1.7928 &  1 & 0.19532 & 18.4265 &   1.1806 & -2.12302e+00 &      p\\ \hline \hline
   pos & 0.0005 & 0.0012 &  1 & 0.99986 & 30.0937 & 1.3311 & -1.0109 & 0.9383 & 1.7928 &  1 & -0.40242 & 86.6501 & -0.40247 & 1.45851e-04\P\  &      p\\ 
   pos & 0.0005 & 0.0012 &  1 & 0.99986 & 30.0937 & 1.3311 & -1.0109 & 0.9383 & 1.7928 &  1 & 0.31316 & 49.6955 &  0.31321 & -1.59219e+00 &      p\\ \hline
     p & 0.9383 & 1.7928 &  1 & 0.40238 & 86.6324 & 1.3309 & 1.0107 & 0.0005 & 0.0012 &  1 & -0.99986 & 30.0967 &  -2.4849 & 3.10250e-05\P\  &    pos\\ 
     p & 0.9383 & 1.7928 &  1 & 0.40238 & 86.6324 & 1.3309 & 1.0107 & 0.0005 & 0.0012 &  1 & 1.00000 & 223.1135 &   2.4852 & 5.06189e-01 &    pos\\ \hline \hline
     p & 0.9383 & 1.7928 &  1 & 0.31304 & 49.6542 & 0.7215 & 0.9337 & 0.9383 & 1.7928 &  1 & -0.22516 & 24.7291 & -0.71928 & -2.00000e+00 &      p\\ 
     p & 0.9383 & 1.7928 &  1 & 0.31304 & 49.6542 & 0.7215 & 0.9337 & 0.9383 & 1.7928 &  1 & 0.31304 & 49.6542 &        1 & -6.00000e-15 &      p\\ \hline
     p & 0.9383 & 1.7928 &  1 & 0.22516 & 24.7281 & 0.3457 & 0.7351 & 0.9383 & 1.7928 &  1 & -0.15313 & 11.1980 & -0.68009 & -2.27831e+00 &      p\\ 
     p & 0.9383 & 1.7928 &  1 & 0.22516 & 24.7281 & 0.3457 & 0.7351 & 0.9383 & 1.7928 &  1 & 0.22516 & 24.7281 &        1 & -8.00000e-15 &      p\\ \hline \hline
     d & 1.8756 & -0.1430 &  1 & 0.18000 & 31.1438 & -0.1115 & 4.2136 & 1.8756 & -0.1430 &  1 & -0.02383 & 0.5326 & -0.13237 & -5.93955e-01 &      d\\ 
     d & 1.8756 & -0.1430 &  1 & 0.18000 & 31.1438 & -0.1115 & 4.2136 & 1.8756 & -0.1430 &  1 & 0.18000 & 31.1438 &        1 & -3.00000e-15 &      d \\ \hline \hline
\end{tabular}
\caption{\label{tbl:full-scale-p-He3}Kinetic parameters for various pairs of polarized beams circulating
in a tunnel designed initially for an AGS having 809\,m circumference. Entries marked \P\ 
in the final column can be tuned exactly to zero.
Double horizontal lines separate matched, four set solutions of Eq.~(\ref{eq:AbbrevFieldStrengths.3-rev}).  
The doubling of solutions is explained in companion paper\ \cite{RT-ICFA}.  
It occurs because $E_0$ appears only as $E_0^2$ in Eq.~(\ref{eq:AbbrevFieldStrengths.3-rev}).
\emph{Every second row in this table shows co- (not contra-) rotating cases---beta1 and beta2 are both positive
in these rows.} i.e. beams of two different momenta can travel in the same direction on the same circular trajectory---for the
higher/lower momentum the bending is constrctive/destructive.
\emph{Reversing the sign of magnetic field $B_0$ (with $E_0$ fixed) would reverse the stable circulation directions 
and interchange constructive/destructive bending. }}
\end{table}

Though reversing the sign of $B_0$ introduces new solutions, reversed-magnetic field solutions obviously cannot apply 
to simultaneously counter-circulating beams.  \emph{Reversed-sign magnetic field  solutions can apply only to consecutive,
not concurrent, runs.}  
The two four-line blocks at the top of Table~\ref{tbl:full-scale-p-He3} represent the only ``doubly magic'' cases known.
The $p,p$ case is ``generic'' in that only the primary beam can be magic, but there is a continuum of singly magic 
counter-circulating cases, two of which are demonstrated in the following four-line blocks.  In each of these cases the primary 
beam is frozen with spin tune zero, so its EDM signal accumulates monotonically, and can therefore be measured,
for example by interpolating measurements taken with Koop wheel rolling forward and backward, as explained in the CYR\ \cite{CYR}.
The bottom two ``deuteron rows'' are repeated in Table~\ref{tbl:CompromiseBending}, which is discussed further in 
Appendix~\ref{sec:Compromise-bending}.  Note that, like all paired rows in this table, both deuteron beams are
traveling in the same direction (though not (typically) at the same time). 

Table~\ref{tbl:full-scale-CESR} contains parameters for a full-scale mutual co-magnetometry EDM ring with 85\,m bending radius 
(matching the Wilson Laboratory tunnel, for which the CESR circumference is 764\,m). This table differs from the previous 
table by displaying $\eta_E$ and $pc/q$ values.  For rough purposes the difference in bending radius for this and 
Table~\ref{tbl:full-scale-p-He3} is not very important.  The ``etaE'' and ``pc/q'' values are useful for estimating the 
significance of the different focusing behavior of counter-circulating beams.  In particular, for the critically important 
proton/helion case, the relevant p1c/q1 ratios are 0.2719 and -0.2356. Differing in sign only because the $p$ factor has
opposite sign, their approximately equal magnitudes indicate that forward and backward optics will be correspondingly close. 
This has been confirmed in Figure~\ref{fig:etaE-dependence}.  As explained previously, and confirmed in the same figure,
the vertical beta functions will be identical for proton an He3 beams.

The order of magnitude momentum imbalance is considerably more serious for the 
doubly-magic proton/positron case.  This, and the absence of positron polarimetry makes the $p,e^+$ EDM difference
measurement problematic.

\setlength{\tabcolsep}{1pt}
\begin{table}[htb]\footnotesize 
\centering
\begin{tabular}{|c|ccccc|cc|ccccccc|c|} \hline 
    bm &     m1 & G1 &     q1 &  etaE1 & p1c/q1 &     E0 &    B0 & m2 &     G2 &     q2  & etaE2 & p2c/q2 & bratio & Qs2 &  bm \\ 
      1 &    GeV &        &    &        &    GeV &   MV/m &  mT  & GeV &        &        &        &    GeV &        &        &      2 \\ \hline
       & $r_0*=$ & 85\,m &  &        &      &       &       &                    &        &        &        &        &        &  &       \\ \hline
     p & 0.9383 & 1.7928 &  1 & 0.69584 & 0.2719 & 0.6194 & 0.9728 & 2.8084 & -4.1842 &  2 & 1.35106 & -0.2356 & -0.59444 & 7.48827e-05\P &      h\\ 
     h & 2.8084 & -4.1842 &  2 & 1.35103 & 0.2356 & 0.6194 & -0.9728 & 0.9383 & 1.7928 &  1 & 0.69585 & -0.2719 &  -1.6822 & -4.62057e-05\P &    p\\ 
   pos & 0.0005 & 0.0012 &  1 & 4.15491 & 0.0306 & 1.4956 & -1.1358 & 0.9383 & 1.7928 &  1 & 0.76593 & -0.4124 & -0.40247 & 1.45851e-04\P &      p\\ 
     p & 0.9383 & 1.7928 &  1 & 0.76596 & 0.4124 & 1.4953 & 1.1355 & 0.0005 & 0.0012 &  1 & 4.15386 & -0.0306 &  -2.4849 & 3.10250e-05\P &    pos\\ 
\hline
\end{tabular}
\caption{\label{tbl:full-scale-CESR}Kinetic parameters for doubly-magic counter-circulating beams in a
tunnel comparable in size to that in Table~\ref{tbl:full-scale-p-He3} with 85\,m bending radius 
(matching the Wilson Laboratory tunnel, for which the CESR circumference is 764\,m).  Entries marked \P\ 
in the final column can be tuned exactly to zero.
}
\end{table}

\clearpage

\section{Experimental investigation of Aharonov-Anandan spin tune shift\label{sec:Aharonov-Anandan} }
The abstract to a paper ``Phase Change during a Cyclic Quantum Evolution'' by 
Aharonov and Anandan\ \cite{Aharonov-Anandan} announced a ``new geometric phase 
defined for any cyclic evolution of a quantum system'' and continued with the statement that ``this phase 
factor is a  gauge-invariant generalization of one found by Berry''.  Paper\ \cite{Aharonov-Anandan} is reprinted,
along with many similarly-motivated papers, in reference\ \cite{Shapere-Wilczek}.

Continuing to quote from Aharonov and Anandan, their first paragraph reads: ``A type of evolution 
of a physical system which is often of interest in physics is one in which the state of the system 
returns to its original state after an evolution.  We [meaning Aharonov and Anandan] shall call this a cyclic 
evolution.  An example is periodic motion such as the precession of a particle with intrinsic spin and 
magnetic moment in a constant magnetic field.  Another example is the adiabatic evolution of a quantum 
system whose Hamiltonian $H$ returns to its original value and the state evolves as an eigenstate of the 
Hamiltonian and returns to its original state.  A third example is the splitting and recombination of 
a beam so that the system may be regarded as going backwards in time along one beam and returning along 
the other beam to its original state at the same time.''

With one minor change, the proposed PTR ring matches all three of these types of 
phase evolution.  The first example needs to have ``constant magnetic field'' made more explicit by 
its replacement by ``constant magnetic field in the particle rest frames''.  This seems harmless.  
A frozen spin beam propagating for billions of turns in a time independent storage ring is 
an example of the second type.  The third Aharonov-Anandan example also fits the PTR configuration of 
beams following time reversed orbits as the ``cyclic evolution of a quantum system''. 

In their Eq.~(3),
\begin{equation}
\beta \equiv \phi + h^{-1} \int_0^t \langle\psi(t)|H|\psi(t)\rangle dt,
 \label{eq:Aharonov-Anondan.3}
\end{equation}
Aharonov and Anandon distinguish a dynamical part $\beta$ from a phase part $\phi$.  Then, 
combining Schroedinger picture and Heisenberg picture equations, and defining $\phi$ by 
\begin{equation}
|\psi((\tau)\rangle = e^{i\phi} |\psi((0)\rangle,
 \label{eq:Aharonov-Anondan.def}
\end{equation}
they obtain 
\begin{equation}
\beta = \int_0^t \langle\bar\psi| i(d|\bar\psi\rangle)/dt) dt,
 \label{eq:Aharonov-Anondan.4}
\end{equation}
as the formula for the ``phase independent part'' $\beta$ of the evolution.

As explained in companion reference\ \cite{RT-ICFA}, it is useful to distinguish between Schroedinger
and Heisenberg pictures even in the purely classical mechanics analysis of storage ring
orbits.  As commented upon in the long footnote in Section~\ref{sec:Wollnik} of the
present paper, a significant phase ambiguity can arise in the description of transverse (i.e. betatron) 
oscillations in a storage ring.  This ambiguity exposes itself, for example, in the comparison
of results obtained in Schroedinger and Heisenberg pictures.  In the footnote ``blame'' for the 
phase ambiguity is placed on the aliasing ambiguity present in any trigonometric expression for a ``phase angle''.

Though expressed as an aliasing issue in classical physics, it seems clear that an equivalent phase
ambiguity makes its appearance in the  quantum mechanical axiom, as stated by Hermann 
Weyl\ \cite{Weyl-QM}, ``All assertions concerning the 
probabilities in a given state $\mathfrak{x}$ are numerically unaltered when $\mathfrak{x}$ 
is replaced by $\epsilon\mathfrak{x}$, where $\epsilon$ is an arbitrary complex number of 
modulus 1''; ``we cannot distinguish between these two cases''.

Berry phases are usually quantified by anholonomic angular phase advances acquired during a single complete 
traversal around a closed path. In a storage ring the same path is traversed billions of times. 
In accelerator jargon the precession rate is expressed as spin tune $Q_s$, which is the precession
occurring each turn as a fraction of $2\pi$.  This makes it appropriate to express excess precession 
of beam polarization as a ``shift $\delta Q_s$ in spin tune''.

In the case of counter-circulating beams in PTR such a tune shift is readily measurable with high 
precision as the \emph{difference between counter-circulating spin tunes}. Formulas for spin 
tunes $Q_s^E$ and $Q_s^M$ have been given previously in Eqs.~(\ref{eq:BendFrac.7}).  
Any measurably large deviation from these equations would represent deviation from current theory.

\section{Worst case scenarios: failure risks from fundamental physics or cost\label{sec:FailureRisk}}
As recommended in the CERN Yellow Report\ \cite{CYR} (CYR), the fundamental reason for building a prototype EDM
ring (referred to here as PTR) was to investigate and reduce the risk of complete failure of the nominal
all-electric EDM proton EDM ring.  Such failure could result from fundamental physics issues or from excessive
cost.  Though these issues are tightly coupled, they are discussed individually in the next two subsections.

\subsection{``Compromise bending'' for deuteron EDM measurement\label{sec:Compromise-bending}}
A risk that is deathly serious is the possibilty that the presence of magnetic bending field (of whatever magnitude) 
superimposed on electrical bending field will lead inexorably to unacceptably high sparking breakdown of the electric field.  
Most of what is known about sparking
breakdown for such electric/magnetic superposition was learned at various labs, including CERN, before 1965, in connection 
with the design of electrostatic beam separators.  Probably correlated with poor vacuum, major advances have
been made  in this area in the meantime.  Though primary beam separation in such separators is due to electric field, 
to cause separated beams to suffer no net bending requires (very weak) compensating magnetic bending. These issues are 
discussed in References~\ \cite{Septier} through\ \cite{RohrbachElectricPrisms}.

It has been shown in Appendix~\ref{sec:Hybrid} that, because of their magnetic quadrupoles, even the ``all-electric'' hybrid 
EDM ring design will require a more or less uniform superimposed magnetic field to cancel the net bending associated with 
establishing a smooth closed orbit. (Without this compensation CW and CCW proton beam energies could not be
simultaneously globally frozen, because of their slightly different energies.)

Lattices with separate electric and bending elements, intended for deuteron EDM measurement have been investigated at COSY, 
especially by Yuri Senichev\ \cite{Senichev-private}.  As it happens, it is understanding of complications in the measurement of 
the deuteron EDM that makes it easiest to understand the compromises imposed by ``compromise bending'' (i.e. physically 
separated magnetic and electric bending elements). 

The caption to Table~\ref{tbl:full-scale-p-He3} promised later discussion of the deuteron rows, which are now repeated in 
Table~\ref{tbl:CompromiseBending}, though with more convenient column headings.
\setlength{\tabcolsep}{1pt}
\begin{table}[htb]\footnotesize 
\centering
\begin{tabular}{|c|ccccc|cc|ccccccc|c|} \hline 
    bm &     m1 & G1 &     q1 &  etaE1 & p1c/q1 &     E0 &    B0 & m2 &     G2 &     q2  & etaE2 & p2c/q2 & bratio & Qs2 &  bm \\  \hline
     1 &    GeV &        &    &        &    GeV &   MV/m &    mT & GeV &        &        &        &    GeV &        &        &      2 \\ \hline
       & $r_0*=$ & 95\,m &  &        &      &       &       &                    &        &        &        &        &        &  &       \\ \hline
     d & 1.8756 & -0.1430 &  1 & -0.17243 & 0.3432 & -0.1253 & 4.7341 & 1.8756 & -0.1430 &  1 & 0.52631 & -0.0447 & -0.13237 & -5.93955e-01 &      d \\ 
     d & 1.8756 & -0.1430 &  1 & -0.17243 & 0.3432 & -0.1253 & 4.7341 & 1.8756 & -0.1430 &  1 & -0.17243 & 0.3432 &        1 & -4.00000e-15 &      d \\ 
\hline
\end{tabular}
\caption{\label{tbl:CompromiseBending}Kinetic parameters for counter-circulating deuteron beams in a tunnel of the same 
size as in Table~\ref{tbl:full-scale-p-He3} with 95.5\,m bending radius, matching the AGS tunnel, for which the 
circumference is 809\,m. 
These deuteron rows are typical of all (or most) atomic nuclei for which only one or the other beam can have vanishing spin tune. 
But the kinematic beam conditions are especially hard to meet in the deuteron case because the ratio, ``bratio'', of frozen spin 
beam velocities, like the ratio of energies, is so different from unity.  As an aside, one notes though, because of the large 
bending radius of the AGS tunnel, that all electric and magnetic field values are conveniently small, which likely prevents 
superimposed electric/magnetic bending from being a ``show stopper''.}
\end{table}
One sees from the small electric bending fraction ``etaE1'' (i.e. $\eta_E=-0.17243$) that global freezing of deuteron 
spins requires a ring with primarily magnetic bending---opposite to freezing proton spins, which requires primarily electric
bending.  Referring again to the lattice layout shown in Figure~\ref{fig:PTR-layout-Toroidal8_102p2-mod} one sees that, with
red sectors representing magnetic bending, and (much shorter) blue sectors representing electron bending, that the
spins of a circulating deuteron beam could be frozen by making the ratio of blue/red sector angles equal to 0.17243/0.82757.
Since these sectors are completely disjoint, this ``compromise bending'' would avoid the need for superimposed electric/magnetic 
bending. 

The reason the title for this section includes ``worst case'' is that such a ring could be used only for deuteron EDM 
measurement---otherwise the requirement for the closed orbit to remain circular would prevent adjusting the beam spin tune.
Even for deuteron EDM measurement there is a defect to segregating electric and magnetic bending.  This defect was clearly explained in
the original Farley et al. paper\ \cite{Farley}.  (Slightly edited for brevity) they state 

``\ 
\noindent
 It might be supposed that the electric and magnetic
fields could be applied to separate sections of the orbit,
with the result that the spin makes small to and fro
movements about the vertical axis but the net
precession is zero over one turn. While this would fulfill
the main requirement, some misalignment errors would
not be perfectly canceled. For example, a harmonic of the
azimuthal field B would cause the spin to
oscillate about the horizontal axis parallel to p.
Because rotations do not commute, 
the combination would generate a net rotation about
the radial axis, leading to a false EDM. This is an 
example of Berry phase. Similarly, a misalignment of
the electric field would generate an oscillating
radial angular velocity as explained previously. The combined
 oscillation would give rise to a false EDM. To minimize these effects, 
the electric and magnetic fields must be located at the same place. 
This error is also canceled by injecting CW and CCW.
``
\subsection{Need for ultra-magnetic shielding and/or cryogenic vacuum?\label{sec:MagShielding-and-vacuum}}
For brevity, and because they are covered in the CYR\ \cite{CYR}, many important technical issues concerning 
the PTR design have not even been mentioned in the present paper.  As recommended in the CYR, the PTR was to be
as inexpensive as possible, consistent with establishing whether ultra-magnetic shielding and/or cryogenic 
vacuum would be required for the nominal all-electric proton EDM ring\ \cite{HybridEDM}.  As one consequence PTR
design assumes only minimal magnetic shielding.

In the two-plus years since completion of the CYR, both PTR and nominal all-electric designs have diverged
significantly.  As presently understood, the PTR circumference will be roughly eight times smaller than
the full scale all-electric ring.  Its cost can be estimated, therefore, to be eight times less.  However, 
this fails to account for the excess cost of cryogenic vacuum and/or ultra-magnetic shielding, either of
which could double the cost.  In this sense the PTR cost can perhaps be assessed as twenty times less than
the full scale ring.   

There has been a more significant divergence between the PTR 
(quadrupole-free, combined function focusing)\footnote{The absence of quadrupoles in the PTR design 
enters here only by chance.  Quadrupoles are eliminated from PTR primarily to maximize spin coherence time
by adopting thick lens optics copied from STEM (scanning transmission electron miscopropy); 
a topic completely unrelated to the current discussion.}
and the ``Hybrid'' (electric bending, separated function, magnetic focusing) full scale proton EDM ring.

The presence of absence of magnetic quadrupoles enters the present discussion only because of
the influence of magnetic quadrupoles on the cost of magnetic shielding of a full scale proton EDM ring.
Though not cheap, it might be practical to magnetically shield a continuous toroidal tube containing 
a truly all-electric ring.  But the presence of stray magnetic fields from the current leads needed to power 
the 100, or so, separated-function quadrupoles would seriously complicate the ultra-magnetic shielding. 
This might, for example require 100 separate ``magnetically shielded rooms'' one for each bend. 
For reducing magnetic fields to the miniscule level needed to render negligible the spurious, MDM-induced,
EDM mimicking, out of plane precession, would be very expensive.

The contention of the present paper is that phase-locked spin control methods, and concentrating on
the doubly-magic proton-He3 case, can avoid the need for ultra-magnetic shielding, while investigating
BSM with sensitivity as good or better than proposed by the nominal all-electric design in the 
Omarov\ \cite{HybridEDM} paper.


\begin{thebibliography}{99}
\bibitem{RT-ICFA}     
R. Talman, \emph{Superimposed Electric/Magnetic Dipole Moment Comparator Lattice Design,} 
ICFA Beam Dynamics Newsletter \#82, Yunhai Cai, editor, Oct, 2021

\bibitem{RT-Positron-q}          
R. Talman, \emph{Is a positron really an electron moving backward in
time? proposed experimental investigation using a
two way toroidal focusing storage ring,} in preparation, available upon request

\bibitem{RT-LHC2}    
R. Talman, \emph{How to double the LHC energy}, in preparation, available upon request

\bibitem{Farley}
F. J. M. Farley et al., \emph{New Method of Measuring Electric Dipole Moments in Storage Rings,}
Phys. Rev. Lett. 93, 5, 2004

\bibitem{Anastassopoul}
D. Anastassopoulos et al., {Search for a permanent electric dipolemoment of the deuteron nucleus 
at the $10^{-27}\,e$ cm level,} AGS proposal, 2011

\bibitem{CYR}     
CPEDM, \emph{Storage ring to search for electric dipole moments of charged particles
Feasibility study, CERN Yellow Reports: Monographs,} CERN-2021-003, 2021

\bibitem{PSI-neutron}    
C. Abel et al., \emph{Measurement of the Permanent Electric Dipole Moment of the Neutron,}
Phys. Rev. Lett. 124, 081803

\bibitem{PTR-Bad-Honnef}      
R. Talman, on behalf of CPEDM, \emph{Lattice Design for a Proton EDM Prototype
Storage Ring (PTR),} WE-Heraeus-Seminar: Towards Storage Ring Electric Dipole Moment Measurements,
Bad Honnef, 2021

\bibitem{Koop-different-particles}            
I. Koop, Asymmetric energy colliding ion beams in the EDM storage ring, Proc. 4th Int. Particle
Accelerator Conf. (IPAC 2013), Shanghai, Ed. Z. Dai et al. (JACoW Conferences, Geneva),
p. 1961, 2013

\bibitem{APS-FSU}     
R. Talman, on behalf of CPEDM, \emph{Comagnetometry Measurement of Electric and
Magnetic Dipole Moments,} APS, DPF2021 FSU Meeting, 2021

\bibitem{Kolya-Snowmass}     
N.N. Nikoliev et al., \emph{Test of the Standard Model and Search for Physics Beyond,} 
Letter of Interest submitted to RF3: Fundamental Physics in Small Experiments, Snowmass planning, 2020 

\bibitem{Hempelmann}      
N. Hempelmann et al., \emph{Phase-locking the spin precession in a storage ring,} P.R.L.
119, 119401, 2017

\bibitem{Eversmann}            
D. Eversmann et al., \emph{New method for a continuous determination of the spin tune
in storage rings and implications for precision experiments,}
Phys. Rev. Lett. {\bf 115} 094801, 2015

\bibitem{Gabrielse-eEDM}     
G. Gabrielse, \emph{The standard model's greatest triumph,} 
Physics Today, p.~24, December, 2013

\bibitem{Koop-SpinWheel}       
I. Koop, \emph{Spin wheel---a new method of suppression of spin decoherence in the EDM storage rings, }
{\tt http://collaborations.fz-juelich.de\
/ikp/jedi/public-files/student\_seminar/ \\
SpinWheel-2012.pdf}

\bibitem{Obukhov:2016vvk}       
Yuri~N. Obukhov, Alexander~J. Silenko, and Oleg~V. Teryaev.
\newblock {Manifestations of the rotation and gravity of the Earth in
  high-energy physics experiments}.
\newblock {\em Phys. Rev.}, D94(4):044019, 2016.

\bibitem{Silenko:2004ad}       
Alexander~J. Silenko and Oleg~V. Teryaev.
\newblock {Semiclassical limit for Dirac particles interaction with a
  gravitational field}.
\newblock {\em Phys. Rev.}, D71:064016, 2005.

\bibitem{OrlovFlanaganSemertzidis}       
Y. Orlov, E. Flanagan, and Y. Semertzidis.
\newblock {Spin Rotation by Earth's Gravitational Field in a "Frozen-Spin"  Ring}.
\newblock {\em Phys. Lett.}, A376:2822--2829, 2012.

\bibitem{Laszlo-Zimboras}       
A. L\' aszl\'o and Z. Zimbor\' as, \emph{Quantification of GR effects in muon g-2, EDM and
other spin precession experiments,}  Class.Quant.Grav 35, 175003, 2018

\bibitem{Silenko:2006er}       
Alexander~J. Silenko and Oleg~V. Teryaev.
\newblock {Equivalence principle and experimental tests of gravitational spin
  effects}.
\newblock {\em Phys. Rev.}, D76:061101, 2007.

\bibitem{Silenko:2015jqa}       
Alexander~J. Silenko.
\newblock {Comparison of spin dynamics in the cylindrical and Frenet-Serret
  coordinate systems}.
\newblock {\em Phys. Part. Nucl. Lett.}, 12(1):8--10, 2015.

\bibitem{Khriplovich:1997ni}       
I.~B. Khriplovich and A.~A. Pomeransky.
\newblock {Equations of motion of spinning relativistic particle in external
  fields}.
\newblock {\em J. Exp. Theor. Phys.}, 86:839--849, 1998.
\newblock [Zh. Eksp. Teor. Fiz. 113, 1537 (1998)].

\bibitem{Pomeransky:2000pb}       
A.~A. Pomeransky, R.~A. Senkov, and I.~B. Khriplovich.
\newblock {Spinning relativistic particles in external fields}.
\newblock {\em Phys. Usp.}, 43:1055--1066, 2000.
\newblock [Usp. Fiz. Nauk 43, 1129 (2000)].

\bibitem{BMT}       
V. Bargmann, L. Michel, and V.L. Telegdi, Phys. Rev. Lett. {\bf 2}, 435, 1959

\bibitem{Wolski}       
A. Wolski, \emph{Beam Dynamics in High Energy Particle Acceleratorsa,} Imperial College Press, 2014

\bibitem{ConformalHandbook}       
P.K. Kythe, \emph{Handbook of Conformal Mappings and Applications,} CRC Press, 2019

\bibitem{AlbrechtToroidal}        
R. Albrecht, \emph{Das Potential in doppelt gekrümmten Kondensatoren,} 
Zeitschrift für Naturforschung A, {\bf 11a}, 156, 1956

\bibitem{Alarcron-EDM}
R. Alarcron et al., \emph{Electric dipole moments and the search for new physics,} 
arXiv, hep-ph > arXiv:2203.08103

\bibitem{ToffolettoLeckeyRiley}        
F. Toffoletto et al., \emph{ Design Criteria for an Angle Resolved Electron
Toroidal Geometry,} Nucl. Instr. and Meth. in Phys. B12 282-297, 1985

\bibitem{CourantSnyder}       
E. Courant and H. Snyder, \emph{Theory of the Alternating Gradient Synchrotron,} 
Annals of Physics, Vol. 3, No. 1, 1958

\bibitem{CourantDirichlet}       
R. Courant, \emph{Dirichlet's Principal,} Interscience Publishers, Inc., 1950

\bibitem{Erni-TEM}      
R. Erni, \emph{Aberration-Corrected Imaging in Transmission Electron Microscopy,} 
Imperial College Press, 2015 

\bibitem{Good}       
B.H. Good, \emph{Classical Equations of Motion for a Polarized Particle in an Electromagnetic Field,}
Phys. Rev., {\bf 125}, 6, 2112, 1962

\bibitem{Magiera}       
A. Magiera, Phys. Rev. Accel. Beams 20 094001. https://doi.org/10.1103/PhysRevAccelBeams.20.094001, 2017

\bibitem{Nehari}       
Zeev Nehari, \emph{Conformal Mapping,} McGraw-Hill Book Company, INC., p. 113, 1951

\bibitem{Wollnik}        
H. Wollnik, \emph{Optics of Charged Particles,} Academic Press Inc. Orlando, FL, USA, 1987

\bibitem{Lebedev}       
A. Lebedev, \emph{500\,m Electric Ring: IBS and Ring Parameters,} COSY Juelich report, December 9-10, 2013

\bibitem{EDM-Challenge}       
R. Talman, \emph{The Electric Dipole Moment Challege,} IOP Publishing, 2017

\bibitem{TaharCarli}
M. Tahar and C. Carli, \emph{Spin tracking studies for EDM,} Unpublished report EDM\_Aug28\_MC.pdf, to CPEDM
group meeting 28 August, 2018 

\bibitem{Haissinski}       
J. Haissinski, \emph{The Orsay Electron-Positron Storage Ring (Status Report),} in \emph{Physics with
Intersecting Storage Rings}, InternationalSchool of Physics ``Enrico Fermi'', 1971

\bibitem{ChromaticitySharing}       
S. Henderson, R. Talman, et. al., \emph{Investigation of Chromaticity Sharing at the Cornell Storage Ring,}
      Proc. of Vancouver PAC, 1997

\bibitem{RoundBeams}       
E, Young, R. Talman, et. al., \emph{Collisions of Resonantly Coupled Round Beams at CESR,}
     Proc. of Vancouver PAC, 1997

\bibitem{RT-Instrumentation-paper}       
M. Plotkin, \emph{The Brookhaven Electron Analogue, } 1953-1957 BNL Report 45058,
December, 1991

\bibitem{RT-JDT-AGS-Analogue}       
R. Talman and J, Talman, \emph{Symplectic orbit/spin tracking code for all-electric storage rings,}
Phys. ReV. Accel. Beams, {\bf 18}, 074003 

\bibitem{COSY-stripper-explained}        
H.J. Stein, et al., \emph{Electron Cooling at COSY-JUELICH, } {\tt www.researchgate.net/publication/48198877,} January 2011

\bibitem{Doubly-frozen}       
R. Talman, \emph{A Doubly-Magic Storage Ring EDM Measurement Metthod,} arXiv:1812.05949, December, 2018

\bibitem{RT-RAST}              
R. Talman, \emph{Prospects for Electric Dipole Moment Measurement Using Electrostatic Accelerators,} Reviews of Accelerator Science and Technology, 
A. Chao and W. Chou, editors, Volume 10, 2018

\bibitem{Septier}
A. Septier, \emph{Focusing of Charged Particles, Vol II,} Academic Press, 1967

\bibitem{Germain}
C. Germain, CERN 59-38, 1959

\bibitem{Germain-Rohrbach}
C. Germain and F. Rohrbach, CERN 63-17, 1963

\bibitem{Germain-Tingley}
C. Germain and R. Tingley Nucl. Instr. Methods, {\bf 20}, 21, 1963

\bibitem{RohrbachElectricPrisms}
F. Rohrbach, \emph{Private communication to A. Septier}.

\bibitem{Senichev-private}
Y. Senichev, private communication.

\bibitem{RT-scaling}
R. Talman, \emph{Scaling Behavior of Synchrotron Radiation Dominated Circular Colliders,} 
World Scientific Journal, IJMPA, Vol. 30, 1544003, 2015

\bibitem{HybridEDM}
Z. Omarov et al., \emph{Comprehensive Symmetric-Hybrid ring design for pEDM experiment at below}1.0e-29\,$e$\,cm, 
Phys. Rev. D, {\bf 105}, 032001, 2022

\bibitem{RT-Universal}
R. Talman, \emph{A Universa Algorithm for Accelerator Correction,} in ``Advanced Beam Dynamics 
Workshop on Effects of Errors in Accelerators, their diagnosis and Correction'',
Conference Proceedings No. 255, Corpus Christi TX, Alex Chao editor AIP, 1991 

\bibitem{Aharonov-Anandan}
Y. Aharonov and J. Anandan, \emph{Phase change during a cyclic quantum evolution}, PRL {\bf 58}, 6, 1987

\bibitem{Shapere-Wilczek}
A. Shapere and F. Wilczek, \emph{Geometric Phases in PHysics,} World Scientic Publishing Co., 1989

\bibitem{Weyl-QM}
H. Weyl, \emph{The Theory of Groups and Quantum Mechanics,} Section II.7; Dover Publications, 1956,
English translation of \emph{Gruppentheorie und Quantenmechanik}, 1931 

\end{thebibliography}
\end{document}